\documentclass[11pt,a4paper]{article}              

\usepackage[margin=25mm]{geometry}
\usepackage{graphicx}
\usepackage{colortbl}
\usepackage{subfig}
\usepackage{fancyhdr}
\usepackage{amssymb,amsfonts,amsmath,amsthm}
\usepackage{natbib}
\usepackage{bstnotations}
\usepackage{algorithm}
\usepackage{algpseudocode}
\usepackage{float}
\usepackage{caption}
\newcommand{\fc}{{\mathfrak c}}

\newcommand{\fw}{f_{\ve{W}}}

\newcommand{\I}{\mathbbm{1}_{\cd_f}}
\newcommand{\dd}{\textrm{d}}

\usepackage{authblk}
	
\definecolor{airforceblue}{rgb}{0.36, 0.54, 0.66}
\definecolor{blush}{rgb}{0.87, 0.36, 0.51}

\definecolor{ballblue}{rgb}{0.13, 0.67, 0.8}

\definecolor{bondiblue}{rgb}{0.0, 0.58, 0.71}
\definecolor{bostonuniversityred}{rgb}{0.8, 0.0, 0.0}
\definecolor{brickred}{rgb}{0.8, 0.25, 0.33}
\definecolor{deepcerise}{rgb}{0.85, 0.2, 0.53}
\definecolor{gold(metallic)}{rgb}{0.83, 0.69, 0.22}
\definecolor{indiagreen}{rgb}{0.07, 0.53, 0.03}

\definecolor{mypink}{rgb}{1.0000,0.8824,0.9961}
\definecolor{myblue}{rgb}{0.8824,0.9882,1.00000}

\graphicspath{{./},{./Figures/}}

\begin{document}
\title{High-dimensional reliability-based design optimization using stochastic emulators  } 

\author[1]{M. Moustapha} \author[1]{B. Sudret}

\affil[1]{Chair of Risk, Safety and Uncertainty Quantification,
  
  ETH Zurich, Stefano-Franscini-Platz 5, 8093 Zurich, Switzerland}

\date{}
\maketitle

\abstract{%
Reliability-based design optimization (RBDO) is traditionally formulated as a nested optimization and reliability problem. Although surrogate models are generally employed to improve efficiency, the approach remains computationally prohibitive in high-dimensional settings. This paper proposes a novel RBDO framework based on a stochastic simulator viewpoint, in which the deterministic limit-state function and the uncertainty in the model inputs are combined into a unified stochastic representation. Under this formulation, the system response conditioned on a given design is modeled directly through its output distribution, rather than through an explicit limit-state function.

Stochastic emulators are constructed in the design space to approximate the conditional response distribution, enabling the semi-analytical evaluation of failure probabilities or associated quantiles without resorting to Monte Carlo simulation. Two classes of stochastic emulators are investigated, namely generalized lambda models and stochastic polynomial chaos expansions. Both approaches provide a deterministic mapping between design variables and reliability constraints, which breaks the classical double-loop structure of RBDO and allows the use of standard deterministic optimization algorithms.

The performance of the proposed approach is evaluated on a set of benchmark problems with dimensionalities ranging from low to very high, including {\color{black} cases} with stochastic excitation. The results are compared against {\color{black} a Kriging-based approach formulated in the full input space and a heteroscedastic Gaussian process model defined solely in the design space.} The proposed method yields substantial computational gains, {especially for problems characterized by high-dimensional random inputs.} While its efficiency is comparable to Kriging for low-dimensional problems, it significantly outperforms Kriging as the dimensionality increases. {\color{black} Finally, unlike heteroscedastic Gaussian process models, GLaM and SPCE do not rely on restrictive assumptions about the conditional response distribution, making them applicable to a broader class of stochastic simulators.} \\[1em] 

}

\maketitle

\section{Introduction}
Design optimization is a crucial step in engineering, which allows for a more efficient use of resources in a context where they are increasingly scarce and valuable. Its importance is further amplified by the inherent uncertainty of the environment in which engineering systems operate, requiring designers to make informed decisions despite incomplete knowledge. Numerous frameworks have been developed to optimize systems while accounting for uncertainties. This paper focuses on reliability-based design optimization (RBDO), where a cost is minimized while ensuring that the probability that the system fails to safely operate is kept below a prescribed threshold.

The safety of an engineering system is commonly assessed through the definition of a limit state, which represents the boundary between safe and failed system behavior. In practice, this concept is formalized through a limit-state function, whose sign indicates whether the system satisfies its performance requirements. The limit-state function in turn depends on a computational model that represents the system response and on a set of uncertain parameters accounting for variability in design and environmental conditions, e.g., manufacturing tolerances, loads or material properties. Within this probabilistic framework, reliability analysis aims at quantifying the probability that the system is in a failed state, commonly referred to as the \emph{probability of failure} \citep{Melchers2018}. Reliability-based design optimization extends this framework by incorporating reliability analysis within the design process, such that the probability of failure is evaluated for each candidate design considered during the search.

The classical formulation of RBDO adopts a two-level approach, in which reliability analyses are nested within a deterministic optimization problem. In this framework, an optimizer explores candidate design variables in an outer loop, while for each {\color{black} design} the associated failure probability is evaluated in an inner loop. Computing these probabilities generally requires solving a full reliability analysis, which can be computationally demanding, particularly when high-fidelity models or high-dimensional inputs are involved. Common reliability analysis techniques include approximation methods such as the first-order reliability method (FORM) \citep{Hasofer1974}, as well as simulation-based approaches ranging from direct Monte Carlo simulation to more advanced variance-reduction techniques, such as importance sampling \citep{Rubinstein2016,Papaioannou2016,Wang2016} and subset simulation \citep{Au2001,Papaioannou2015,Au2016b}.

To reduce the computational burden associated with the nested formulation, three main classes of approaches have been proposed in the literature \citep{Chateauneuf2008}. The first class replaces simulation-based reliability analyses in the inner loop with approximation techniques such as FORM or inverse FORM. This strategy has led to well-known formulations, including the reliability index approach (RIA, \citet{Nikolaidis1988}) and the performance measure approach (PMA, \citet{Tu1999}). While these methods substantially reduce computational costs, their accuracy deteriorates for strongly nonlinear limit-state functions or in the presence of complex uncertainty structures.

A second class of approaches seeks to reformulate the RBDO problem itself into a more tractable form, or into a sequence of simpler problems. Single loop approaches eliminate the explicit computation of failure probabilities by replacing probabilistic constraints with optimality conditions, hence solving the RBDO problem within a single optimization loop. Decoupled loop methods, on the other hand, transform the original nested RBDO formulation into a sequence of deterministic optimization problems, in which optimization and reliability assessment are performed alternately. Representative methods in this category include the sequential optimization and reliability assessment (SORA, \citet{Du2004}) and the single loop approach (SLA, \citet{Chen1997,Liang2004}). Comprehensive reviews of these methods can be found in \citet{Aoues2010,MoustaphaSMO2019}.

The third class of approaches, which has proven particularly effective for complex problems, relies on surrogate modeling. In this framework, inexpensive approximations of the limit-state function are constructed and used within the reliability analysis loop. A wide variety of surrogate models have been employed, including polynomial response surfaces \citep{Agarwal2004}, Gaussian process models (a.k.a Kriging) \citep{Dubourg2011,MoustaphaSMO2016,Thompson2025}, polynomial chaos expansions \citep{Zhou2019AIAA,Lee2022}, support vector machines \citep{Boroson2017,Ling2021}, and artificial neural networks \citep{Thedy2023,Lee2025}. More recently, machine learning and deep learning–based surrogates have been introduced to address increasingly complex RBDO problems \citep{Li2022,Zhang2024}.

Surrogate-based RBDO methods have been extensively investigated, and several reviews highlight their effectiveness \citep{Valdebenito2010,MoustaphaSMO2019}. Among the various surrogate models, Kriging appears to be the most widely used approach, largely due to its built-in uncertainty measure, which naturally enables the development of active learning strategies. These strategies, where the surrogate is built adaptively by smartly selecting training points through a learning function, have been widely adopted in reliability analysis and subsequently extended to RBDO. \citet{Thompson2025} recently reviewed and benchmarked popular learning functions used in RBDO.

Following the classification proposed by \citet{MoustaphaSMO2019}, surrogate-based active learning approaches can be broadly divided into local and global approximation strategies. In \emph{local approximation} approaches, multiple surrogate models are constructed locally around moving design points encountered during the optimization process \citep{Zhang2017,Zhang2021}. In contrast, \emph{global approximation} approaches rely on a  single surrogate model built in an augmented reliability space and used to perform reliability analyses for the various designs explored during optimization \citep{Kim2021,Park2023}. Global approaches are generally more efficient than local ones. However, it remains challenging to construct a surrogate that accurately represents the limit-state surface over the entire augmented space. 

The difficulty substantially increases in high-dimensional settings. Kriging, for instance, is known to suffer from the curse of dimensionality, with performance degrading significantly when the number of input parameters exceeds, say $20$. To address this issue, several methods have been proposed. \citet{Jia2013} proposed reducing the input and output dimensionality for wave and surge modeling in hurricane assessment using principal component analysis prior to constructing a Kriging approximate. \citet{Li2019} combined Kriging with high-dimensional model representation (HDMR), decomposing the limit-state function into low-order component functions to mitigate dimensionality effects. \citet{Li2022} proposed a dimensionality reduction strategy based on autoencoders, mapping the high-dimensional input space into a low-dimensional latent space on which a Kriging surrogate is constructed.

A particularly challenging class of high-dimensional problems arises in RBDO under stochastic excitations, as commonly encountered in earthquake and wind engineering. In such settings, the structural response depends not only on uncertain design and material parameters, but also on complex random processes representing the excitation. A comprehensive review of existing methodologies in this context is provided by \citet{Jerez2022}. However, the approaches surveyed therein do not rely on surrogate modeling, which remains relatively uncommon in this context. The few surrogate-based strategies that have been proposed generally adopt a common principle, that is, for a fixed set of design variables, the system response is modeled through its conditional distribution, which captures the sources of uncertainty associated to the stochastic excitation. This conditional perspective provides a natural way to reduce the effective dimensionality of the problem, while retaining the essential probabilistic features of the response.

An early contribution in this direction is due to \citet{Clark2020}, who proposed a nondeterministic Kriging framework to approximate the conditional structural response at fixed {\color{black} support} points. Their approach assumes Gaussian input uncertainties and a response that is linear or nearly linear with respect to the random variables. Although an extension to non-Gaussian inputs is proposed, it still requires partial reliance on Monte Carlo simulation. Building on a similar conceptual framework, \citet{Xiao2022} approximated the engineering demand parameter of structures subjected to earthquake excitation using two Gaussian process models, one for the conditional mean and one for the conditional variance of the response. More recently, \citet{Kim2024} proposed an RBDO framework in which the structural response is expressed solely as a function of basic random variables directly linked to the design parameters. By assuming a lognormal conditional response for a given design, they constructed a heteroskedastic Kriging model in the design space, enabling the evaluation of conditional failure probabilities without resorting to Monte Carlo simulation. 

Related but conceptually distinct approaches focus on reformulating the reliability problem itself. \citet{Jiang2024} addressed computational challenges for structures under random excitations by introducing a mapping between an operator norm and the reliability index, thereby transforming the RBDO problem into a deterministic optimization problem. This idea was proposed earlier by \citet{Faes2020}, who formulated the problem as the direct minimization of the failure probability.

{\color{black}Finally, in a closely related performance-based setting, \citet{Rodrigues2026} proposed a noisy Kriging metamodel for performance-based risk optimization, in which the conditional distribution of the engineering demand parameters is approximated at each support design while explicitly accounting for the statistical uncertainty arising from the limited number of ground motions used in nonlinear time-history analyses. This conditional modeling again enables the expected consequences (e.g.,\ repair costs) to be optimized at a fraction of the computational cost of a full Monte Carlo treatment.}

In this paper, we propose an approach that is conceptually related to surrogate-based RBDO methods in which conditional response distributions are approximated in a reduced design space. The proposed framework, however, is formulated for general high-dimensional RBDO problems and is not tied to a specific class of uncertainties such as stochastic excitations. In addition, no parametric assumptions are made on the type of the conditional response distribution, e.g., that they are Gaussian or lognormal, since the latter are rarely encountered in realistic engineering problems. As observed in case studies developed for wind turbine simulation \citep{Zhu2020}, earthquake engineering \citep{ZhuPEM2023}, or tornado simulation \citep{Kroetz2026}, the shape of the conditional response distribution may vary significantly across the input design space.

The main contributions of this work are twofold. First, we introduce the framework of stochastic simulators, which provides a general modeling perspective for constructing surrogates that approximate conditional distributions rather than scalar responses. Second, we propose two stochastic emulators that not only approximate these conditional distributions efficiently for different designs, but also enable the semi-analytical evaluation of failure probabilities, thereby significantly reducing the computational cost of RBDO.

The remainder of the paper is organized as follows. Section~\ref{sec:ProblemFormulation} formulates the RBDO problem. Section~\ref{sec:RBDOStoSimu} introduces the concept of stochastic simulators and shows how the RBDO problem can be reformulated through their incorporation. Section~\ref{sec:RBDOStoEmu} presents two stochastic emulators, namely stochastic polynomial chaos expansions and generalized lambda models, and demonstrates how they can be used to efficiently solve {\color{black} RBDO problems with high-dimensional stochastic input spaces}. Section~\ref{sec:Applications} validates the proposed approach on a set of benchmark problems, ranging from low-dimensional examples for visualization purposes to high-dimensional cases involving stochastic processes in the input space. The efficiency of the proposed framework is compared with that of a Kriging-based surrogate constructed using a static experimental design in an augmented reliability space.

\section{Problem formulation}\label{sec:ProblemFormulation}

We consider the reliability-based design optimization (RBDO) problem of the form \citep{Dubourg2011}:
\begin{equation}\label{eq:RBDO_pf}
	\ve{d}^\ast = \arg \min_{\ve{d} \in \mathbb{D}} \fc \prt{\ve{d}} \quad \text{subject to: } 
	\left\{ \begin{array}{ll}
		\mathfrak{f}_j \prt{\ve{d}} \leq 0, \quad &  j = 1,\dots,n_s, \\[0.5em]
		\Prob{g_k \prt{\ve{X}\prt{\ve{d}},\ve{Z}} \leq 0} \leq \bar{p}_{f_k}, \quad & k = 1,\dots,n_h,
	\end{array} \right.
\end{equation}
where $\ve{d} \in \mathbb{D} \subset \mathbb{R}^{n_{\ve{d}}}$ are design parameters associated to the cost function $\fc$ to be minimized under two types of constraints. The first ones, $\mathfrak{f}_j$, are deterministic functions defining the feasible design space through simple analytical functions, e.g., assembly or geometric constraints. The second ones are probabilistic constraints accounting for different sources of uncertainty that affect the system. They depend on a set of random variables which can be split into two categories: $\ve{X}\prt{\ve{d}}$ are random variables associated to the design parameters, which model for instance manufacturing tolerances around nominal design dimensions (length, thickness, etc.). $\ve{Z}$ are environmental variables that directly impact the response of the system but cannot be controlled by the designer, such as loadings or variability in material properties. The probabilistic constraint restricts the failure probability of each identified failure mode $k$ to remain below a target threshold $\bar{p}_{f_k}$. Gathering the random parameters in the vector $\ve{W} = \prt{\ve{X}(\ve{d}),\ve{Z}}$, where $\ve{W} \in \mathbb{R}^{n_{\textrm{tot}}}$ denotes the concatenation of $\ve{X}\prt{\ve{d}}$ and $\ve{Z}$, the failure probability associated with a given limit-state function $g_k$ can be written as
\begin{equation}\label{eq:Pf_final}
	p_{f_k}\prt{\ve{d}} = \Prob{g_k\prt{\ve{X}\prt{\ve{d}},\ve{Z}} \leq 0} = \int_{\mathcal{D}_f} \fw\prt{\ve{w}}\, \textrm{d}\ve{w},
\end{equation}
where $\mathcal{D}_f = \acc{\ve{w} \in \Xx \times \Zz: g_k\prt{\ve{w}} \leq 0}$ represents the failure domain and $\fw$ is the joint distribution of $\ve{W}$ given a particular value of $\ve{d}$. 

Equivalently, the probabilistic constraints can be expressed in terms of quantiles, leading to the quantile-based RBDO formulation \citep{MoustaphaSMO2016}:
\begin{equation}\label{eq:RBDO_q}
	\ve{d}^\ast = \arg \min_{\ve{d} \in \mathbb{D}} \fc \prt{\ve{d}} \quad \text{subject to: }
	\left\{ \begin{array}{ll}
		\mathfrak{f}_j\prt{\ve{d}} \leq 0, \quad &  j = 1,\dots,n_s, \\[0.5em]
		Q_{\alpha_k}\prt{\ve{d}; g_k} \leq 0, \quad & k = 1,\dots,n_h,
	\end{array} \right.
\end{equation}
with $\alpha_k = \bar{p}_{f_k}$ and
\begin{equation}
	Q_{\alpha_k}\prt{\ve{d}; g_k} = \inf\acc{q \in \mathbb{R}: \Prob{g_k \prt{\ve{X}(\ve{d}),\ve{Z}} \leq q} \geq \alpha_k }.
\end{equation}
For convenience, we drop the subscript $k$ in the remainder of this paper.

Surrogate-assisted techniques are among the most efficient approaches to solve this problem. They are generally used in a two-level scheme consisting of an outer optimization loop and an inner reliability analysis loop. However, these approaches suffer from two fundamental limitations.

The first limitation arises because surrogate models are usually built in an augmented random variable space, which is often high-dimensional. In such cases, surrogate models face the curse of dimensionality and quickly become inefficient. Active learning can mitigate this issue, but only when the dimensionality remains small to medium (say, $n_{\textrm{tot}} \approx 1–20$). Beyond this range, traditional surrogates, such as Kriging, become ineffective.

The second limitation is the computational cost associated to repeated reliability analyses, which remains significant even when surrogates are employed. Kriging, arguably the most widely used surrogate in RBDO, is particularly slow when evaluated on large sample sets, and even more so as the experimental design grows. Moreover, reliability analysis in surrogate-assisted RBDO often relies on Monte Carlo simulation. Estimating small failure probabilities therefore requires very large sample sizes. When compounded with the fact that reliability analyses must be repeated hundreds or even thousands of times, the overall computational cost of solving an RBDO problem becomes prohibitive.

We propose a novel RBDO solution scheme that addresses these two limitations. The main idea is to reformulate the deterministic limit state $g_{\textrm{det}}\prt{\ve{X},\ve{Z}}$ as a stochastic limit state $g_{\textrm{sto}}\prt{\ve{d}}$ with latent variables, which will then allow us to break the double loop. By constructing suitable approximations of the resulting stochastic simulator, the failure probability can be estimated without resorting to costly Monte Carlo simulations, thereby expediting the design process. 
Moreover, in certain cases, this reformulation helps mitigate the curse of dimensionality by enabling the construction of the surrogate model in a reduced design space, making the proposed approach particularly suitable for high-dimensional RBDO problems.

\section{RBDO formulation using stochastic simulators}\label{sec:RBDOStoSimu}

\subsection{Stochastic simulators in short}
Most simulators used in practice are deterministic by nature, meaning they yield the same output quantity of interest (QoI) when they are evaluated multiple times at the same input. This is typically the case in RBDO, where uncertainty in the output stems solely from randomness in the input parameters. In contrast to such deterministic simulators, stochastic simulators possess an intrinsic source of variability which yield different outputs when the simulator is evaluated multiple times. They can be defined as follows \citep{LuethenThesis,ZhuThesis}:
\begin{equation} 
	\begin{split} 
		\cm_s : D_{\ve{X}} \times \Omega \quad \to \quad & \mathbb{R}, \\
		(\ve{x}, \omega) \quad \mapsto \quad & \cm_s(\ve{x}, \omega),
	\end{split}
	\label{eq_map_definition_stochastic_models}
\end{equation}
where $\ve{x} \in D_{\ve{X}} \subset \Rr^{n_{\ve{X}}}$ is an $n_{\ve{X}}$-dimensional input vector and $\omega$ is a random event in a probability space $(\Omega, \mathcal{F}, \mathbb{P})$ that
captures the inherent stochasticity of the simulator. For a fixed $\ve{x} \in \cd_{\ve{X}}$, the response
$Y_{\ve{x}} = \cm_s\prt{\ve{x}, \cdot}: \; \Omega \to \Rr$ is a random variable. More precisely, for a given input vector
$\ve{x}_0$, each run of the simulator corresponds to a different $\omega_i$ and produces a realization
$\cm_s(\ve{x}_0, \omega_i)$.

In practice, the intrinsic stochasticity  is represented explicitly by introducing a vector of latent random variables $\ve{\Lambda}$ of dimension $n_{\ve{\Lambda}}$. With this formulation, the stochastic simulator can be seen as a deterministic simulator $\cm_d$ acting on both the physical inputs $\ve{x}$ and the latent variables {\color{black} $\ve{\Lambda}$}:
\begin{equation}
	\label{eq:002}
	Y_{\ve{x}} = \cm_s(\ve{x}, \omega) \eqdef \cm_d(\ve{x} , \ve{\Lambda} (\omega)).
\end{equation}
Examples of such simulators are numerous in the literature. In earthquake engineering, for instance, ground motion models can be formulated as stochastic simulators. Macroscopic parameters such as magnitude, distance, and site conditions define the overall intensity and frequency content. Yet even for fixed values of the latter, the resulting ground motion remains inherently random. This variability is modeled through a latent random vector, often made of hundreds to thousands of standard normal variables, so that each realization yields a different yet statistically consistent ground motion record. For wind turbine design, so-called wind boxes are generated to represent a three-dimensional wind velocity field over, say $10$ minutes, based on a handful of wind climate parameters (mean velocity, turbulence intensity, shear exponent) and again thousands of normal variables.

\subsection{Reformulation of the RBDO problem using stochasitc simulators}

We reformulate the RBDO problem by recasting the deterministic limit-state function as a stochastic simulator whose only explicit inputs are the design parameters:
\begin{equation}
	g\prt{\ve{X}(\ve{d}),\ve{Z}} = g\prt{\ve{X}(\omega)\mid \ve{d},\ve{Z}(\omega)} \equiv g_s(\ve{d}; \omega),
\end{equation}
where $\omega$ denotes latent random effects.

\begin{figure}[!ht]
	\centering
	\includegraphics[width=0.8\textwidth]{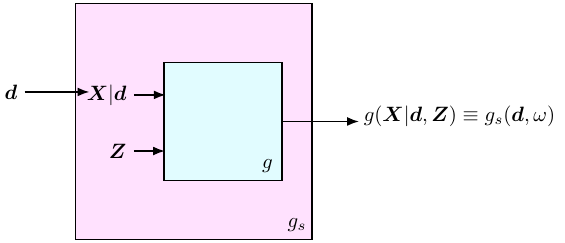}
	\caption{Schematic illustration of the \emph{stochasticization} transform. The original deterministic limit-state function is represented by the blue box with two random inputs $\ve{X}|\ve{d}$ and $\ve{Z}$. The stochastic limit-state function is schematized in pink and encapsulates both the deterministic limit-state and its random inputs.}
	\label{fig:RBDO_StoSIm}
\end{figure}

Figure~\ref{fig:RBDO_StoSIm} illustrates the transformation further called \emph{stochasticization}.  The original deterministic limit-state $g$ is represented by the blue box and has two random input vectors, namely $\ve{X}|\ve{d}$ and $\ve{Z}$. In contrast, the stochastic limit-state $g_s$, represented by the pink box, encapsulates these sources of uncertainty within the simulator itself. As a result, its only explicit input is the deterministic design vector $\ve{d}$.

\begin{figure}
	\centering
	\includegraphics[width=1.0\textwidth]{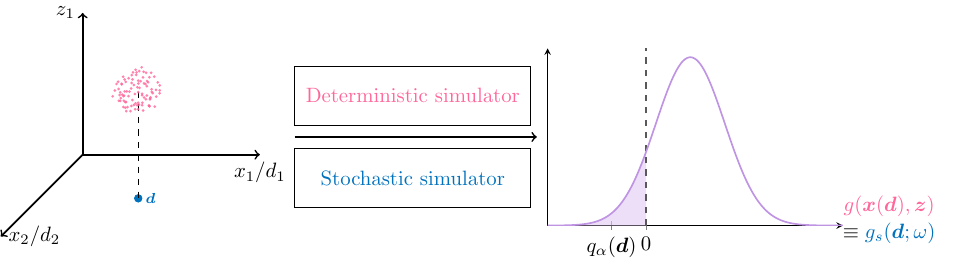}
	\caption{Uncertainty propagation using a deterministic simulator versus a stochastic simulator.}
	\label{fig:StoEmuRBDO}
\end{figure}

Through this reformulation, the uncertainty originally associated with the random inputs $\ve{X}|\ve{d}$ and $\ve{Z}$ is transferred to the latent space represented by $\omega$. When this latent space is made explicit through a variable $\Lambda$, the latter effectively plays the role of both $\ve{X}|\ve{d}$ and $\ve{Z}$. The resulting response, i.e., the limit-state output for a given design, remains random but is now characterized by a stochastic simulator whose randomness is entirely controlled by the latent variables. This is illustrated on a three-dimensional example in Figure~\ref{fig:StoEmuRBDO}, where the input consists of two design parameters $\ve{d} = \prt{d_1,\,d_2}$, their associated random variables $\ve{X} = \prt{X_1,\,X_2}$ and a single environmental variable $Z_1$. In a standard approach, uncertainties are generally propagated by first generating realizations of the three-dimensional random vector $W \sim f_{\ve{X}|\ve{d}} \times f_{Z_1}$, represented by the pink cloud of points in the left panel, and then evaluating the deterministic simulator on these points. Instead, our approach bypasses this Monte Carlo step entirely as we evaluate the deterministic design point (blue) directly with a stochastic simulator that internally embeds the effect of the uncertainty in $\ve{W}$. Both approaches are theoretically equivalent in terms of the induced response distribution.

In the context of RBDO, however, the full distribution is not required. Instead, the sole quantity of interest is the failure probability (or an associated quantile) corresponding to a given design $\ve{d}^{(i)}$:
\begin{equation}\label{eq:Pf_1Df}
	p_f\prt{\ve{d}^{(i)}} 
	= \Prob{g\prt{\ve{X}(\ve{d}^{(i)}), \ve{Z}} \leq 0} = \int_{\mathbb{R}^{n_{\textrm{tot}}}}  \I(\ve{x}, \ve{z}) f_{\ve{X}\mid \ve{d}^{(i)}}\prt{\ve{x}} f_{\ve{Z}}\prt{\ve{z}} \, \dd\ve{x}\, \dd\ve{z}. 
\end{equation}

Traditionally, estimating Eq.~\eqref{eq:Pf_1Df} requires Monte Carlo or other advanced simulation techniques applied to the deterministic limit-state. For each candidate design, uncertainty is propagated through repeated evaluations of $g$, which can become computationally demanding when the simulator is expensive.

In the stochastic simulator formulation, the random response is represented as $g_s\prt{\ve{d}^{(i)}; \omega}$, where $\omega$ encapsulates all sources of uncertainty. Consequently, the conditional distribution of the limit-state response, and therefore the associated failure probability, is entirely characterized by the distribution of the latent variables. 

By expressing the failure probability in Eq.~\eqref{eq:RBDO_pf} in terms of the stochastic simulator, the integration over the original variables $\ve{X}$ and $\ve{Z}$ is implicitly replaced by an integration in the latent space. We thus obtain the following equivalent formulation of the RBDO problem:
\begin{equation}\label{eq:RBDO_StoSim}
	\ve{d}^\ast = \arg \min_{\ve{d} \in \mathbb{D}} \fc \prt{\ve{d}} \quad \text{subject to: } 
	\left\{ \begin{array}{ll}
		\mathfrak{f}_j \prt{\ve{d}} \leq 0, \quad &  j = 1,\dots,n_s, \\[0.5em]
		\mathbb{P}_{\omega}\prt{g_{s_{k}} \prt{\ve{d}; \omega} \leq 0} \leq \bar{p}_{f_k}, \quad & k = 1,\dots,n_h,
	\end{array} \right.
\end{equation}

The main difficulty, however, is that $g_s \prt{\ve{d}; \omega}$ and the corresponding failure probability are generally not available in closed form. If an internal sampling mechanism for the latent variables $\ve{\Lambda}$ exists, these quantities can in principle be obtained by Monte Carlo simulation in the latent space. However, this would require extensive evaluations of the stochastic simulator. Because such simulators are often computationally expensive, directly estimating the conditional distribution or the associated failure probability in this manner is also impractical.

Instead, we resort to surrogate models specifically designed for stochastic outputs, known in the literature as \emph{stochastic emulators}. When these emulators are chosen appropriately, conditional failure probabilities and quantiles can even be obtained semi-analytically, as we will demonstrate in the next section. This constitutes precisely one of the key advantages of our approach, i.e., expensive simulation methods such as Monte Carlo can be replaced by inexpensive semi-analytical evaluations of the failure probability.

Another major advantage of the proposed approach is dimensionality reduction. Using a traditional surrogate-based approach with a deterministic simulator, the surrogate model is generally built in an augmented space of dimension $n_{\textrm{tot}} = n_{\ve{d}} + n_{\ve{z}}$, where $n_{\ve{d}}$ and $n_{\ve{z}}$ are the dimensions of $\ve{X}$ (or $\ve{d}$) and $\ve{z}$, respectively. In contrast, the stochastic emulator is constructed on the design space of dimension $n_{\ve{d}}$. The dimensionality is therefore reduced by $n_{\ve{z}}$, which can be $\mathcal{O}\prt{10^{-2}-10^{3}}$ in applications involving random fields or time series, such as wind or earthquake engineering. For instance, in some applications related to the analysis of structures subject to stochastic earthquake excitation, representing ground motion time-series would require hundreds of Gaussian random variables. When these are combined with design parameters and additional environmental variables (e.g., material properties), the overall dimensionality can become prohibitively large, easily exceeding $n_{\textrm{tot}}>100$. Building accurate surrogate models in such a high-dimensional input space is extremely challenging. In particular, when time series are involved, specific methods such as nonlinear autoregressive (NARX) models are necessary \citep{Billings2013,SchaerMSSP2024,schaer_fnarx,schaer_mnarxp}. With the stochastic simulator approach, however, the emulator is built solely on the design variables, which is relatively small in most applications, usually $n_{\ve{d}} <10$. Although constructing a stochastic emulator is generally more data-intensive than building a deterministic surrogate, once $n_{\ve{z}}$ becomes large enough, the difficulty of training the stochastic emulator is outweighed by the challenge of building an accurate deterministic surrogate model in a high-dimensional augmented input space.

\section{Stochastic emulators}\label{sec:RBDOStoEmu}
In this paper, we consider two types of stochastic emulators that have proven effective for reliability analysis \citep{PiresSS2025b,PiresSS2025a}, namely the generalized lambda models (GLaM) \citep{ZhuSIAM2021} and the stochastic polynomial chaos expansions (SPCE) \citep{ZhuStoPCE2023}. They are now briefly reviewed. 

\subsection{Generalized lambda models}\label{sec:GLaM}
Generalized lambda models (GLaM) are surrogate models for stochastic simulators whose goal is to approximate the conditional distribution of the output $Y_{\ve{d}} = g_s\prt{\ve{d};\omega}$. They rely on the generalized lambda distribution (GLD), a highly flexible parametric family capable of representing a wide range of unimodal distributional shapes, including usual ones such as Gaussian, Weibull, and uniform distributions, as well as many others that are not captured by classical parametric families. The GLD is defined through its quantile function, i.e., the inverse of the cumulative distribution function. Among the several existing parametrizations, GLaM adopts the widely used Freimer--Kollia--Mudholkar--Lin (FKML) formulation, whose quantile function for $u\in \bra{0,\,1}$ reads:
\begin{equation}\label{eq:GLDquantile}
	Q\prt{u ; \ve{\lambda}} = \lambda_1 + \frac{1}{\lambda_2}\prt{\frac{u^{\lambda_3}-1}{\lambda_3}-\frac{(1-u)^{\lambda_4}-1}{\lambda_4}},
\end{equation}
where $\lambda_1$ is the location parameter, $\lambda_2>0$ is the scale parameter, and $\lambda_3$ and $\lambda_4$ are shape parameters related to skewness and kurtosis, respectively. 

In the GLaM framework, it is assumed that the conditional distribution of the simulator output follows a generalized lamba distribution. More specifically, for each input $\ve{d}$, we model
\begin{equation}
	Y_{\ve{d}} = g_s\prt{\ve{d};\omega} \sim \textrm{GLD}\prt{\lambda_1\prt{\ve{d}},\, \lambda_2\prt{\ve{d}},\, \lambda_3\prt{\ve{d}},\, \lambda_4\prt{\ve{d}}}
\end{equation}
By evaluating the stochastic simulator at several {\color{black} points} $\ve{d}$ and estimating the associated GLD parameters, we obtain a dataset from which we can construct a mapping $\ve{d} \;\longmapsto\; \ve{\lambda}(\ve{d})$. Each component of this mapping is then approximated using polynomial chaos expansions (PCE). Specifically, for $l = \acc{1,\, 2, \,4}$, we write
\begin{equation}
	\lambda_l\prt{\ve{d}} \approx \lambda_l^{\textrm{PC}}\prt{\ve{d};\ve{c}} = \sum_{\ve{\beta} \in \mathcal{B}_l} c_{l,\ve{\beta}}\psi_{\ve{\beta}}\prt{\ve{d}}, 
\end{equation}
while for the scale parameter $\lambda_2$, the expansion is constructed in the log-space to enforce positivity:
\begin{equation}
	\lambda_2\prt{\ve{d}}  \approx \lambda_2^{\textrm{PC}}\prt{\ve{d};\ve{c}} = \exp\prt{\sum_{\ve{\beta} \in \mathcal{B}_2} c_{2,\ve{\beta}}\psi_{\ve{\beta}}\prt{\ve{d}}}.
\end{equation}
Here, $\psi_{\ve{\beta}}\prt{\ve{d}}$ denotes the multivariate polynomial basis functions associated to the multi-index $\ve{\beta}$ and  $\mathcal{B}_l$ is the corresponding truncation set defining the polynomials retained in the expansion of $\lambda_l$. Further details on polynomial chaos expansions and their construction can be found in \citet{Xiu2002,Sudret2015a}.

To build the GLaM, we first construct an experimental design
\begin{equation}\label{eq:ED}
	\mathcal{D} = \acc{\prt{\ve{d}^{(i)},y^{(i)}}, \, i = 1, \ldots, N_{\textrm{ED}}}, \quad \textrm{with} \quad y^{(i)} = g_s\prt{\ve{d}^{(i)};\omega^{(i)}},
\end{equation}
where $\ve{d}$ is sampled uniformly in $\mathbb{D}$ and the stochastic simulator is evaluated \emph{once}, i.e., \emph{without replication}. The expansion coefficients are then estimated by maximizing the likelihood function:
\begin{equation}\label{eq:GLaMLikelihood}
	\hat{\ve{c}}=\arg \max _{\ve{c} \in \cc} \ell(\ve{c}),
\end{equation}
\noindent where
\begin{equation}
	\ell(\ve{c})=\sum_{i=1}^{N_{\textrm{ED}}} \log \left(f^{\textrm{GLD}}\left(y^{(i)} ; \ve{\lambda}^{\mathrm{PC}}\left(\ve{d}^{(i)} ; \ve{c}\right)\right)\right),
\end{equation}
and $f^{\textrm{GLD}}$ denotes the probability density function (PDF) of the generalized lambda distribution, which is obtained numerically as the reciprocal of the derivative of the quantile function defined in Eq.~\eqref{eq:GLDquantile}. Additional details on the estimation procedure can be found in \citet{ZhuSIAM2021,UQdoc_22_120}. 

Once the GLaM model has been constructed, it can be used to estimate conditional failure probabilities or quantiles required for design optimization. In particular, the quantile function of the GLD allows us to obtain conditional quantiles for any design $\ve{d}$ directly. For a target reliability level $\bar{p}_f$, the estimated conditional quantile therefore reads:
\begin{equation}\label{eq:GLaM:q}
	\hat{q}_{\alpha}\prt{\ve{d}} = Q_\alpha\prt{\bar{p}_f;\ve{\lambda}^{\mathrm{PC}}\prt{\ve{d}}} = \lambda_{1}^{\mathrm{PC}}\prt{\ve{d}} + \frac{1}{\lambda_{2}^{\mathrm{PC}}\prt{\ve{d}}} \prt{\frac{{\bar{p}_f}^{\lambda_{3}^{\mathrm{PC}}\prt{\ve{d}}} - 1}{\lambda_{3}^{\mathrm{PC}}\prt{\ve{d}}} - \frac{\prt{1-\bar{p}_f}^{\lambda_{4}^{\mathrm{PC}}\prt{\ve{d}}} - 1}{\lambda_{4}^{\mathrm{PC}}\prt{\ve{d}}}}.
\end{equation}
By relying on this closed-form quantile expression, we avoid the repeated calls to the stochastic emulator that would otherwise be required for a Monte Carlo-based estimation. Consequently, when employing GLaM, we exploit this property and solve the RBDO problem using the quantile-based formulation introduced in Eq.~\eqref{eq:RBDO_q}.

\subsection{Stochastic polynomial chaos expansions}\label{sec:SPCE}
Stochastic polynomial chaos expansions (SPCE) are surrogate models designed to emulate stochastic simulators \citep{ZhuStoPCE2023}.  
Unlike GLaM, which assumes a parametric family for the conditional distribution, SPCE introduces an explicit latent variable to represent the inherent variability of the stochastic simulator. The central idea is to construct a mapping from this latent variable to the conditional distribution of interest as a function of $\ve{d}$, and to approximate this mapping using polynomial chaos expansions.

Let $F_{Y_{\ve{d}}}(y \mid \ve{d})$ denote the cumulative distribution function (CDF) of the conditional response $Y_{\ve{d}}$ and consider an auxiliary random variable $\Xi$ with CDF $F_{\Xi}$. The mapping between the two random variables can be expressed using the probability integral transform:
\begin{equation}
	Y_{\ve{d}} \stackrel{\mathrm{d}}{=} 
	F^{-1}_{Y_{\ve{d}}}\prt{F_{\Xi}(\Xi) \mid \ve{d}},
	\label{eq_PIT}
\end{equation}
where $\stackrel{\mathrm{d}}{=}$ indicates equality in distribution.

In the SPCE framework, the auxiliary random variable $\Xi$ is considered latent. Building on this, SPCE approximates this mapping using a PCE model in the joint space of the design parameters and latent variable:
\begin{equation}
	Y_{\ve{d}}
	\stackrel{\mathrm{d}}{\approx}
	\tilde{Y}_{\ve{d}}
	=
	\sum_{\beta \in \mathcal{B}} c_{\ve{\beta}} \, \psi_{\ve{\beta}}(\ve{d}, \Xi)
	\;+\; \epsilon,
\end{equation}
where $\epsilon \sim \mathcal{N}(0,\sigma^2)$ is an additive noise term introduced to ensure numerical stability, see details in \citet{ZhuStoPCE2023}. By convoluting this Gaussian distribution with the PCE expansion and integrating out the latent variable, we can obtain the conditional probability density function (PDF) of $Y_{\ve{d}}$ as follows:
\begin{equation}
	f_{\tilde{Y}_{\ve{d}}}(y) =\int_{\mathcal{D}_{\Xi}} \frac{1}{\sigma} \varphi\left(\frac{y-\sum_{\ve{\beta} \in \mathcal{B}} c_{\ve{\beta}} \psi_{\ve{\beta}}(\ve{d}, \xi)}{\sigma}\right) f_{\Xi}(\xi) \mathrm{d} \xi,
	\label{eq_PDF_SPCE}
\end{equation}
where $\mathcal{D}_{\Xi}$ is the support of the latent variable $\Xi$ and $\varphi$ is the standard Gaussian PDF.

To construct the SPCE surrogate, an experimental design is generated in the same manner as in Eq.~\eqref{eq:ED}. The model parameters, namely the coefficients $\ve{c}_{\beta}$ and the noise standard deviation $\sigma$, are then obtained by maximum likelihood. Further details on the SPCE fitting procedure are provided in \citet{ZhuStoPCE2023,UQdoc_22_121}.

Once the model is built, SPCE can be used to calculate conditional failure probabilities for the optimization.  
As shown in \citet{PiresSS2025a}, SPCE enables the semi-analytical computation of conditional failure probabilities.  
This follows from the fact that the CDF of the conditional distribution can be written as the following double integral:
\begin{equation}\label{eq:SPCE_double_integral}
	F_{\tilde{Y}_{\ve{d}}}(y) = \int_{-\infty}^{y}  f_{\tilde{Y}_{\ve{d}}}(t) \, \mathrm{d}t = \int_{-\infty}^{y} \int_{\mathcal{D}_{\Xi}}
	\frac{1}{\sigma}
	\,
	\varphi\!\left(
	\frac{
		t - \sum_{\ve{\beta}\in\mathcal{B}} c_{\ve{\beta}} \psi_{\ve{\beta}}(\ve{d},\xi)
	}{\sigma}
	\right)
	f_{\Xi}(\xi)\, \mathrm{d}\xi\,\mathrm{d}t.
\end{equation}
The inner integral is one-dimensional and can be efficiently approximated by Gaussian quadrature as follows:
\begin{equation}
	f_{\tilde{Y}_{\ve{d}}}(y) \approx \sum_{j=1}^{N_Q} \frac{1}{\sqrt{2 \pi} \sigma} \exp \left(-\frac{\left(y-\sum_{\ve{\beta} \in \mathcal{B}} c_{\ve{\beta}} \psi_{\ve{\beta}}\left(\ve{d}, \xi_j\right)\right)^2}{2 \sigma^2}\right) w_j,
\end{equation}
where $(\xi_j, w_j)$ are the quadrature nodes and weights associated with the latent variable density $f_{\Xi}$ and $N_Q$ is the number of quadrature points.

By inverting the order of integration and integrating the Gaussian PDF analytically, we obtain:
\begin{equation}\label{eq:cdf_spce}
	F_{\tilde{Y}_{\ve{d}}}(y) \approx  \sum_{j=1}^{N_Q} w_j \Phi\prt{\frac{y-\sum_{\ve{\beta} \in \mathcal{B}} c_{\ve{\beta}} \psi_{\ve{\beta}}\left(\ve{d}, \xi_j\right)}{\sigma}},
\end{equation}
where $\Phi$ is the standard Gaussian CDF.

The conditional failure probability for a given design $\ve{d}$ follows by evaluating the conditional CDF at $y=0$:
\begin{equation}\label{eq:SPCE:Pf}
	\hat{p}_f\prt{\ve{d}} \approx  \sum_{j=1}^{N_Q} w_j \Phi\prt{-\frac{\sum_{\ve{\beta} \in \mathcal{B}} c_{\ve{\beta}} \psi_{\ve{\beta}}\left(\ve{d}, \xi_j\right)}{\sigma}}. 
\end{equation}

\section{Solving RBDO problems with stochastic simulators}

The proposed method is now summarized together with all algorithmic details. Algorithm~\ref{Alg:1} outlines the main steps of the approach, which are divided into three sequential stages.

The first stage consists in generating an experimental design to train the stochastic emulators. During this phase, the effect of uncertainties on the model response is captured implicitly through a limited number of model evaluations. First, {\color{black} support} points are sampled in the design space $\mathbb{D}$ using Latin hypercube sampling:
\[
\{\ve{d}^{(i)} \in \mathbb{D} \subset \mathbb{R}^{n_{\ve{d}}}, \, i=1, \ldots, N_{\mathrm{ED}} \}.
\]
For each {\color{black} support} point, realizations of the random inputs are generated using crude Monte Carlo simulation:
\[
\ve{x}^{(i)} \sim f_{\ve{X} \mid \ve{d}^{(i)}}\prt{\bullet},
\qquad
\ve{z}^{(i)} \sim f_{\ve{Z}}\prt{\bullet},
\qquad i = 1, \ldots, N_{\mathrm{ED}}.
\]
Uncertainty is then propagated through the model by evaluating the deterministic limit-state at the sampled points:
\[
y^{(i)} = g\prt{\ve{x}^{(i)},\ve{z}^{(i)}}.
\]

In the second stage, the effects of uncertainties are learned through a stochastic emulator. The conditional response $Y_{\ve{d}} = g_s(\ve{d};\omega)$ is approximated using a reduced dataset that retains only the design–response pairs:
\[
\{(\ve{d}^{(i)}, y^{(i)}), \quad i=1, \ldots, N_{\mathrm{ED}}\},
\]
while the corresponding random samples $\ve{x}^{(i)}$ and $\ve{z}^{(i)}$ are discarded. Based on this dataset, a stochastic emulator is constructed using the techniques described in Section~\ref{sec:RBDOStoEmu}. Once trained, the emulator enables the analytical evaluation of the conditional failure probability
$\hat{p}_f(\ve{d})$ when using SPCE (See Eq.~\eqref{eq:SPCE:Pf}), or of the conditional quantile
$\hat{q}_{\alpha}(\ve{d})$ when using GLaM (See Eq.~\eqref{eq:GLaM:q}), for any candidate design $\ve{d}$.

The third and final stage consists in solving the RBDO problem. With the stochastic emulator in place, both the objective function and the reliability constraints become deterministic functions of the design variables. The resulting optimization problem can therefore be solved using standard deterministic optimization algorithms. In this work, a gradient-based solver is employed, with gradients computed numerically via finite differences. The associated computational cost is negligible, since all quantities involved in the optimization process are obtained analytically.

Moreover, the availability of analytical expressions significantly improves numerical stability. In the original formulation based on Monte Carlo simulation, some designs may yield $\hat{p}_f(\ve{d}) = 0$ due to limited Monte Carlo sample sizes, which leads to inaccurate or ill-defined finite-difference gradients. In contrast, the continuous and differentiable nature of the surrogate-based constraints ensures reliable gradient estimates and robust convergence of gradient-based optimization schemes.

\begin{algorithm}[H]
	\caption{RBDO using stochastic emulators}
	\label{Alg:1}
	\footnotesize
	\begin{algorithmic}[1]
		\Require Number of training points $N_{\mathrm{ED}}$, Emulator type: GLaM or SPCE, Initial design $\ve{d}^{(0)}$
		
		\vspace{0.2cm}
		\Comment{  {\color{bondiblue} \textbf{I. Generation of the experimental design}}}
		\State Sample $N_{\mathrm{ED}}$ {\color{black} support} points in the design space 
		
		$\{\ve{d}^{(i)}\}_{i=1}^{N_{\mathrm{ED}}} \subset \mathbb{D}$ using Latin hypercube sampling
		\For{$i = 1,\dots,N_{\mathrm{ED}}$}
		\State Draw one realization of the random inputs:
		\[
		\ve{x}^{(i)} \sim f_{\ve{X} \mid \ve{d}^{(i)}}\prt{\bullet},
		\qquad
		\ve{z}^{(i)} \sim f_{\ve{Z}}\prt{\bullet}
		\]
		\State Evaluate the deterministic limit-state:
		\[
		y^{(i)} = g\prt{\ve{x}^{(i)},\ve{z}^{(i)}}
		\]
		\EndFor
		\State Form the reduced dataset $\{(\ve{d}^{(i)},y^{(i)})\}_{i=1}^{N_{\mathrm{ED}}}$ by discarding 
		$\ve{x}^{(i)}$ and $\ve{z}^{(i)}$
		
		\vspace{0.2cm}
		\Comment{ {\color{bondiblue} \textbf{II. Construction of the stochastic emulator}}}
		\State Build a stochastic emulator (GLaM or SPCE) that approximates the conditional response
		$Y_{\ve{d}} = g_s(\ve{d};\omega)$ from the dataset 
		$\{(\ve{d}^{(i)},y^{(i)})\}_{i=1}^{N_{\mathrm{ED}}}$
		
		\vspace{0.2cm}
		\Comment{ {\color{bondiblue} \textbf{III. Design optimization}}}
		\State Set $i=0$
		\While{not converged}
		\State Calculate cost $\mathfrak{c}\prt{\ve{d}^{(i)}}$
		\State Evaluate the soft constraint $f\prt{\ve{d}^{(i)}}$
		\If{using GLaM}
		\State Evaluate the conditional quantile $\hat{q}_\alpha\prt{\ve{d}^{(i)}}$ \Comment{{\color{indiagreen} Eq.~\eqref{eq:GLaM:q}}} 
		\ElsIf{using SPCE}
		\State Evaluate the conditional failure probability $\hat{p}_f\prt{\ve{d}^{(i)}}$ \Comment{{\color{indiagreen} Eq.~\eqref{eq:SPCE:Pf}}}
		\EndIf
		\State Compute gradients of cost and constraints numerically
		\State Update design: $\ve{d}^{(i)} \gets \ve{d}^{(i)} + \nu^{(i)}$ \Comment{ {\color{indiagreen} $\nu^{(i)}$ are update directions}}
		\State $i \gets i+1$
		\EndWhile
		
		\Return Optimal design $\ve{d}^\star$ and associated cost $\mathfrak{c}\prt{\ve{d}^\star}$
	\end{algorithmic}
\end{algorithm}

\section{Applications}\label{sec:Applications}
The proposed approach is now validated on {\color{black}four} examples. We compare the results with those obtained using a double-loop Monte Carlo based approach with Kriging as surrogate. The quantile-based RBDO formulation in Eq.~\eqref{eq:RBDO_q} is adopted for the Kriging-based solution, as it is more robust when relying on Monte Carlo simulation. In all cases, the optimization problems are solved using a gradient-based solver implemented in \textsc{Matlab}'s \texttt{fmincon} function. For the Monte-Carlo based solutions, common random numbers are considered to enable the use of the general-purpose deterministic solver.

{\color{black} For the three high-dimensional examples, we additionally compare against the heteroscedastic Gaussian process (HGP) model introduced for RBDO by \citet{Kim2024}, which is built directly in the reduced design space like the stochastic emulators. It assumes a Gaussian conditional response with design-dependent mean $\mu(\ve{d})$ and variance $\sigma^2(\ve{d})$, both represented by Gaussian processes through the variational heteroscedastic formulation of \citet{Lazaro-Gredilla2011}, whose predictive moments are available in closed form. The failure probability then follows semi-analytically, without any inner sampling, as
	\begin{equation}\label{eq:HGP:Pf}
		p_f(\ve{d}) = \Phi\!\left(-\,\mu(\ve{d})/\sigma(\ve{d})\right).
	\end{equation}
	For the seismic frame, we additionally consider a log-transformed variant, HGP$_{\log}$, which instead assumes a lognormal conditional response.}

For all examples, we use the same surrogate settings. In GLaM, the location parameter $\lambda_1$  is approximated by a sparse PCE model with candidate bases of degrees varying between $1$ and $15$, the scale parameter $\lambda_2$ by an expansion with degree varying between $0$ and $5$ and the shape parameters $\lambda_3$ and $\lambda_4$ with degree $0$ or $1$ (constant or linear). We also consider an adaptive $q$-norm strategy with values between $0.6$ and $1$, using increments of $0.1$. For SPCE, we consider an adaptive degree and $q$-norm between $1$ and $15$ and $0.8$ and $1$, respectively. The Kriging surrogates are built using a Mat\'ern 5/2 auto-correlation family with a constant but unknown trend, which corresponds to ordinary Kriging. All the analyses are carried out in \textsc{Matlab} using \textsc{UQLab} \citep{Marelli2014a} with the appropriate modules \citep{UQdoc_22_120,UQdoc_22_121,UQdoc_22_105,UQdoc_22_115}.

The analyses are repeated $15$ times to assess the robustness of the benchmarked methods. While different experimental designs are considered, the accuracy is assessed using the relative error:
\begin{equation}\label{eq:error}
	\varepsilon_{\mathfrak{c}} = \frac{\left|\mathfrak{c}\prt{\ve{d}^\ast} - \mathfrak{c}\prt{\ve{d}^{\ast,\textrm{ref}}}\right|}{\mathfrak{c}\prt{\ve{d}^{\ast,\textrm{ref}}}},
\end{equation}
where $\ve{d}^{\ast,\textrm{ref}}$ and $\ve{d}^\ast$ denote the reference optimal solution and the optimum obtained by one of the {\color{black} four} benchmarked approaches, respectively.

\subsection{Column buckling}
This first benchmark example, originally introduced in \citet{Dubourg2011}, considers a column with a rectangular cross-section of dimensions $b \times h$  subjected to a service compressive load $F_{\mathrm{ser}}$. The objective is to minimize the cross-sectional area while ensuring sufficient resistance against buckling.

Structural failure is defined through the following limit-state function:
\begin{equation}
	g(\ve{d},\ve{z}) = F_{\mathrm{ser}} - F_{\mathrm{buck}}\prt{\ve{d},\ve{Z}} = 
	F_{\mathrm{ser}} -
	k \, \frac{\pi^2 E b h^3}{12 L^2},
\end{equation}
where $\ve{d} = \prt{b,\,h}$ denotes the vector of design variables, $\ve{z} = \prt{k,\,E,\,L}$ is the vector of environmental variables, $F_{\mathrm{buck}}$ is the buckling load, $E$ is the constitutive material's Young's modulus, $L$ is the column length, and $k$ is a model correction factor accounting for geometrical imperfections and modeling uncertainties in the Euler buckling formulation.

The above expression is valid under the assumption that the height $h$ of the cross-section is smaller than or equal to its width $b$ (buckling occurs about the weak (minor) axis of the rectangular cross-section). This modeling requirement 
is enforced through the following soft constraint:
\begin{equation}
	f(\ve{d}) = h - b \leq 0 .
\end{equation}

All environmental variables are assumed to follow lognormal distributions, as
summarized in Table~\ref{tab:Ex1:Z}. Under these assumptions, an analytical
solution to the associated reliability-based design optimization problem can be
derived:
\begin{equation}
	b^* = h^* =
	\frac{12 F_{\mathrm{ser}}}{\pi^2}
	\exp\!\left(
	\lambda_k + \lambda_E - 2\lambda_L
	+ \Phi^{-1}(P_f)
	\sqrt{\zeta_k^2 + \zeta_E^2 + 4\zeta_L^2}
	\right),
\end{equation}
where
\begin{equation}
	\zeta_\bullet = \sqrt{\ln(1+\delta_\bullet^2)}, \qquad
	\lambda_\bullet = \ln(\mu_\bullet) - \tfrac{1}{2}\zeta_\bullet^2,
\end{equation} 
are respectively the scale and location parameters of the lognormal distribution.
For a target probability of failure $\bar{p}_f = 5\%$, the analytical solution yields the optimal dimensions $b^* = h^* = 238.45~\mathrm{mm}$.
\begin{table}[ht]
	\centering
	\begin{tabular}{lccc}
		\hline
		Parameter & Distribution & Mean ($\mu$) & CoV ($\delta \%$) \\
		\hline
		Correction factor \( k \) [-] & Lognormal & $0.6$ & $10$ \\
		Young's modulus \( E \) [MPa] & Lognormal & $10{,}000$ & $5$ \\
		Column length \( L\) [mm] & Lognormal & $3{,}000$ & $1$ \\
		Service load \(  F_{\mathrm{ser}} \) [N] & Constant & $1.4622 \times 10^6$ & $--$ \\
		\hline
	\end{tabular}
	\caption{Column buckling: Environmental variables $\ve{Z}$.}
	\label{tab:Ex1:Z}
\end{table}
By considering all variables but $b$ and $h$ as latent (see Table~\ref{tab:Ex1:Z}), we can compute analytically the conditional distribution of the buckling load. It follows a lognormal distribution with parameters:
\begin{equation}
	\begin{split}
		\lambda_{\textrm{buck}} = & \ln\prt{\frac{\pi^2 b h^3}{12}}
		\prt{\lambda_k + \lambda_E - 2\lambda_L}, \\
		\zeta_{\textrm{buck}} = & 
		\sqrt{\zeta_{k}^{2} + \zeta_{E}^{2} + 4 \zeta_{L}^{2}}.
	\end{split}
\end{equation}

\begin{figure}[H]
	\centering
	\includegraphics[width=0.6\textwidth]{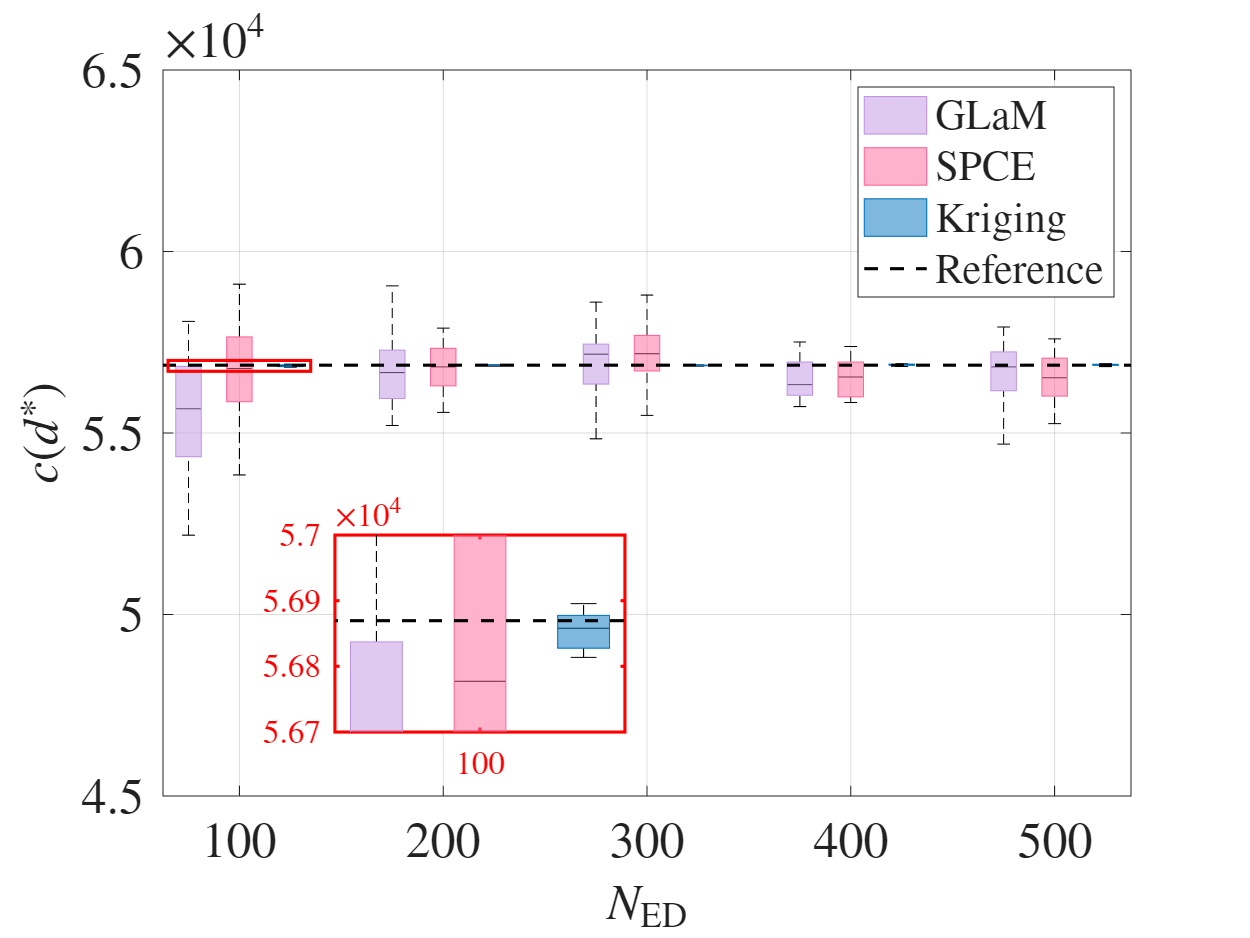}
	\caption{Column buckling: Boxplots of the optimal costs $\mathfrak{c}\prt{b^{\ast},\,h^{\ast}}$ for different methods and experimental design sizes.}
	\label{fig:Ex1:Fstar}
\end{figure}

We consider experimental designs of sizes $N_{\textrm{ED}} = \acc{100, \, 200, \, 300, \, 400, \, 500}$. Figure~\ref{fig:Ex1:Fstar} shows boxplots of the optimal cost corresponding to each method for $15$ repetitions of the analysis. In the boxplots, the box represents the interquartile range, the horizontal line indicates the median, while the whiskers show the extent of the data range, excluding outliers. 
For this example, Kriging is known to provide a very accurate approximation of the limit-state function, even with a small number of training points, as already reported in \citet{MoustaphaSMO2016}. As a result, it exhibits fast and stable convergence toward the optimal solution for all the experimental design sizes considered. GLaM and SPCE also converge toward the reference solution, with SPCE being accurate on average with  $100$ training points, while GLaM requires $200$ points. Nevertheless, both methods display a larger variability in the estimated optimum compared to Kriging.

Table~\ref{tab:Ex1:error} reports the relative errors on the optimal cost $\mathfrak{c}\prt{b^\ast,h^\ast}$, as computed in Eq.~\eqref{eq:error}, associated with the median solutions. With the exception of GLaM for $N_{\textrm{ED}} = 100$, all configurations yield median relative errors on the order of $10^{-3}$. Overall, SPCE performs slightly better than GLaM, while Kriging clearly outperforms both methods for this particular problem.
\begin{table}[ht]
	\centering
	\begin{tabular}{lccccc}
		\hline
		$N_{\textrm{ED}}$ & $100$ & $200$ & $300$ & $400$ & $500$ \\
		\hline
		GLaM 
		& $2.1 \cdot 10^{-2}$ 
		& $3.6 \cdot 10^{-3}$ 
		& $5.3 \cdot 10^{-3}$ 
		& $9.4 \cdot 10^{-3}$ 
		& $8.2 \cdot 10^{-4}$ \\
		SPCE 
		& $1.6 \cdot 10^{-3}$ 
		& $8.4 \cdot 10^{-4}$ 
		& $5.5 \cdot 10^{-3}$ 
		& $5.7 \cdot 10^{-3}$ 
		& $6.1 \cdot 10^{-3}$ \\
		Kriging 
		& $2.3 \cdot 10^{-4}$ 
		& $1.9 \cdot 10^{-4}$ 
		& $3.1 \cdot 10^{-5}$ 
		& $1.1 \cdot 10^{-4}$ 
		& $8.9 \cdot 10^{-5}$ \\
		\hline
	\end{tabular}
	\caption{Column buckling: Relative error on the optimal cost $\mathfrak{c}\prt{b^{\ast},\,h^{\ast}}$ after Eq.~\eqref{eq:error} for the median solution (over $15$ repetitions) for each method and experimental design size $N_{\textrm{ED}}$}
	\label{tab:Ex1:error}
\end{table}

\begin{figure}[H]
	\centering
	\subfloat[$N_{\textrm{ED}} = 100$]{
		\includegraphics[width=0.48\textwidth]{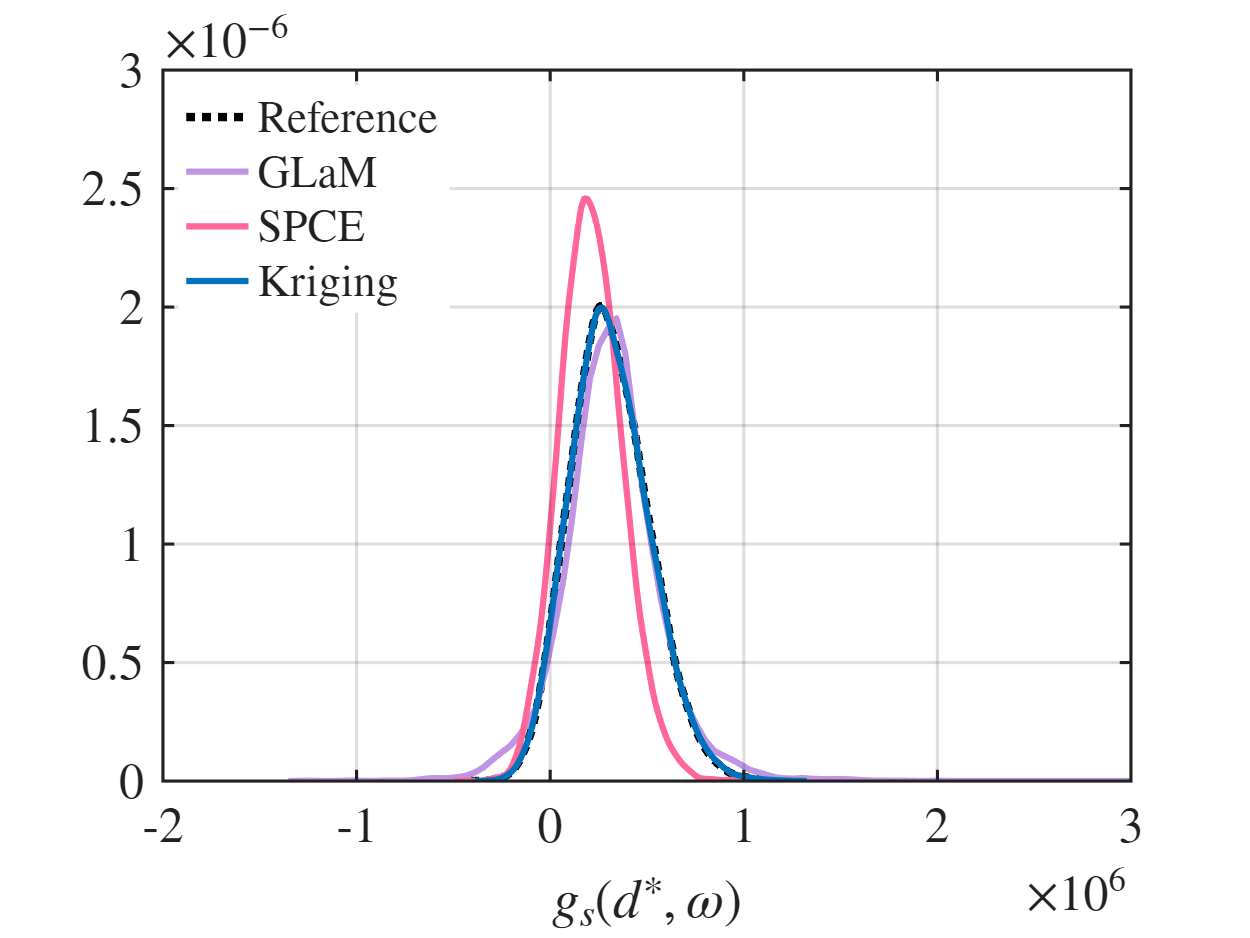}}
	\hfill
	\subfloat[$N_{\textrm{ED}}  = 200$]{
		\includegraphics[width=0.48\textwidth]{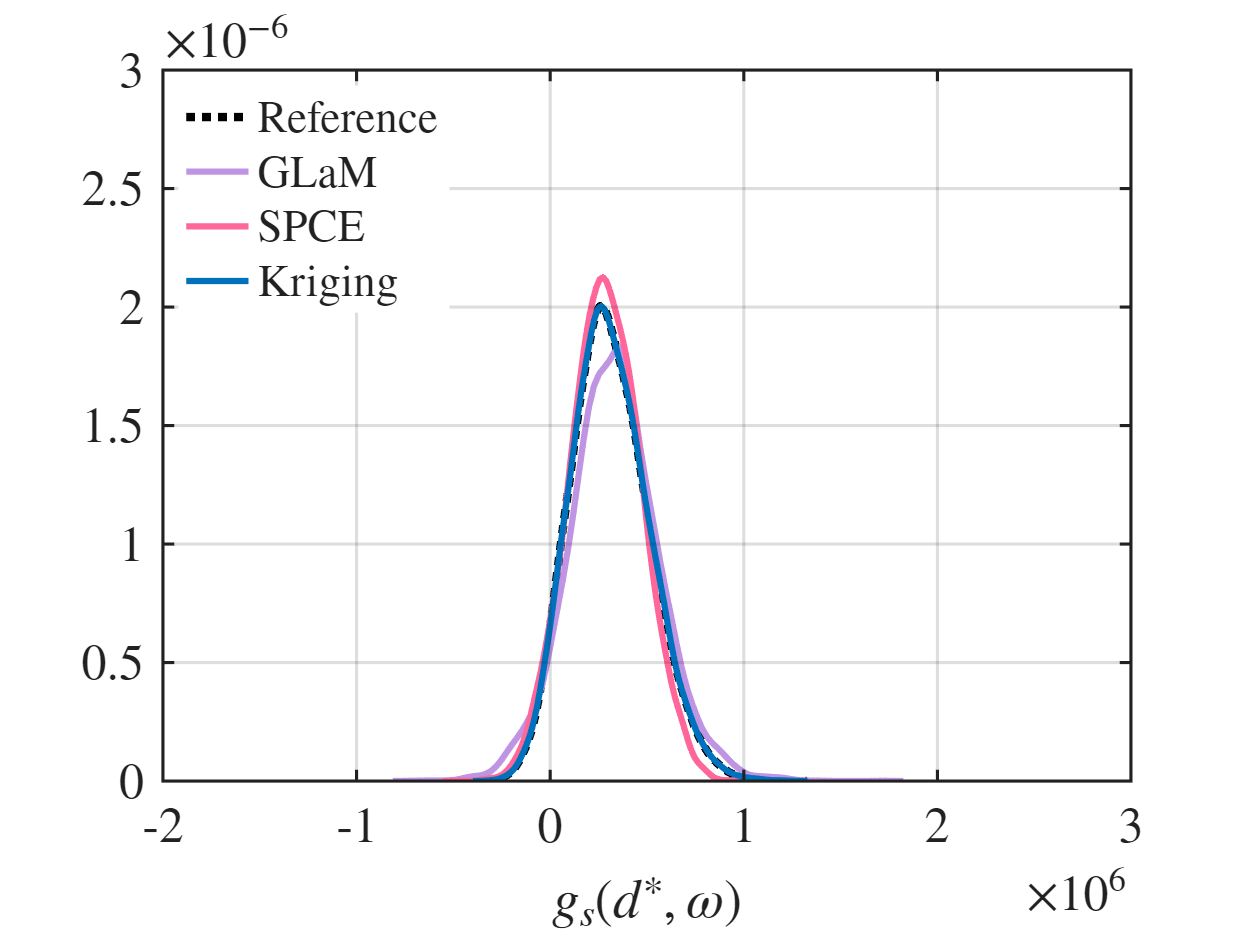}} \\
	\subfloat[$N_{\textrm{ED}}  = 300$]{
		\includegraphics[width=0.48\textwidth]{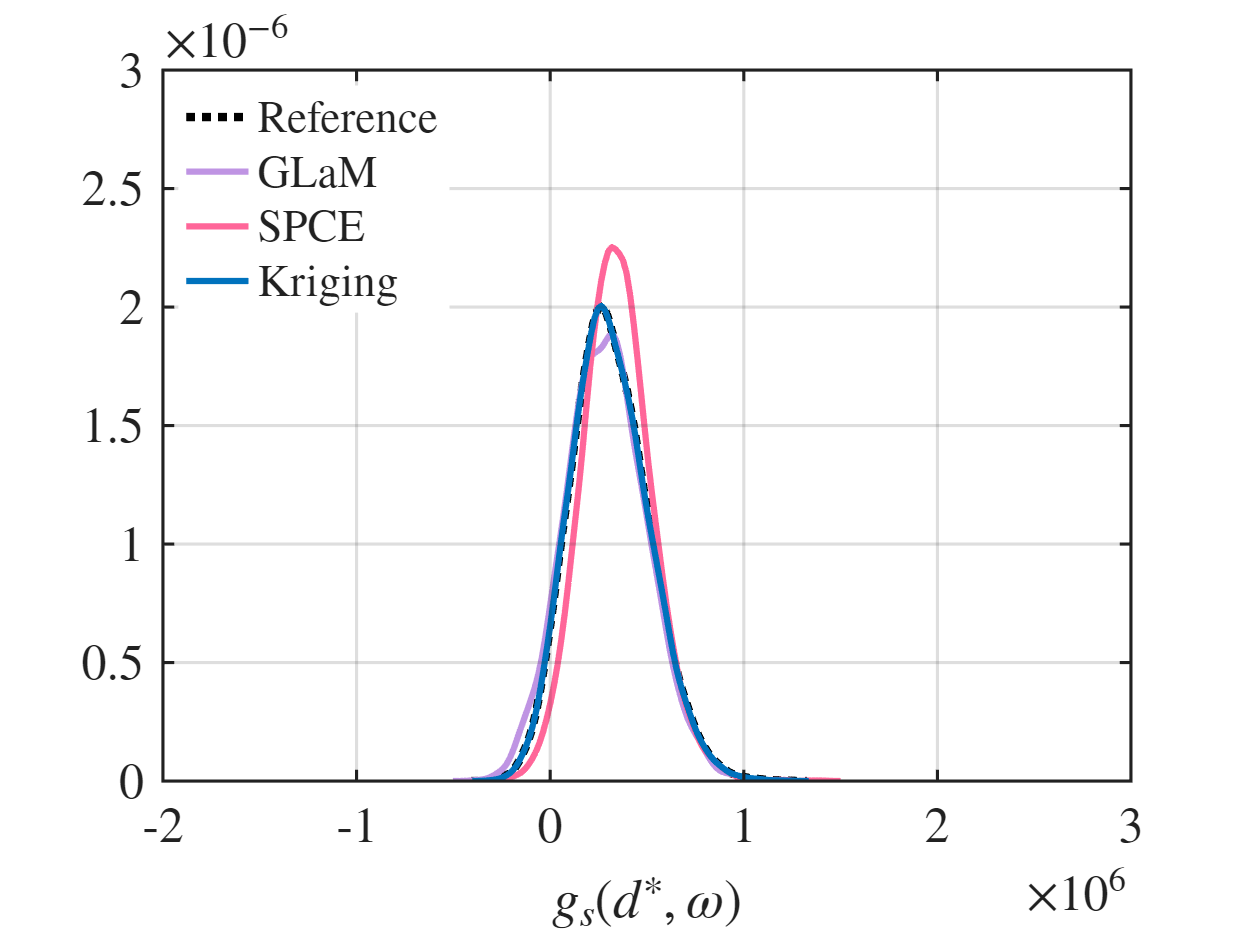}}
	\hfill
	\subfloat[$N_{\textrm{ED}}  = 400$]{
		\includegraphics[width=0.48\textwidth]{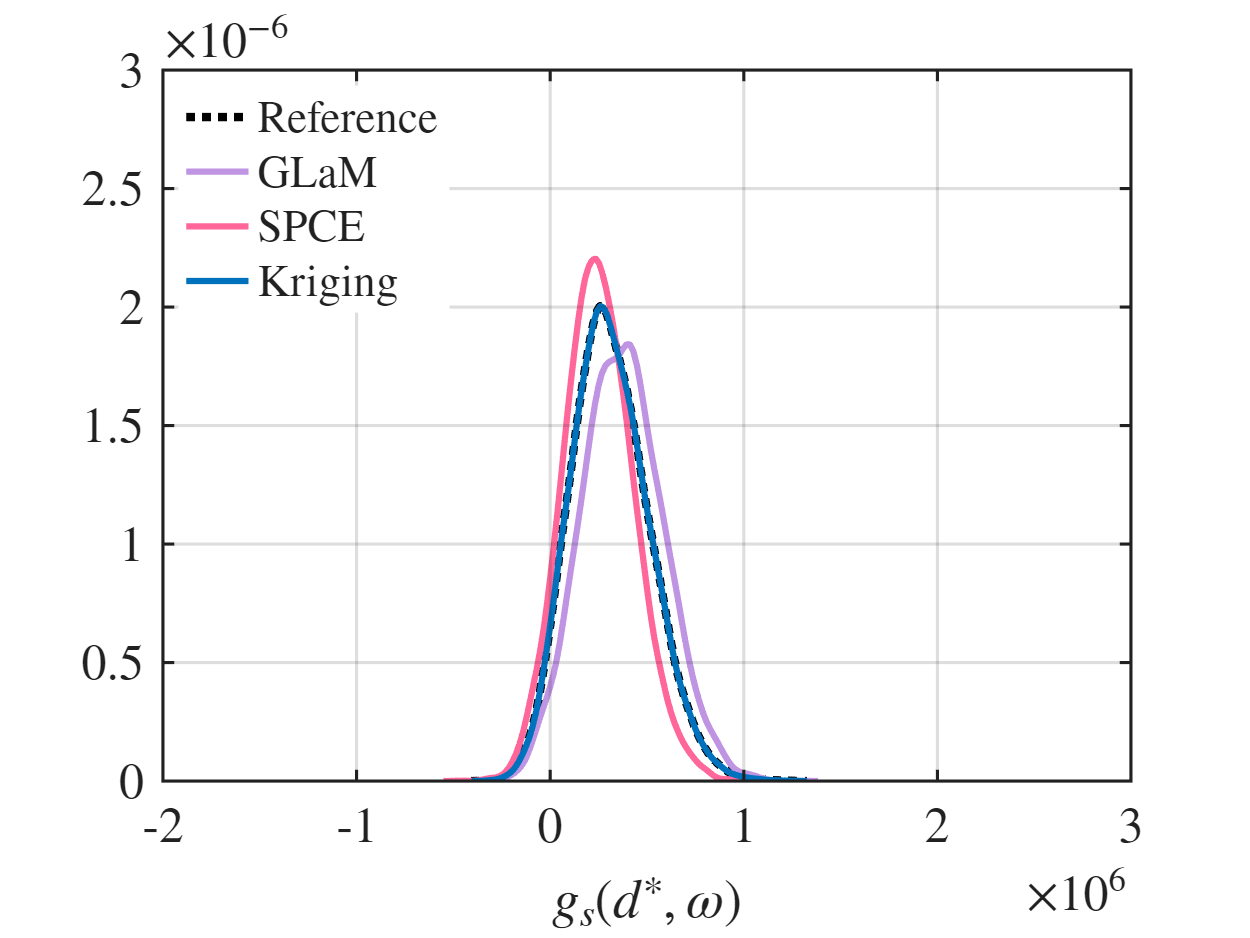}} \\
	\subfloat[$N_{\textrm{ED}}  = 500$]{
		\includegraphics[width=0.48\textwidth]{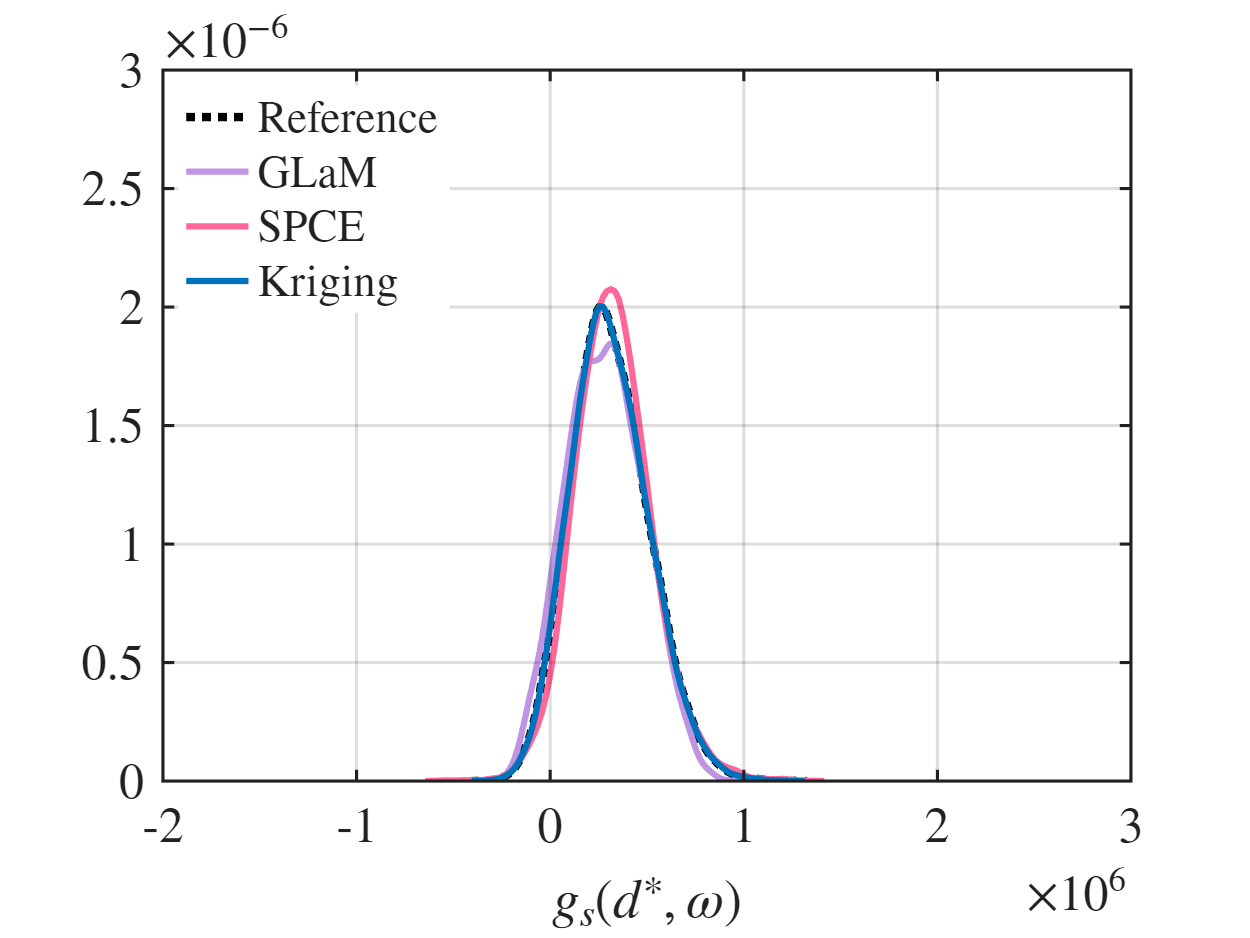}} \\
	\caption{Column buckling: Conditional probability density functions of the model response at the optimal reference solution obtained from evaluating $\hat{g}_s\prt{\ve{d}^{\ast,\textrm{ref}},\omega}$ (for GLaM and SPCE) and $\hat{g}\prt{\ve{d}^{\ast,\textrm{ref}},\ve{Z}}$ (for Kriging).}
	\label{fig:Ex1:Density}
\end{figure}

Figure~\ref{fig:Ex1:Density} presents the conditional distributions of the performance function at the reference {\color{black} optimal design} point $\ve{d}^{\ast,\textrm{ref}}$ using the analytical PDF of $F_{\textrm{buck}}$ on the one hand, and using the surrogates on the other hand. For each experimental design (ED) size, we report the surrogate model corresponding to the median error over 15 independent repetitions of the full analysis. For the stochastic emulators GLaM and SPCE, the conditional response is obtained by directly evaluating the surrogate models $\hat{g}_s(\ve{d}^{\ast,\textrm{ref}}, \omega)$ over $10^4$ independent realizations of the auxiliary variable $u \in \bra{0,\,1}$ for GLaM and $\Xi \sim \mathcal{N}\prt{0,\,1}$ for SPCE. In contrast, for Kriging and for the original model, $10^4$ realizations of the environmental random vector $\ve{Z}$ are first sampled, and the deterministic surrogate or the original model is then evaluated at the points $\acc{(\ve{d}^{\ast,\textrm{ref}}, \ve{z}^{(i)}), i = 1, \ldots, N}$. In all cases, the resulting response values are post-processed using kernel density estimation to produce the distributions shown in Figure~\ref{fig:Ex1:Density}. The plots confirm the results shown in Table~\ref{tab:Ex1:error}. Kriging provides a near-perfect approximation: for all sample sizes, the conditional distributions closely match those of the original model. In contrast, GLaM and SPCE, when considering the median surrogate, are less accurate in the bulk of the distribution. However, they exhibit a better fit in the tails. As a result, the estimated failure probabilities and quantiles remain sufficiently accurate overall.

Although Kriging is already sufficiently accurate in this example and cannot be outperformed by our stochastic emulator framework, the latter exhibits a clear advantage in terms of computational efficiency. Table~\ref{tab:Ex1:computtime} reports the average running time of the RBDO analysis, measured on a standard laptop, over five independent repetitions for experimental design sizes $N_{\textrm{ED}} \in \acc{100,\,300,\,500}$. Only the time spent in the optimization process is reported, i.e., the stage III of the pseudo-algorithm presented in Algorithm~\ref{Alg:1}. 
\begin{table}[H]
	\centering
	\begin{tabular}{lccccc}
		\hline
		$N_{\textrm{ED}}$ & $100$ & $300$ & $500$  \\
		\hline
		GLaM 
		& $0.03$ 
		& $0.04$ 
		& $0.03$ \\
		SPCE 
		& $0.15$ 
		& $0.16$ 
		& $0.18$ \\
		Kriging 
		& $1.27$ 
		& $3.79$ 
		& $7.13$ \\
		\hline
	\end{tabular}
	\caption{Column buckling: Computational time (in seconds) of the optimization procedure for different experimental design sizes and surrogate modeling methods.}
	\label{tab:Ex1:computtime}
\end{table}

GLaM is extremely fast in this setting, with an almost instantaneous runtime, while SPCE also exhibits very low computational cost. Although Kriging remains relatively fast, it is still one to two orders of magnitude slower than GLaM and SPCE. Moreover, unlike the stochastic emulators, the computational time of Kriging is strongly affected by the experimental design size, increasing from approximately $1$~seconds to more than $7$~seconds as $N_{\textrm{ED}}$ grows from $100$ to $500$. These runtimes correspond to a Monte Carlo sample size of $10^5$ for the evaluation of $p_f$. They would increase substantially for larger sample sizes, that would be needed, for intance, when the target failure probability is very small. In contrast, the computational cost of GLaM and SPCE is only weakly sensitive to the experimental design size. This behavior follows directly from Eqs.~\eqref{eq:GLaM:q} and~\eqref{eq:SPCE:Pf}, which show how quantiles and failure probabilities are computed analytically at each optimization iteration. The computational complexity is primarily governed by the order of the underlying polynomial chaos expansions, which does not scale directly with the experimental design size. Furthermore, problems with smaller target failure probabilities $\bar{p}_f$ are not necessarily more computationally expensive when using stochastic emulators. Thus our proposed method is expected to scale well as we increase the complexity of the analysis through higher-dimensional problems or smaller target failure probabilities.

{\color{black}
	Beyond the optimization stage, we compare the \emph{training} cost of the surrogates. Figure~\ref{fig:Ex1:traintime} reports the mean training time over five replications as a function of the experimental design size. GLaM and Kriging have comparable and low training costs over the considered range, whereas SPCE is more expensive to train. For such low-dimensional problems, SPCE is therefore not competitive in terms of training cost compared with Kriging and GLaM. However, the training cost for SPCE does not increase significantly for larger training sets.
	\begin{figure}[H]
		\centering
		\includegraphics[width=0.6\textwidth]{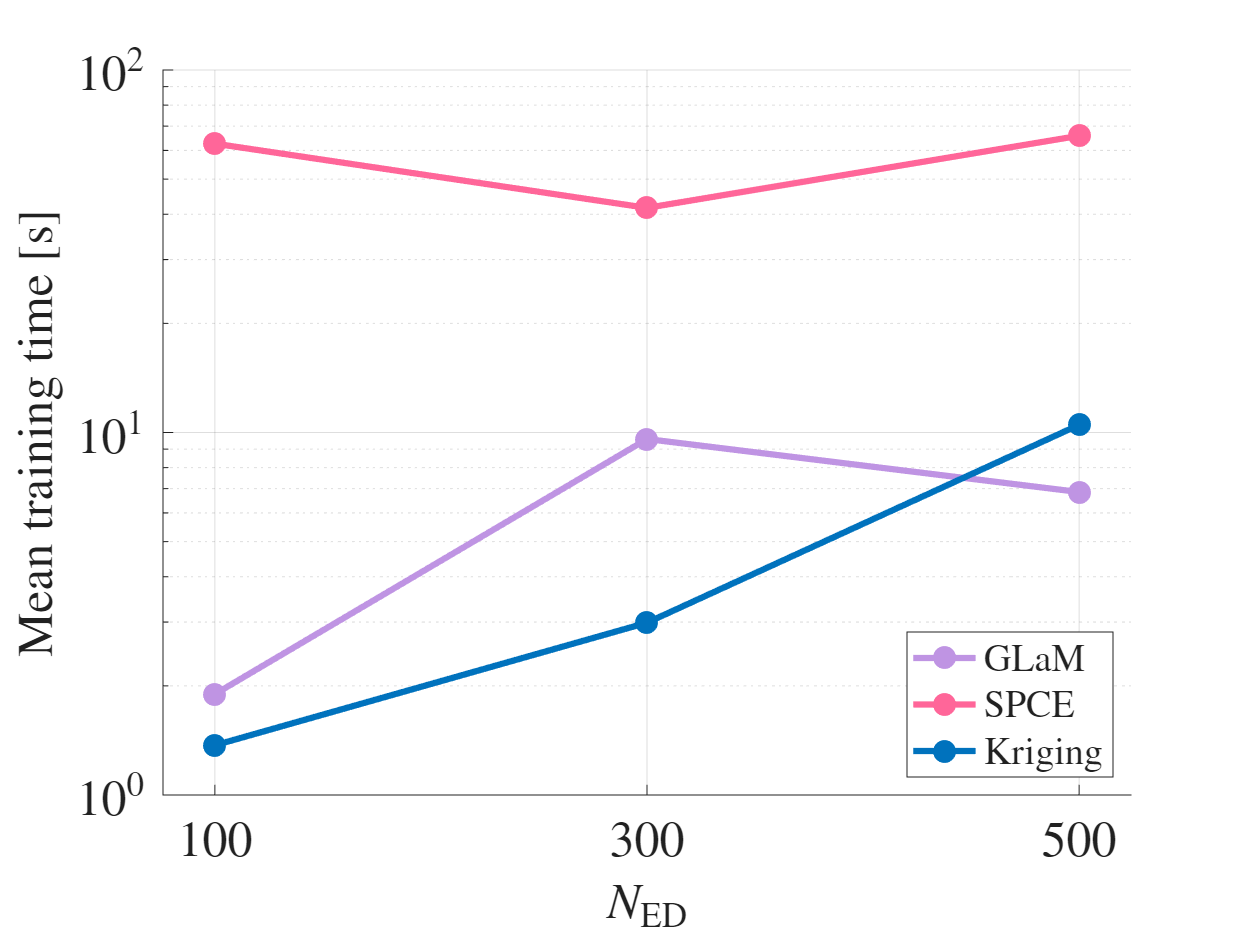}
		\caption{\color{black} Column buckling: Mean surrogate training time (over $5$ repetitions) versus the experimental design size $N_{\textrm{ED}}$ for three surrogate models.}
		\label{fig:Ex1:traintime}
	\end{figure}
}

\subsection{Corroded beam under stochastic excitation}
We consider a simply supported steel beam in bending, with rectangular cross-section $b_0 \times h_0$ and length $L = 5$ m. The beam is subjected to self-weight, modeled as a uniformly distributed dead load $\rho\, b_0 h_0$, where $\rho$ denotes the steel mass density, and to a pinpoint stochastic load $F(t,\omega)$ applied at midspan. In addition, the beam undergoes corrosion, assumed to reduce the cross-sectional dimensions uniformly on all faces. The corrosion depth $d_c$ is assumed to increase linearly with time, i.e., $d_c(t) = \kappa\, t$, where $\kappa$ denotes the corrosion rate. The corroded material is assumed to have lost all mechanical stiffness.
\begin{figure}[H]
	\centering
	\includegraphics[width=0.5\textwidth]{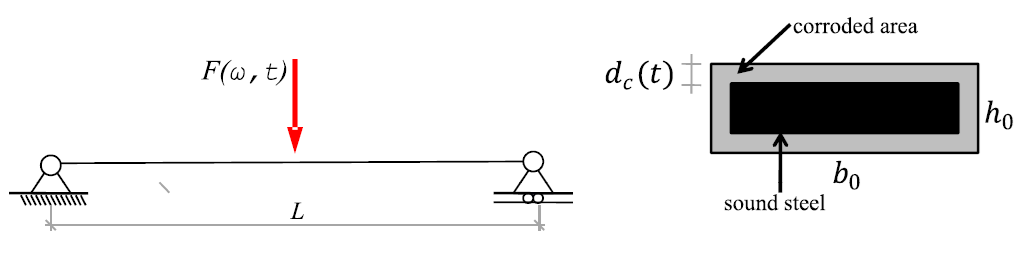}
	\caption{Corroded beam: Schematic representation of a beam under corrosion, from \citet{KroetzRESS2020}.}
	\label{fig:Ex1}
\end{figure}

Failure is defined by the formation of a plastic hinge at midspan. The corresponding limit-state function reads
\begin{equation}\label{eq:Ex2:LS}
	g(\ve{d},\ve{Z};t)
	= \frac{(b_0 - 2\kappa t)(h_0 - 2\kappa t)^2 f_y}{4}
	- \left( \frac{F(t) \, L}{4} + \frac{\rho b_0 h_0 L^2}{8} \right),
\end{equation}
where $f_y$ denotes the yield stress of the steel. The analysis is carried out over a service time interval $t \in [0, \, 120]$ months and the target failure probability is $\bar{p}_f = 0.05$.

The load $F(t,\omega)$ is modeled as a Gaussian random process with mean $12$~kN and coefficient of variation $0.25$. Its temporal variability is characterized by a Gaussian autocorrelation function with a correlation length of one month. The process is discretized using a Karhunen--Lo\`eve expansion.

For the RBDO problem, the optimization parameters are the initial dimensions of the cross-section, i.e., $\ve{d} = \prt{b_0,\,h_0}$. The objective function is defined as the initial mass of the beam:
\begin{equation}
	\mathfrak{c}(\ve{d}) = \rho L b_0 h_0, \quad \textrm{with} \quad \mathbb{D} = \bra{0.03, \, 0.15}^2
\end{equation}
and failure is assumed to occur if
\begin{equation}
	\max_{t \in [0,120]} g(\ve{d}, \ve{Z}; t) \leq 0.
\end{equation}
The vector of environmental random variables is given by $\ve{Z} = \prt{f_y,\, \kappa,\, \rho, \, \ve{\theta}}$, and their probabilistic descriptions are summarized in Table~\ref{tab:Ex2:Z}. The vector $\ve{\theta} = \prt{\theta_1,\ldots,\theta_{100}}$ consists of independent standard Gaussian random variables arising from the Karhunen--Lo\`eve discretization of the random field $F$. These $100$ variables account for $99.06\%$ of the total variance of the load process.

Overall, the problem involves $n_{\ve{Z}} = 103$ random variables and $n_{\ve{d}} = 2$ design variables. The GLaM and SPCE surrogate models are constructed in the reduced design space of dimension $n_{\ve{d}} = 2$, whereas the Kriging model is built in the full space of dimension $n_{\textrm{tot}} = 105$.
\begin{table}[H]
	\centering
	\begin{tabular}{lccc}
		\hline
		Parameter & Distribution & Mean ($\mu$) & CoV ($\delta \%$)\\
		\hline
		Yield stress \( f_y \) [MPa] & Lognormal & $355$ & $3$ \\
		Corrosion rate \( \kappa \) [mm/month] & Gaussian & $1/12$ & $10$ \\
		Density \( \rho \) [kN/m\textsuperscript{3}] & Lognormal & $78.5$ & $3$ \\
		Random field \( \theta_1, \ldots, \theta_{100} \) [--] & Gaussian & $0$ & $1^{\ast}$ \\
		\hline
	\end{tabular}
	\caption{Corroded beam: Environmental variables $\ve{Z}$. $^\ast$corresponds to the standard deviation, rather than the CoV.}
	\label{tab:Ex2:Z}
\end{table}

The reference solution is obtained using the quantile RBDO formulation in a double-loop with the original limit-state, where a Monte Carlo set of size $10^6$ is used. This corresponds to a coefficient of variation $4.35 \cdot 10^{-3}$ when estimating the target failure probability. Experimental designs of sizes $N_{\textrm{ED}} = \acc{250, \, 500, \, 1{,}000, \, 1{,}500}$ are considered.

\begin{figure}[H]
	\centering
	\includegraphics[width=0.6\textwidth]{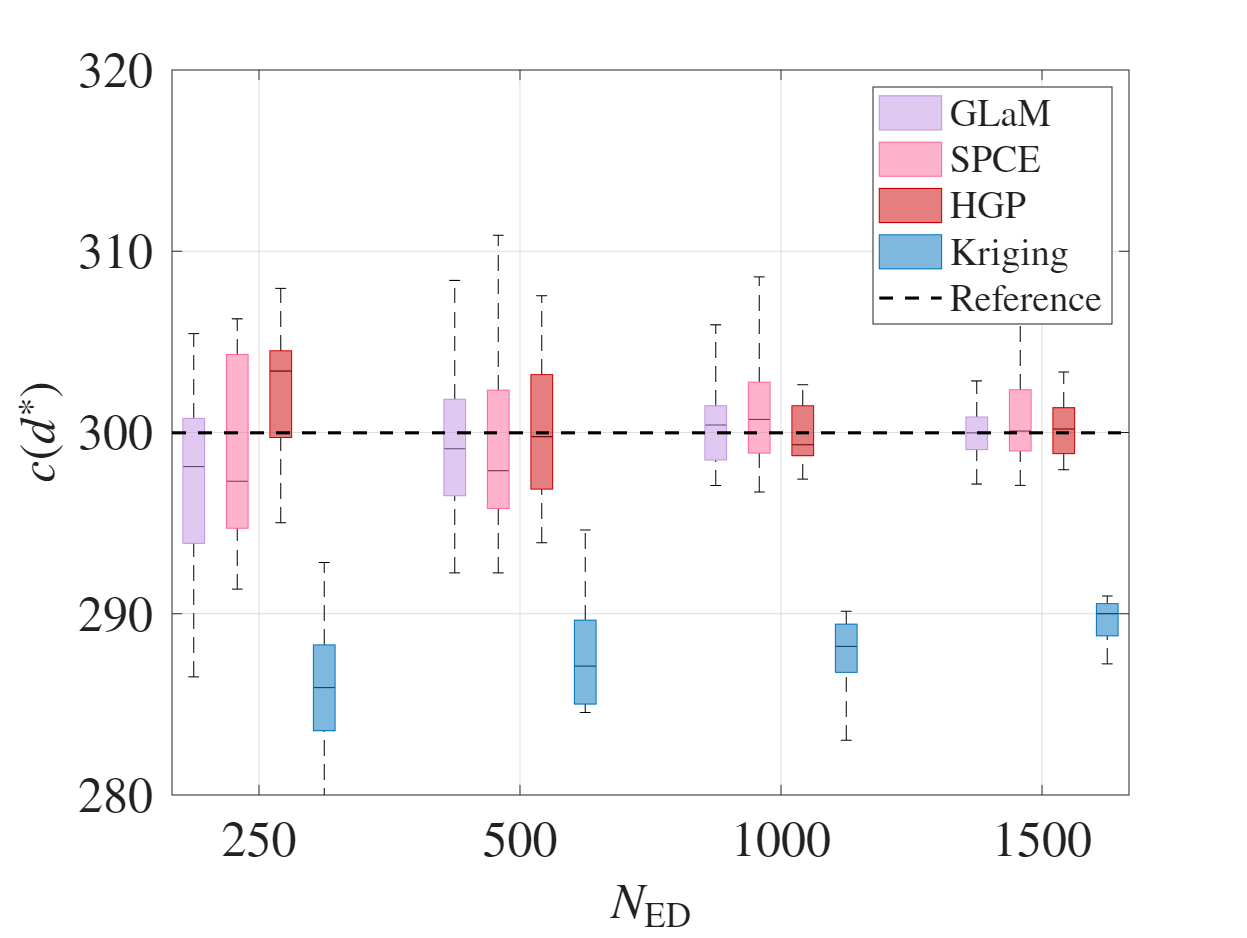}
	\caption{\color{black}Corroded beam: Boxplots of the optimal costs $\mathfrak{c}\prt{b_{0}^{\ast},\,h_{0}^{\ast}}$ for different methods and experimental design sizes.}
	\label{fig:Ex2:Fstar}
\end{figure}

Figure~\ref{fig:Ex2:Fstar} shows boxplots of the optimal costs for the $15$ repetitions of the analysis considering the three approaches and all the ED sizes. The figure shows that both GLaM and SPCE outperform the Kriging approach. The latter even converges to a biased solution. In contrast, GLaM and SPCE yield solutions close to the optimal solution, with experimental design sizes as small as $250$ points. The increase in $N_{\textrm{ED}}$ provides even more accurate results, but more importantly, reduces variability of the obtained optimum over the $15$ repetitions. 

Table~\ref{tab:Ex2:Error} shows the relative errors on the optimal cost $\mathfrak{c}\prt{b_{0}^{\ast},\,h_{0}^{\ast}}$, as computed using Eq.~\eqref{eq:error}. GLaM and SPCE yield the best results for $N_{\textrm{ED}} = 1{,}500$.
\begin{table}[H]
	\centering
	\begin{tabular}{lcccc}
		\hline
		$N_{\textrm{ED}}$ & $250$ & $500$ & $1{,}000$ & $1{,}500$ \\
		\hline
		GLaM & $6.2 \cdot 10^{-3}$ & $2.9 \cdot 10^{-3}$ & $1.4 \cdot 10^{-3}$ & $8.1 \cdot 10^{-5}$  \\
		SPCE & $8.9 \cdot 10^{-3}$ & $6.9 \cdot 10^{-3}$ & $2.4 \cdot 10^{-3}$ & $3.2 \cdot 10^{-4}$  \\
		\textcolor{black}{HGP} & \textcolor{black}{$1.2 \cdot 10^{-2}$} & \textcolor{black}{$1.1 \cdot 10^{-2}$} & \textcolor{black}{$4.8 \cdot 10^{-3}$} & \textcolor{black}{$4.8 \cdot 10^{-3}$}  \\
		Kriging & $4.7 \cdot 10^{-2}$ & $4.3 \cdot 10^{-2}$ & $3.9 \cdot 10^{-2}$ & $3.3 \cdot 10^{-2}$  \\
		\hline
	\end{tabular}
	\caption{\color{black} Corroded beam: Relative error on the optimal cost $\mathfrak{c}\prt{b_{0}^{\ast},\,h_{0}^{\ast}}$ after Eq.~\eqref{eq:error} for the median solution (over $15$ repetitions) for each method and experimental design size $N_{\textrm{ED}}$.}
	\label{tab:Ex2:Error}
\end{table}

\begin{figure}[H]
	\centering
	\subfloat[$d_1$]{
		\includegraphics[width=0.48\textwidth]{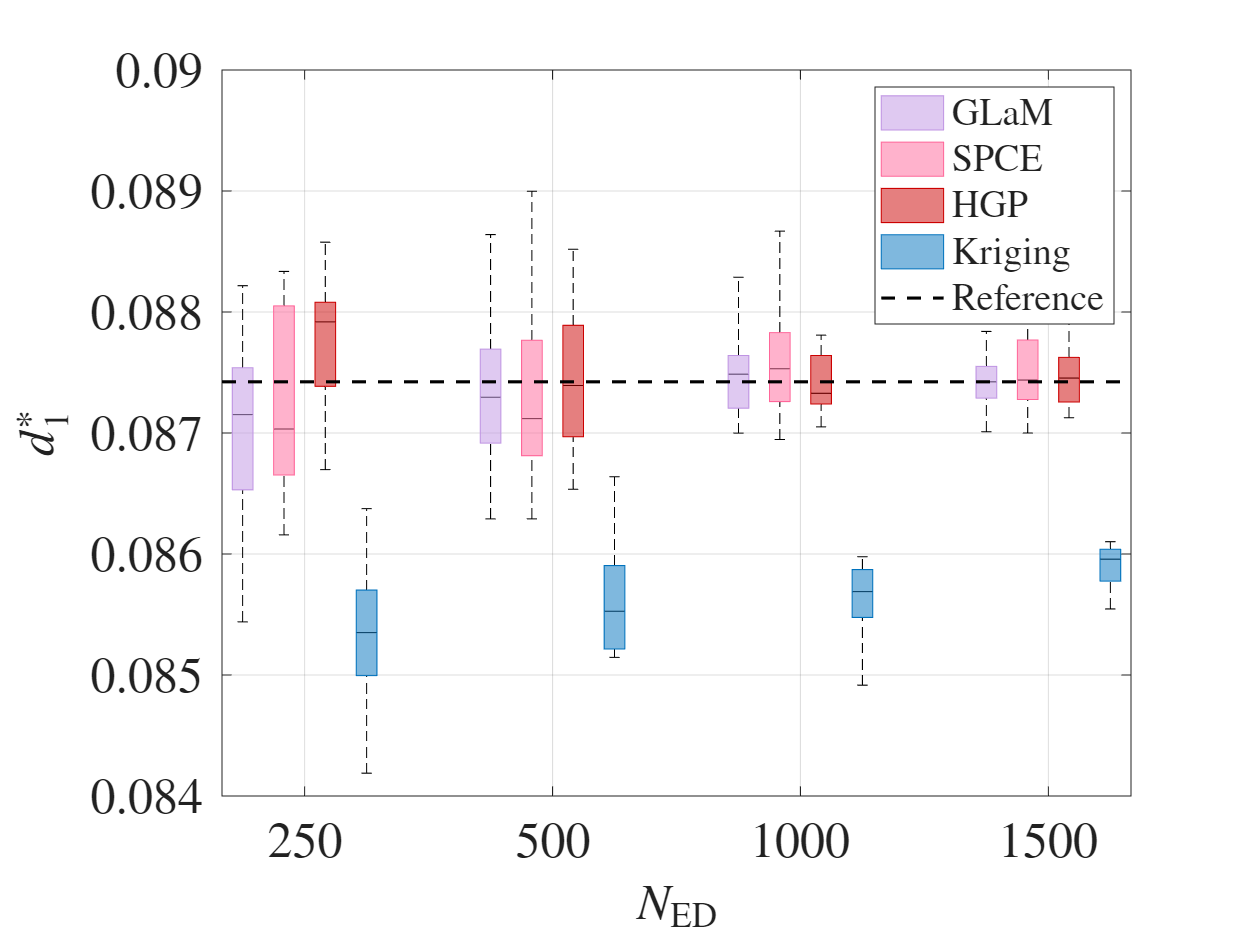}}
	\hfill
	\subfloat[$d_2$]{
		\includegraphics[width=0.48\textwidth]{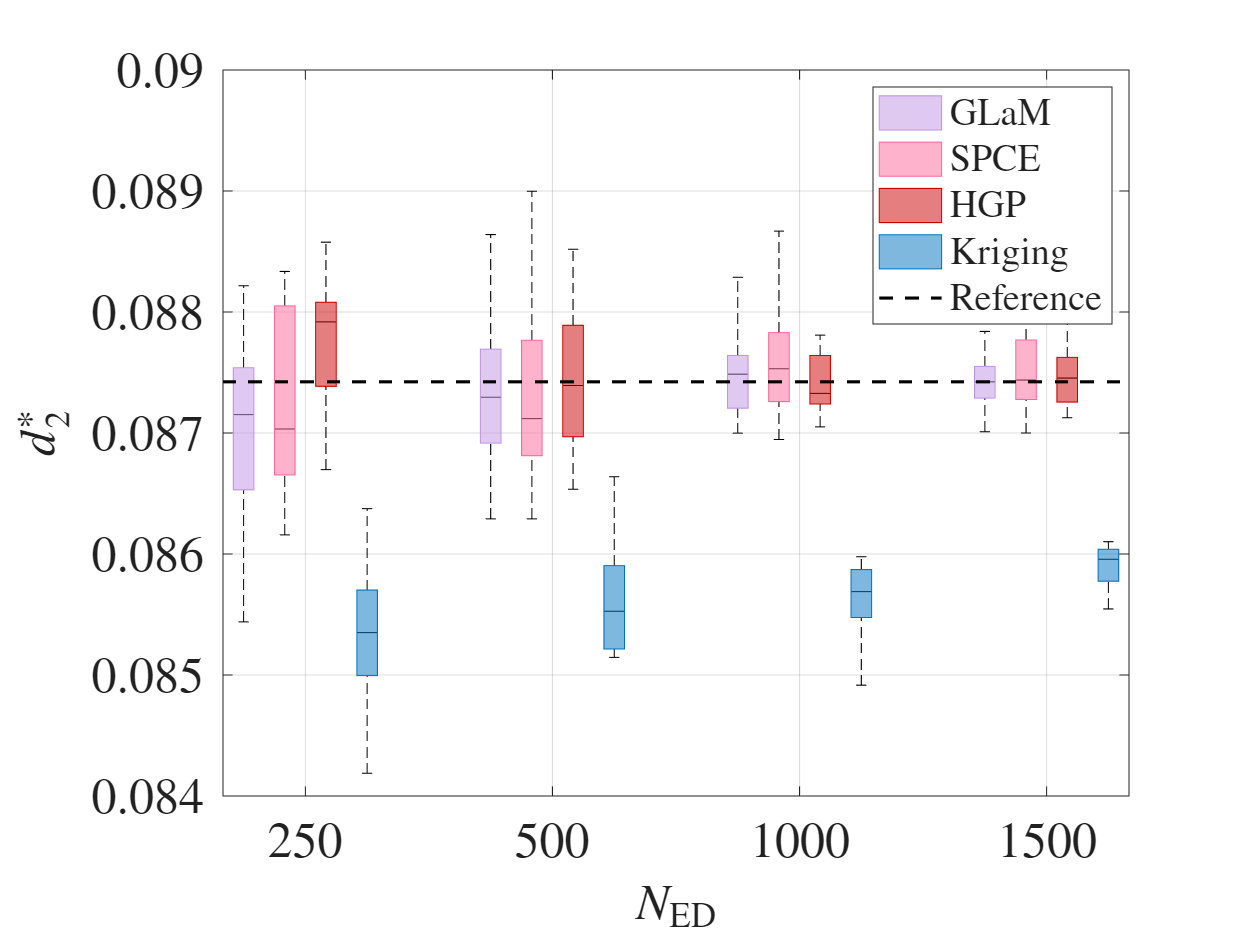}}
	\caption{\color{black}Corroded beam: Boxplots of the optimal designs $b_{0}^{\ast}$ and $h_{0}^{\ast}$ for different methods and experimental design sizes.}
	\label{fig:Ex2:Dstar}
\end{figure}

Figure~\ref{fig:Ex2:Dstar} shows the optimal designs obtained by each of the methods. For this example, all methods yield $d_{1}^{\ast} = d_{2}^{\ast}$, that is, $b_{0}^{\ast} = h_{0}^{\ast}$, showing that the soft constraint is saturated. In both cases, the optimal design converges to the reference solution as the sample size is increased. 

Figure~\ref{fig:Ex2:Density} presents the conditional distributions of the performance function at the reference {\color{black} solution} $\ve{d}^{\ast,\textrm{ref}}$, as obtained using the different models {\color{black}(GLaM, SPCE, HGP and Kriging)}. For each experimental design size, we report the surrogate corresponding to the median error over the 15 independent repetitions. {\color{black}For conciseness, only the smallest and largest experimental designs, $N_{\textrm{ED}} = 250$ and $1{,}500$, are shown.}
The results confirm the conclusions drawn from the boxplots. Both GLaM and SPCE yield conditional distributions that are very remarkably close to the reference distribution obtained from the original model. For small ED sizes, minor discrepancies are observed, but GLaM achieves an almost perfect match as the ED size increases. Although the SPCE approximation remains slightly less accurate overall, it still exhibits very good agreement with the reference distribution, particularly in the tails, which are critical for reliability analysis. In contrast, Kriging consistently fails to reproduce the correct conditional distributions for all ED sizes, exhibiting noticeable discrepancies both in the central region and in the tails. {\color{black}The HGP model, in contrast, reproduces the reference distribution accurately, on par with the stochastic emulators and markedly better than Kriging.}
{\color{black}
	\begin{figure}[H]
		\centering
		\subfloat[GLaM -- $N_{\textrm{ED}} = 250$]{\includegraphics[width=0.33\textwidth]{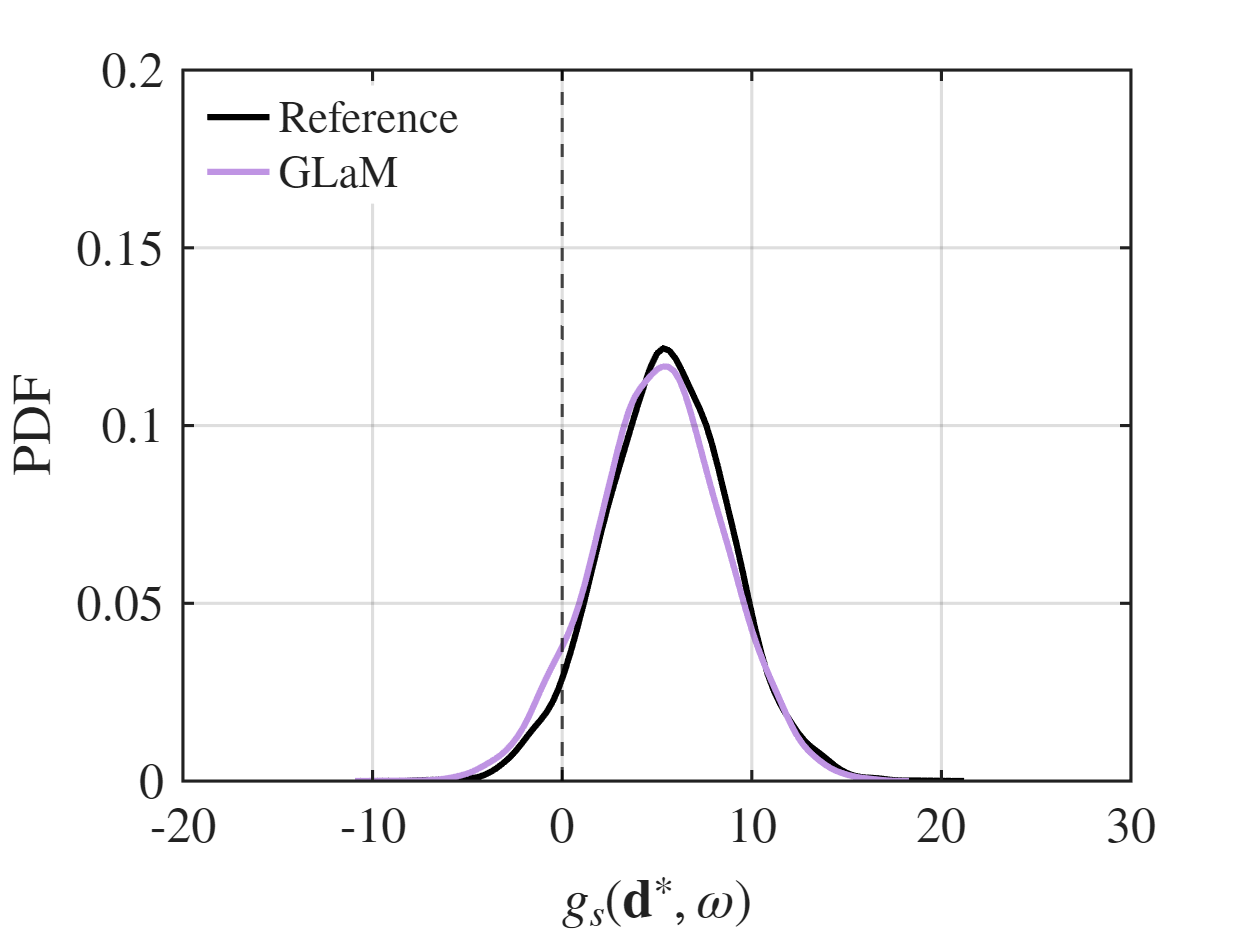}}
		\subfloat[SPCE -- $N_{\textrm{ED}} = 250$]{\includegraphics[width=0.33\textwidth]{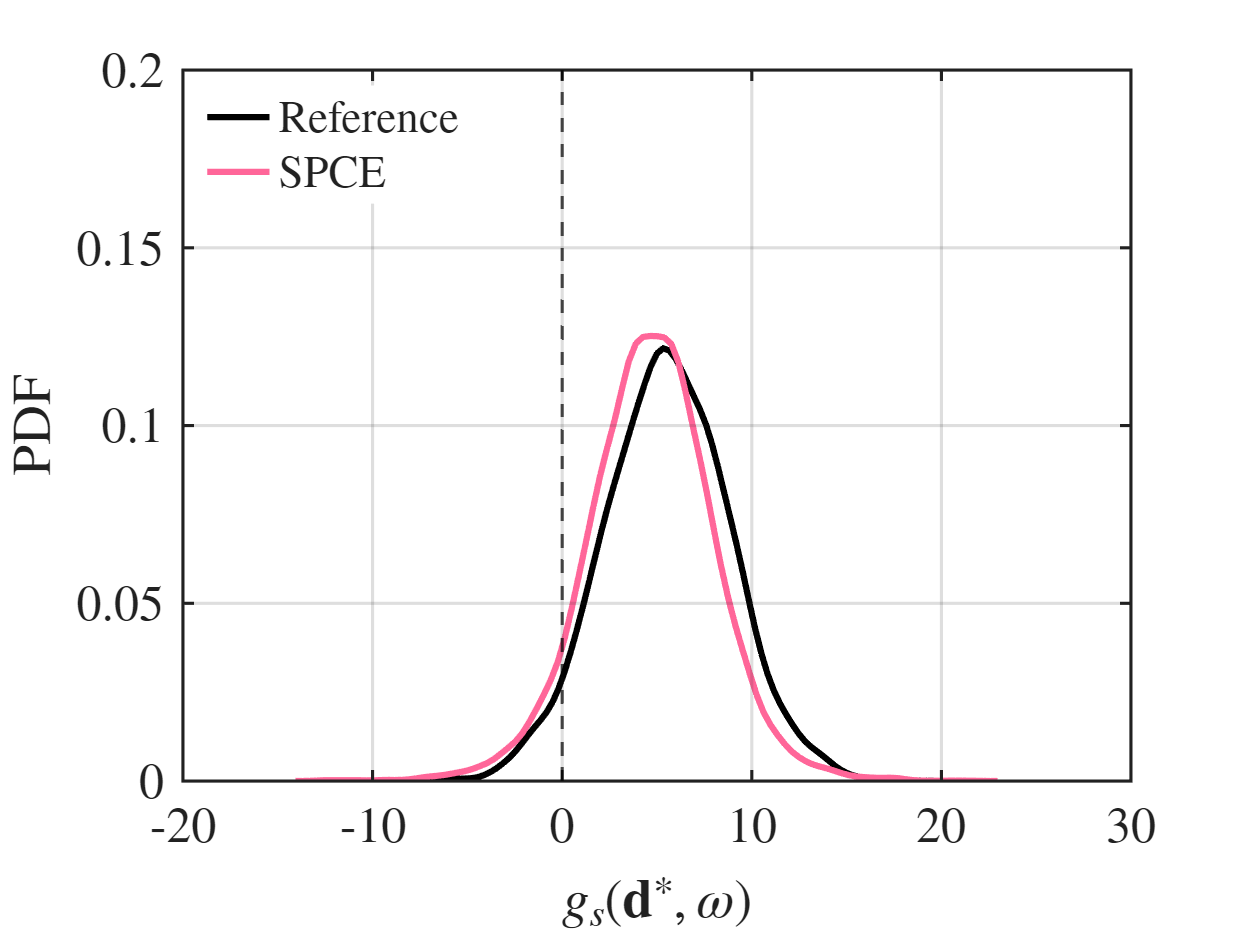}}
		\subfloat[HGP -- $N_{\textrm{ED}} = 250$]{\includegraphics[width=0.33\textwidth]{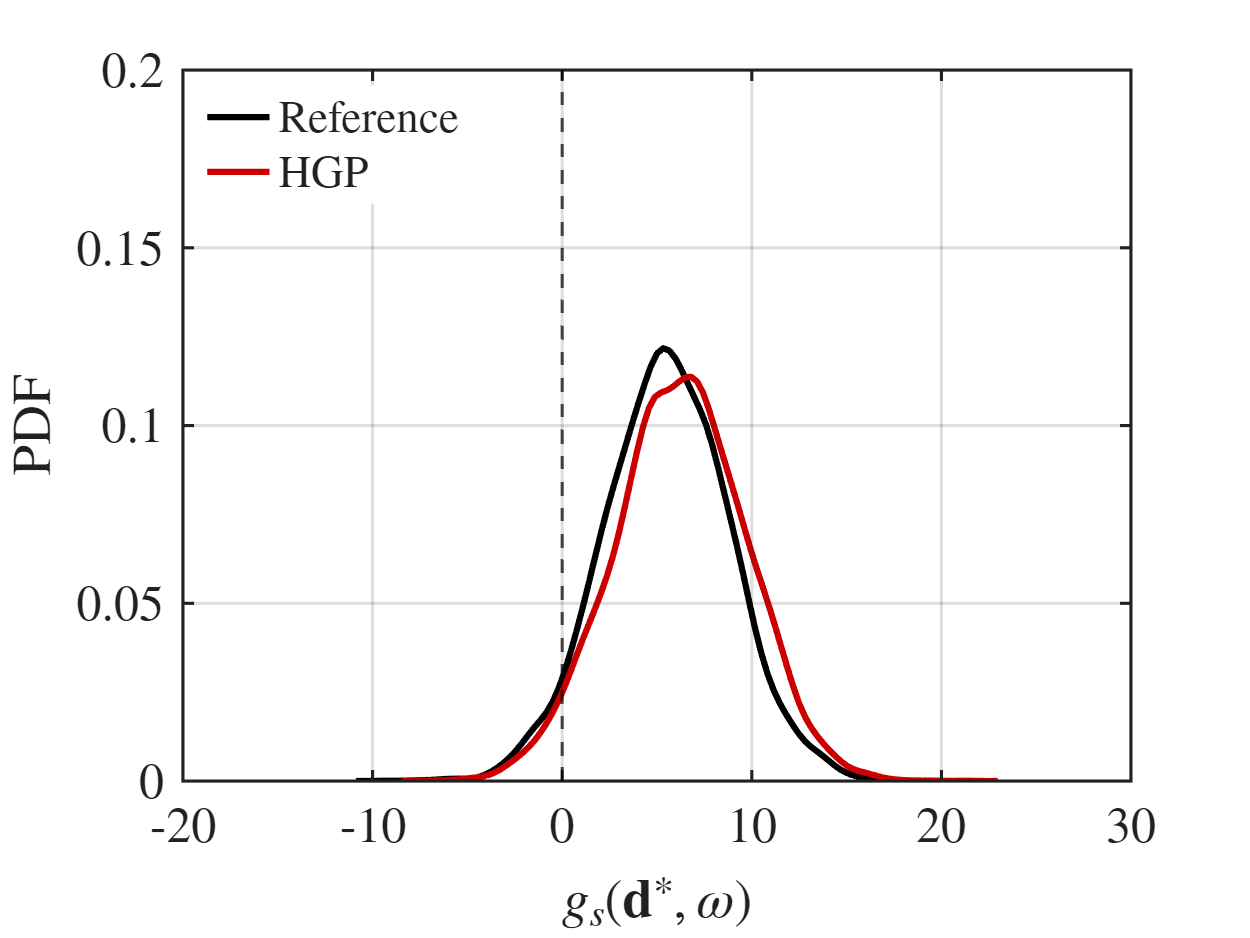}} \\
		%
		\subfloat[GLaM -- $N_{\textrm{ED}} = 1{,}500$]{\includegraphics[width=0.33\textwidth]{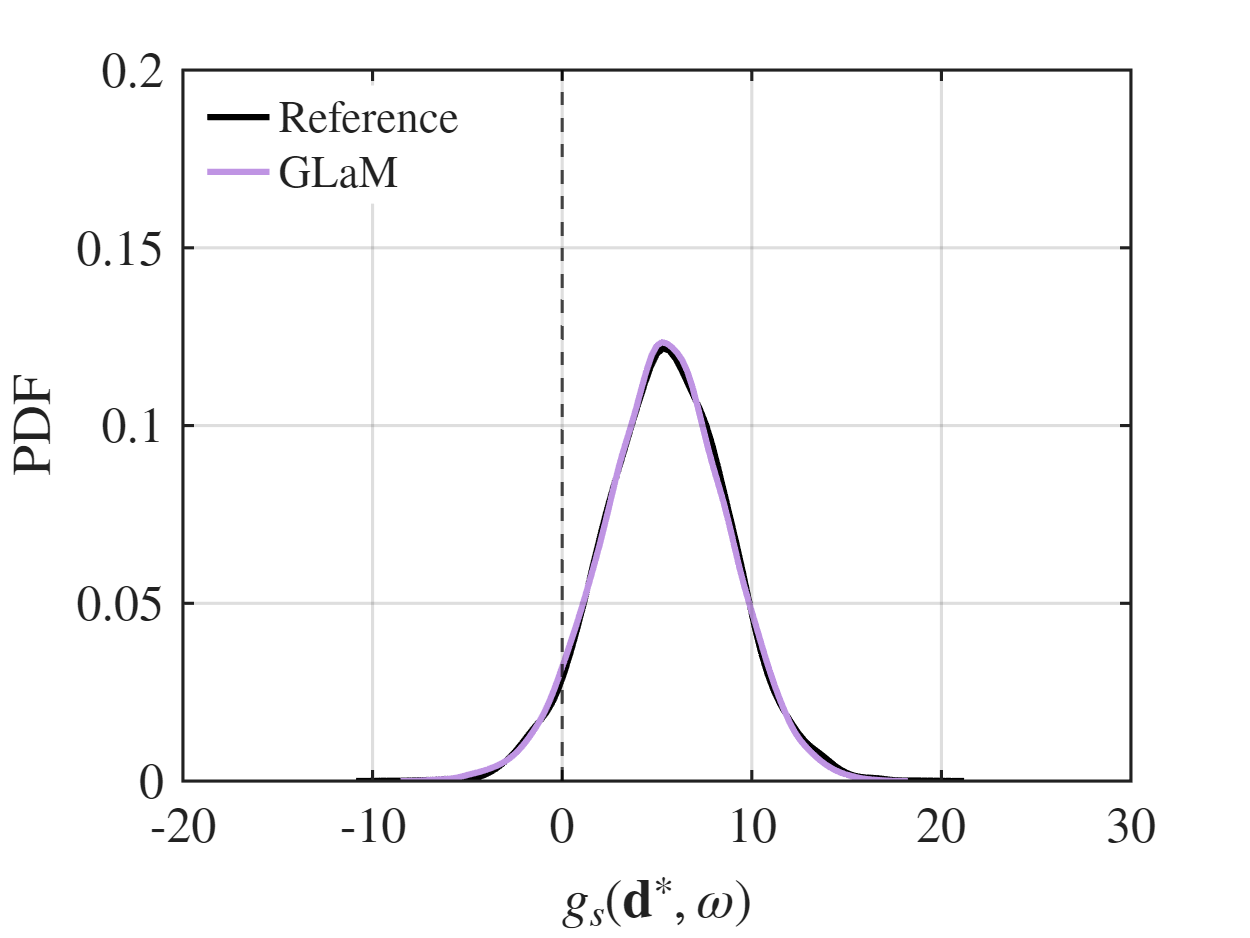}}
		\subfloat[SPCE -- $N_{\textrm{ED}} = 1{,}500$]{\includegraphics[width=0.33\textwidth]{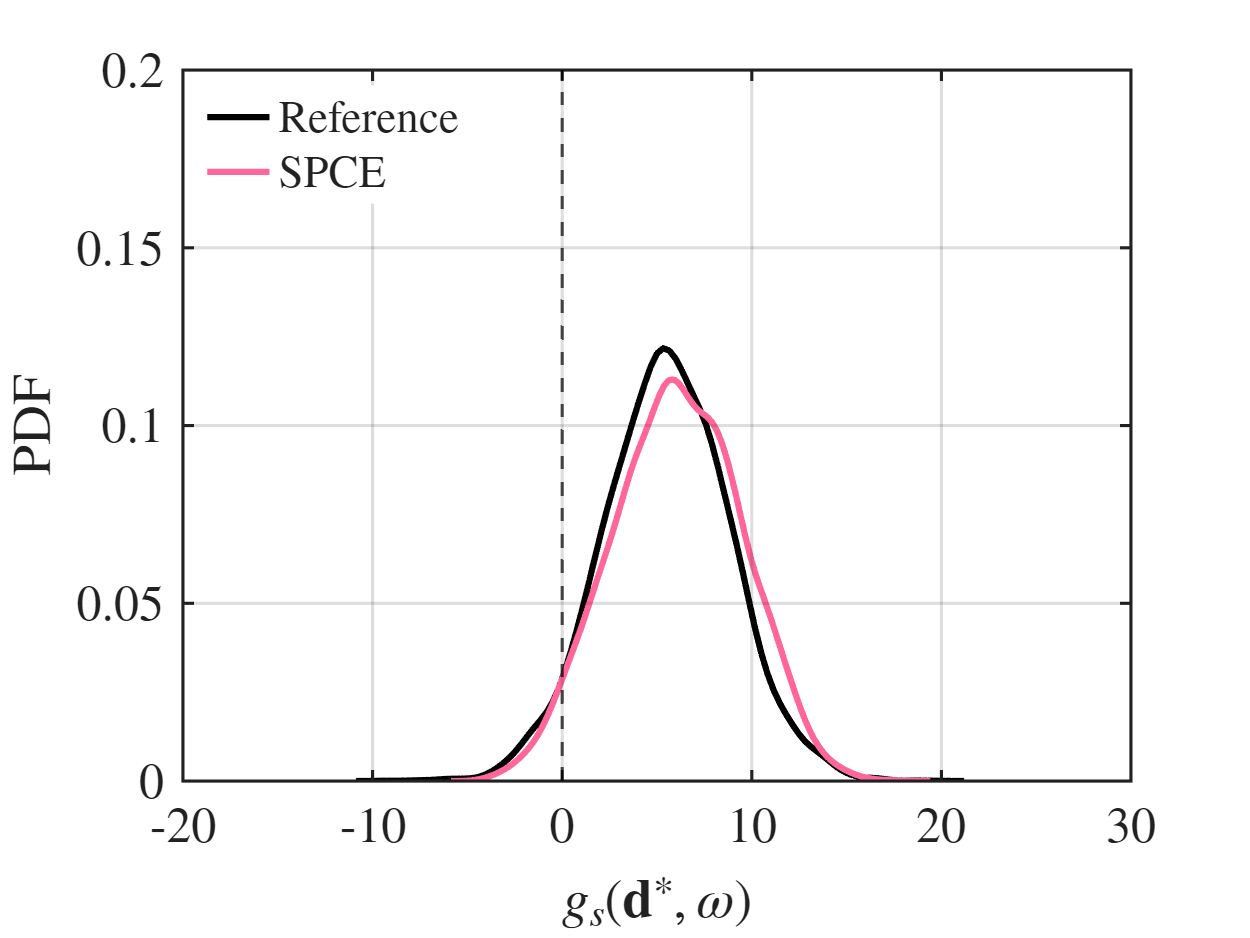}}
		\subfloat[HGP -- $N_{\textrm{ED}} = 1{,}500$]{\includegraphics[width=0.33\textwidth]{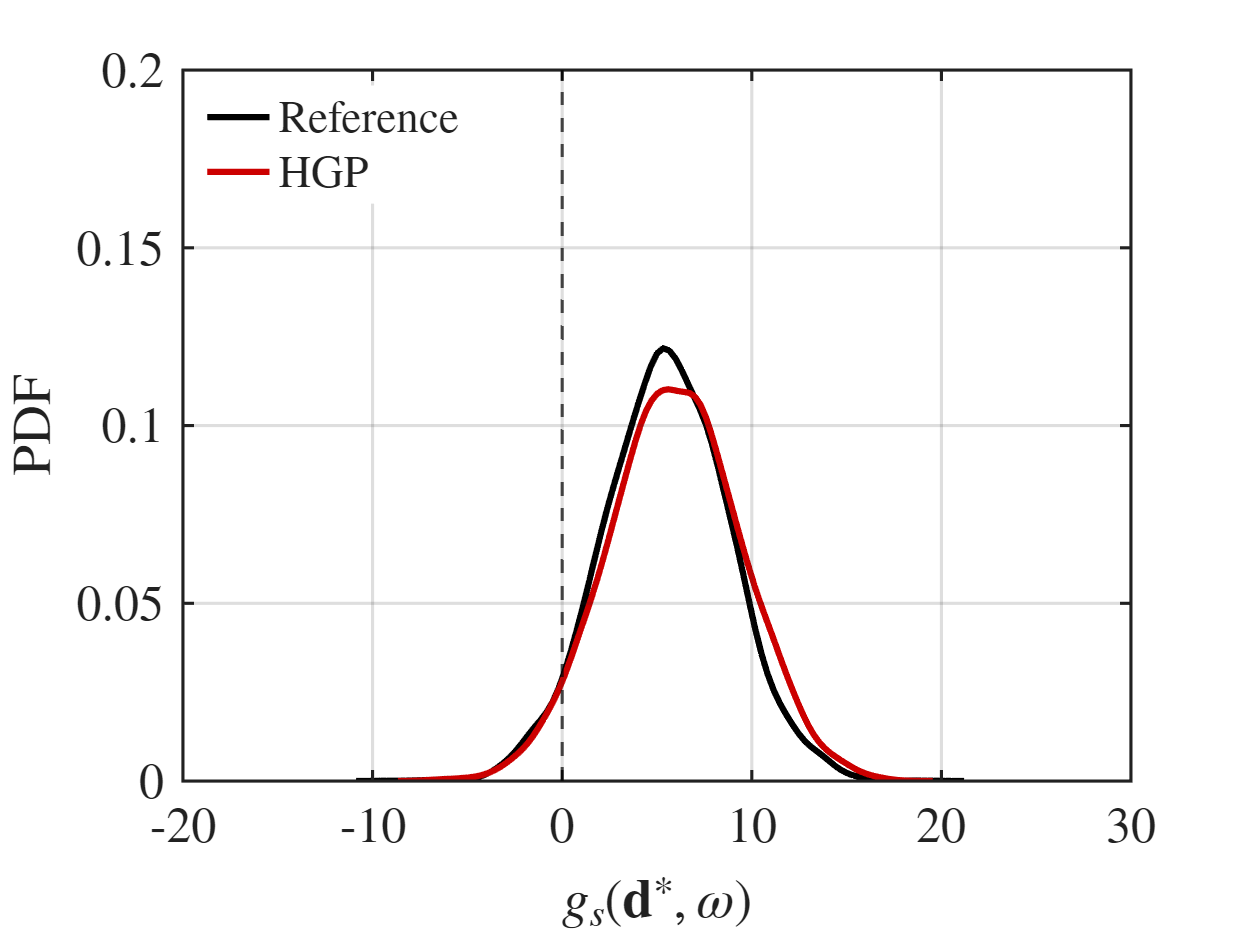}} \\
		\subfloat[Kriging -- $N_{\textrm{ED}} = 250$]{\includegraphics[width=0.33\textwidth]{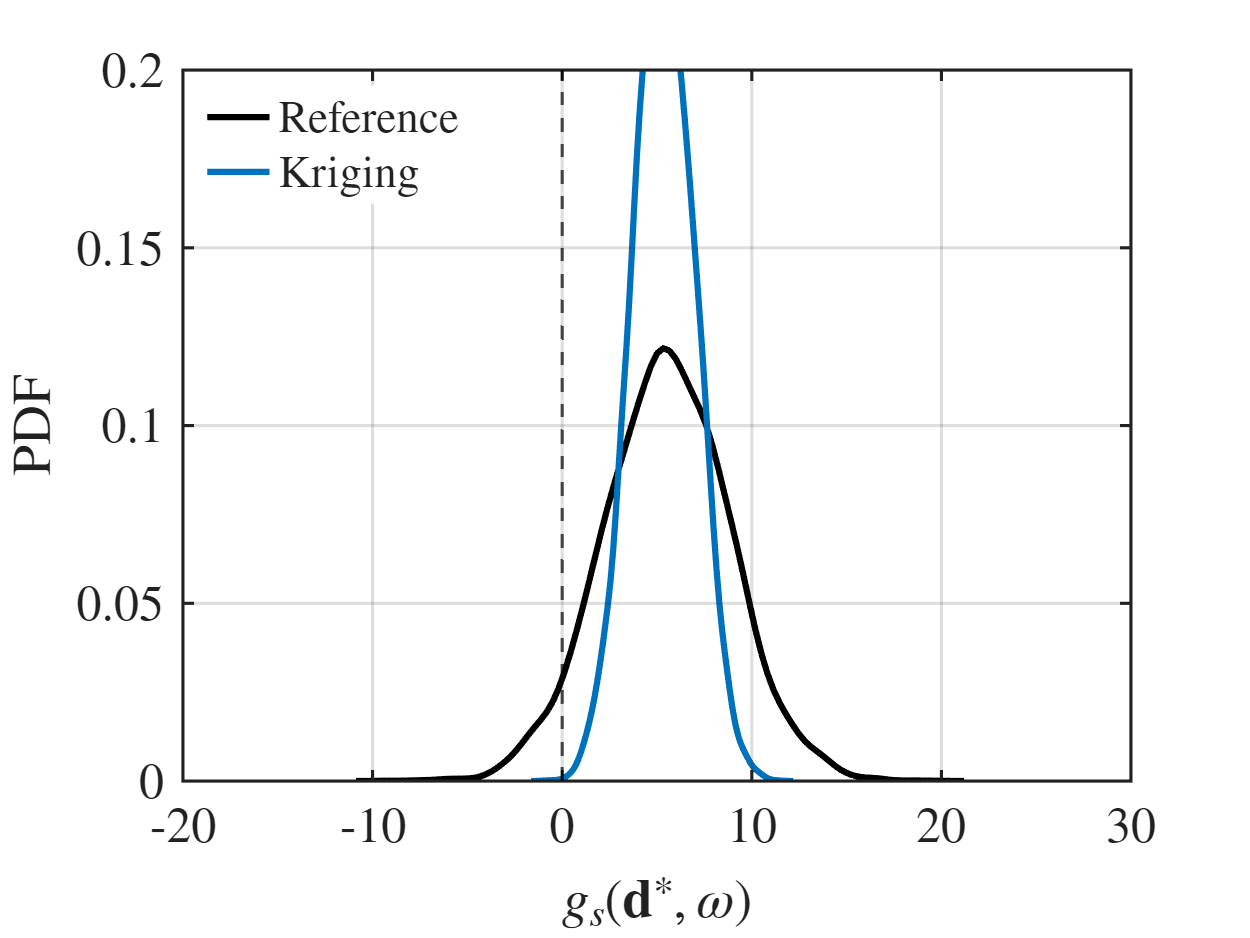}}
		\subfloat[Kriging -- $N_{\textrm{ED}} = 1{,}500$]{\includegraphics[width=0.33\textwidth]{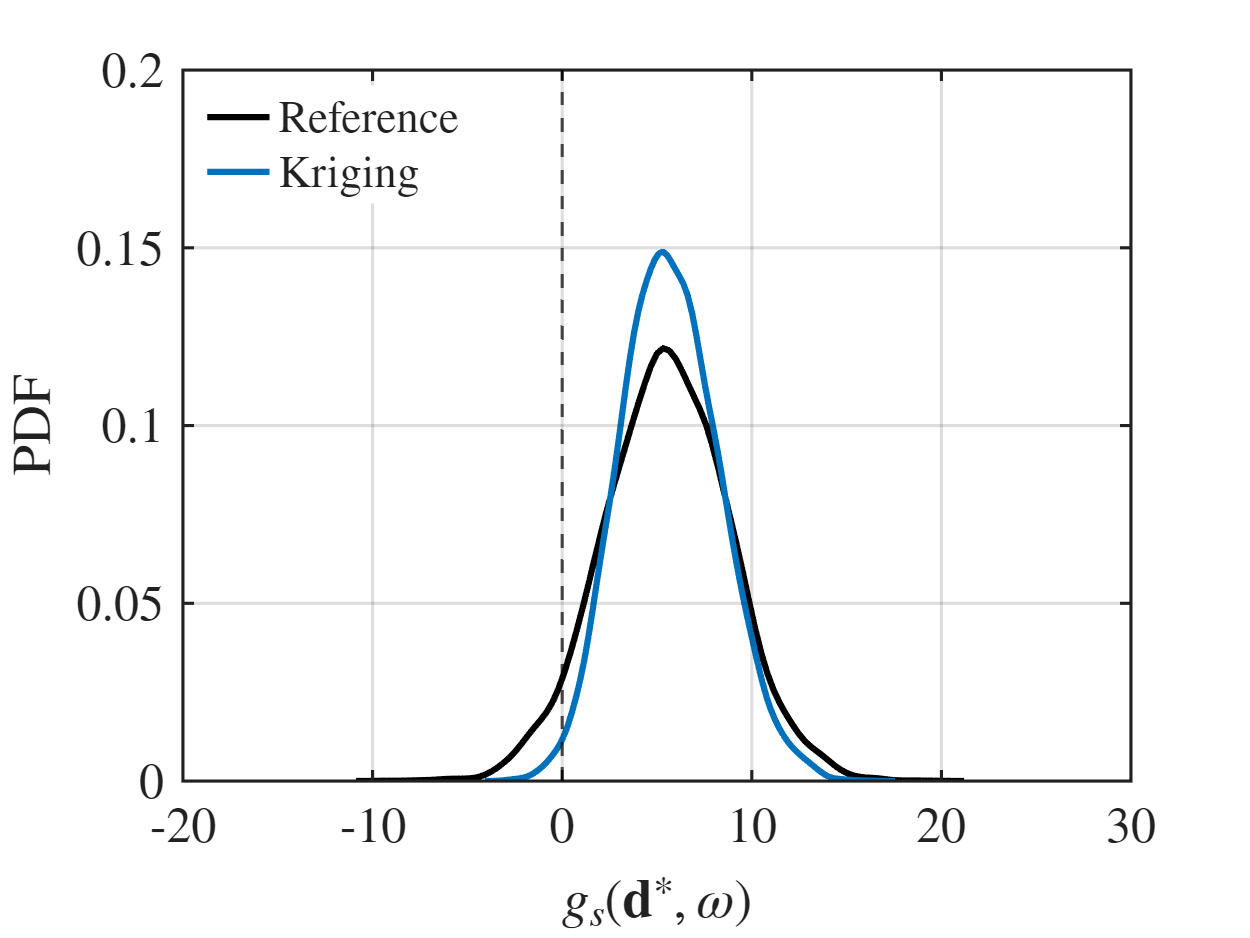}} \\
		\caption{\color{black} Corroded beam: Conditional probability density functions of the stochastic limit-state function at the reference optimal design $\ve{d}^{\ast,\textrm{ref}}$, obtained with each surrogate (median-error model) for $N_{\textrm{ED}} = 250$ and $N_{\textrm{ED}} = 1{,}500$. The reference distribution obtained from the original model is shown in black.}
		\label{fig:Ex2:Density}
	\end{figure}
}

Finally, we compare the estimated quantile surfaces obtained with the different surrogate models, denoted by $\hat{q}_{\alpha}(\ve{d})$ for $\ve{d} \in \mathbb{D}$, against the reference quantiles $q_{\alpha}(\ve{d})$ computed using the original model.  

For the GLaM surrogate, the quantile is obtained analytically using Eq.~\eqref{eq:GLDquantile}. In the case of SPCE, the quantile is computed by numerically inverting the conditional cumulative distribution function given in Eq.~\eqref{eq:cdf_spce}. For Kriging and for the original model, the quantiles are estimated via Monte Carlo simulation by sampling $N = 10^4$ realizations of the environmental random vector $\ve{Z}$ and evaluating the corresponding deterministic limit-state functions. Common random numbers \citep{Spall2003,Taflanidis2008} are used to avoid noise due to sampling variability.

\begin{figure}[H]
	\centering
	\includegraphics[width=0.65\textwidth]{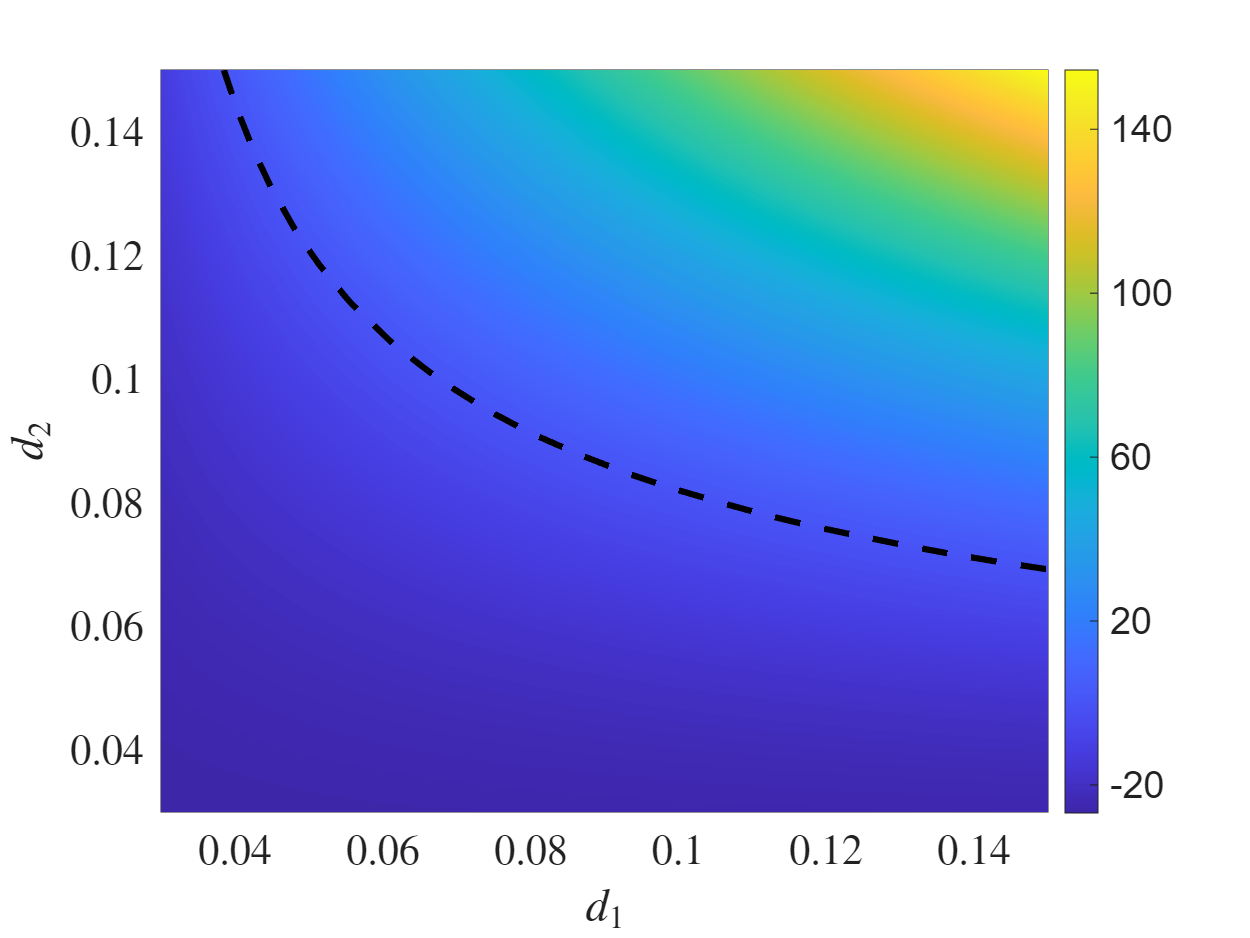}
	\caption{Corroded beam: Quantiles $q_{0.05}\prt{\ve{d}}$ of the stochastic limit-state function evaluated over the design space using the original model. The dashed black line corresponds to the zero value.}
	\label{fig:Ex2:Quantiles:Ori}
\end{figure}

\begin{figure}[H]
	\centering
	\subfloat[GLaM - $N_{\textrm{ED}} = 250$]{
		\includegraphics[width=0.48\textwidth]{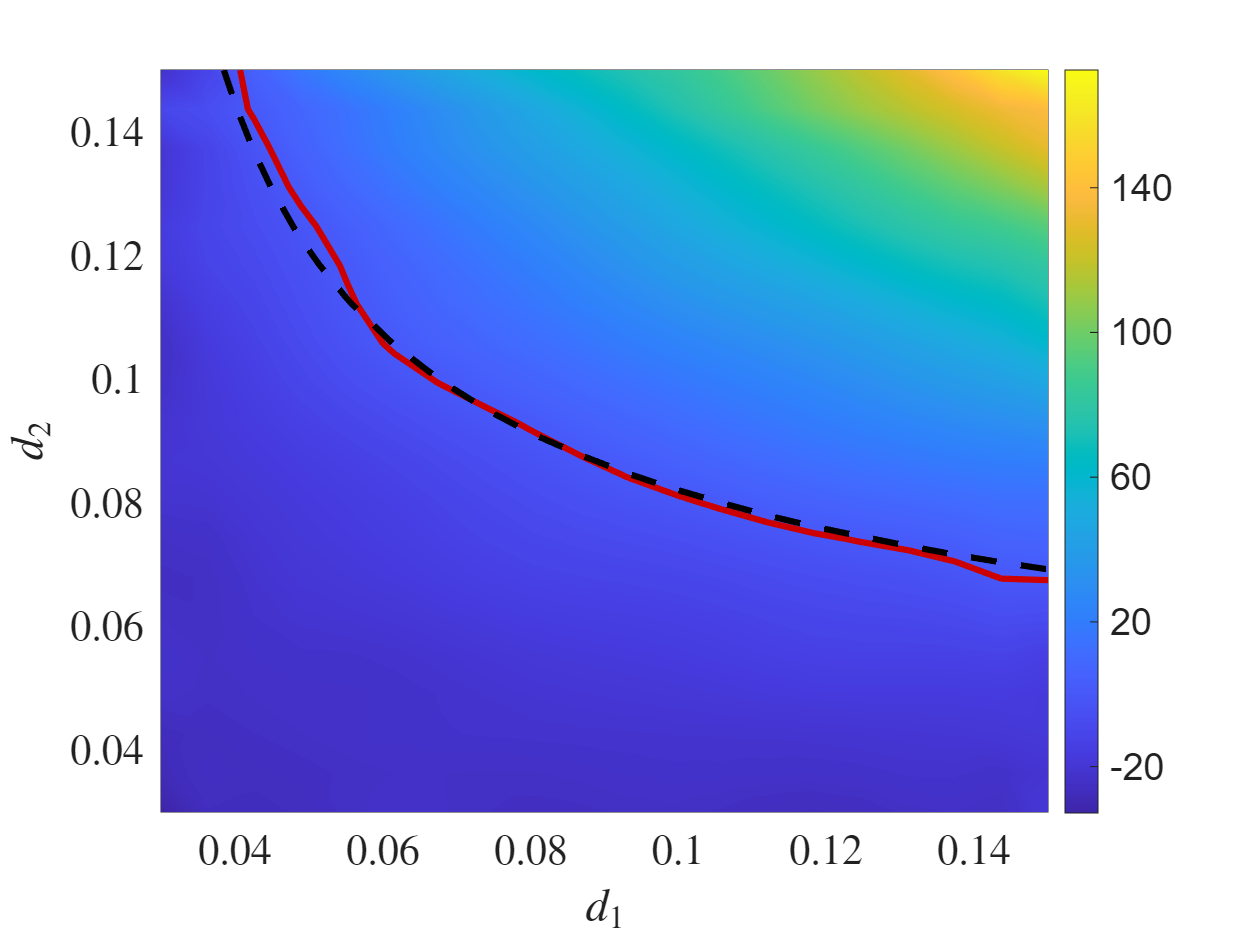}}
	\hfill
	\subfloat[GLaM - $N_{\textrm{ED}} = 1{,}500$]{
		\includegraphics[width=0.48\textwidth]{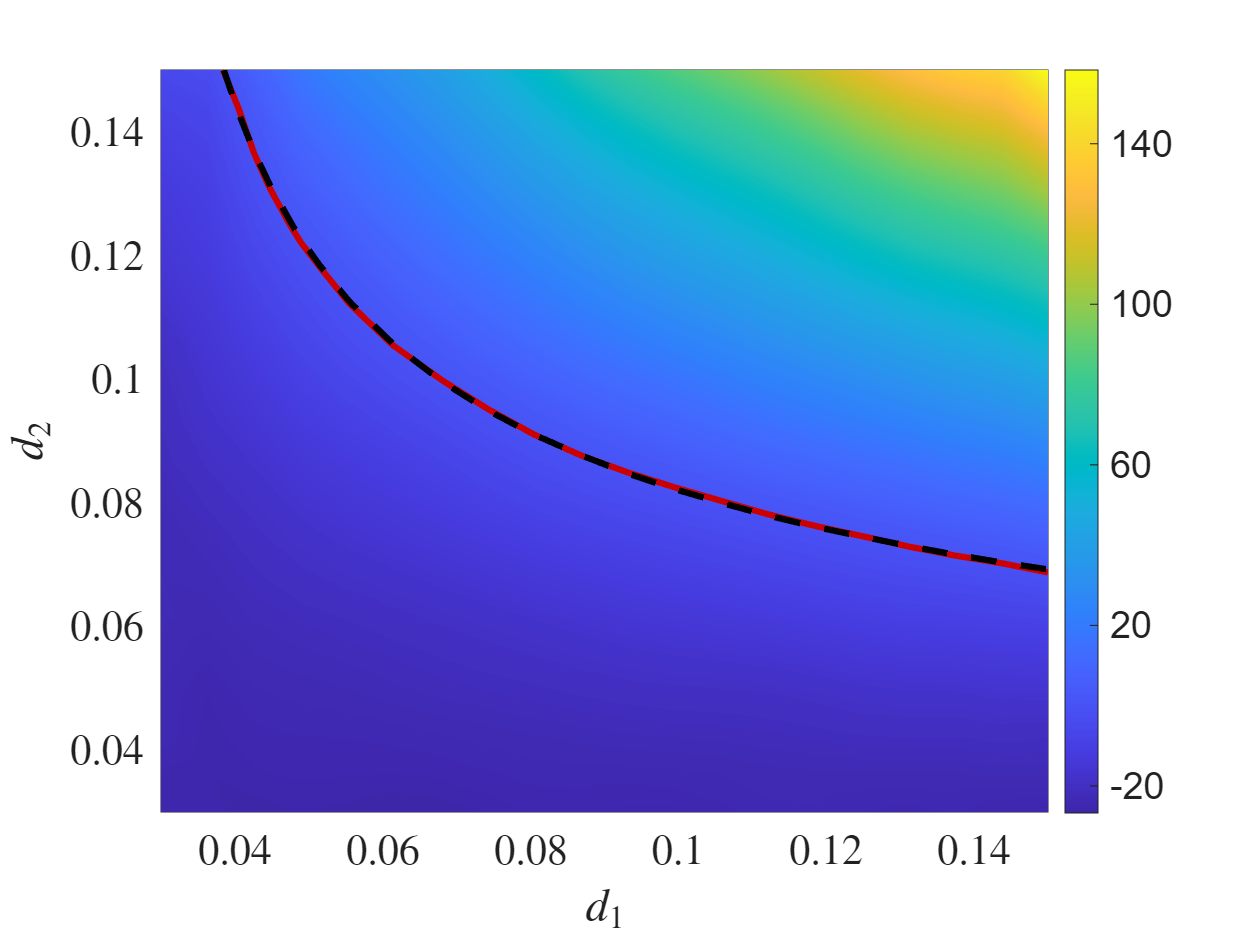}} \\
	\subfloat[SPCE - $N_{\textrm{ED}} = 250$]{
		\includegraphics[width=0.48\textwidth]{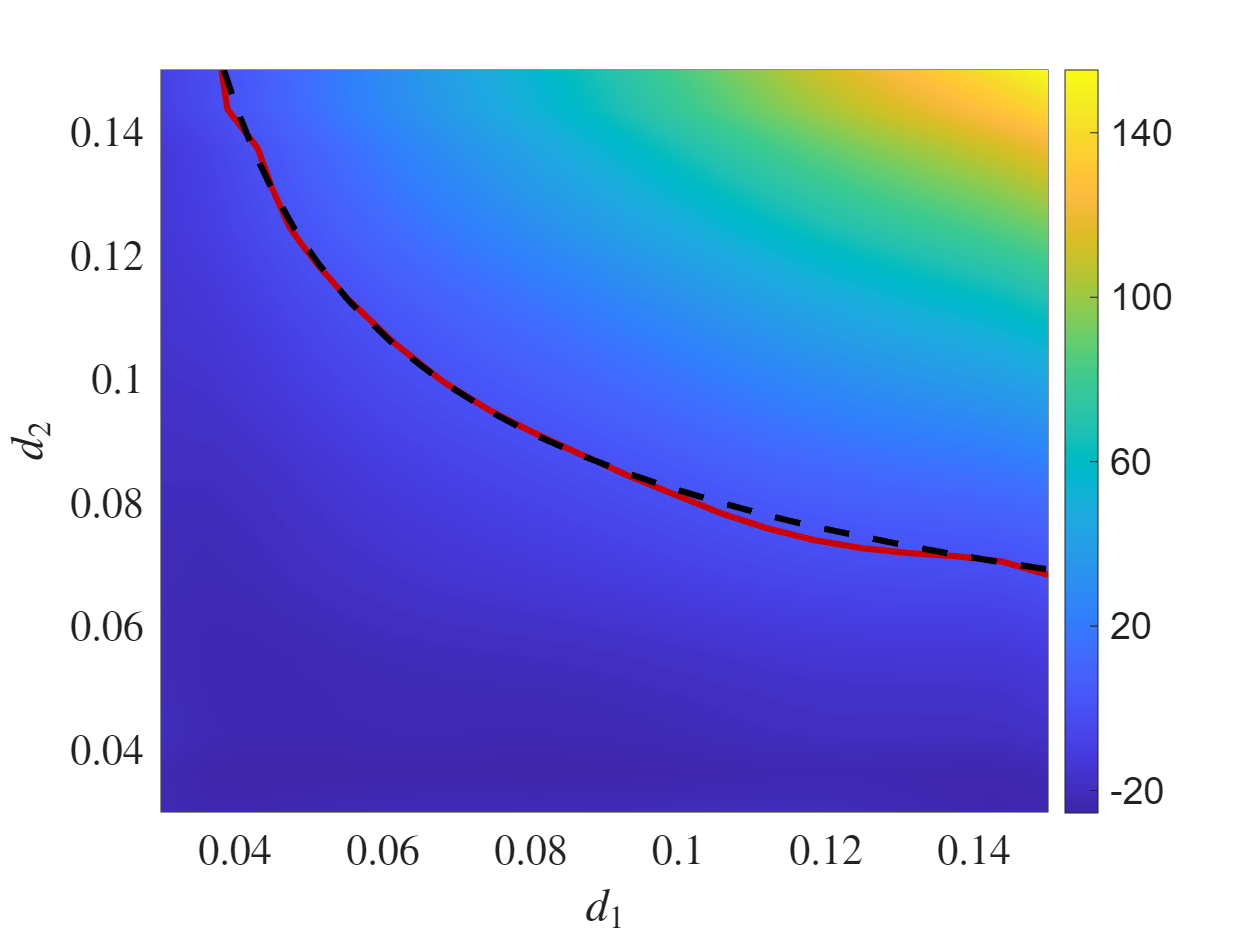}}
	\hfill
	\subfloat[SPCE - $N_{\textrm{ED}} = 1{,}500$]{
		\includegraphics[width=0.48\textwidth]{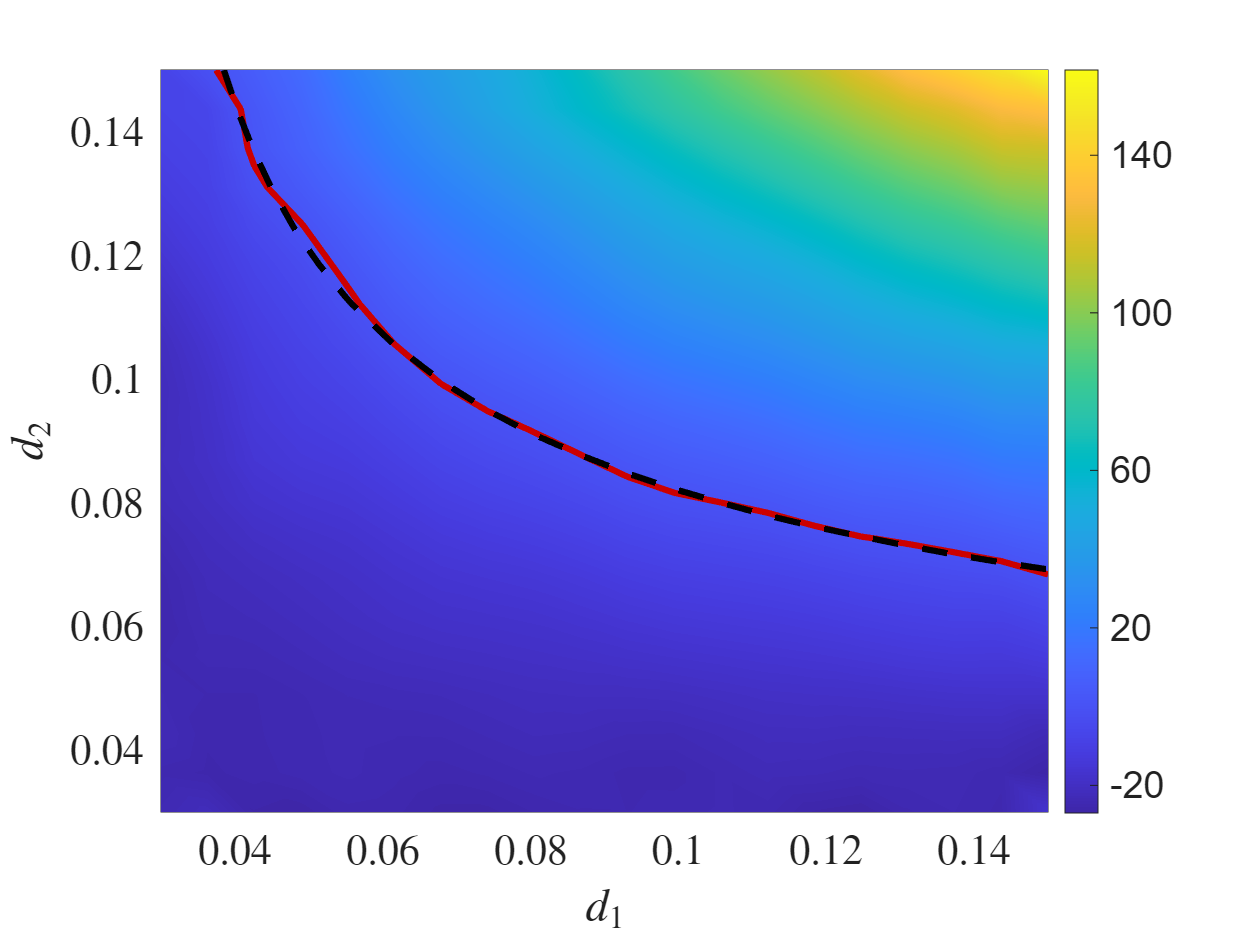}} \\
	\caption{Corroded beam:  Quantiles $q_{0.05}\prt{\ve{d}}$ of the stochastic limit-state function evaluated over the design space using the {\color{black}GLaM and SPCE} surrogate models. The dashed black line corresponds to the zero value obtained from the original limit-state function. The red line corresponds to the surrogate.}
	\label{fig:Ex2:Quantiles:Sur}
\end{figure}

Figure~\ref{fig:Ex2:Quantiles:Ori} shows the reference quantiles obtained from the original model, while the quantile surfaces estimated using the surrogate models are presented in Figures~\ref{fig:Ex2:Quantiles:Sur} {\color{black}and~\ref{fig:Ex2:Quantiles:Sur2}}. In all cases, the quantiles are evaluated on a regular grid of size $20 \times 20$ built on the design space $\mathbb{D}$. The black dotted lines represent the set
\[
\left\{ \ve{d} \in \mathbb{D} : q_{\alpha}(\ve{d}) = 0 \right\},
\]
which corresponds to the designs that exactly satisfy the probabilistic constraint and where the optimal solution is expected to lie. The analogous sets obtained from the surrogate models,
\[
\left\{ \ve{d} \in \mathbb{D} : \hat{q}_{\alpha}(\ve{d}) = 0 \right\},
\]
are shown by the red curves.

Overall, GLaM and SPCE yield quantile surfaces that are in excellent agreement with the reference solution throughout the design space. For smaller experimental design sizes, herein $N_{\textrm{ED}} = 250$, minor discrepancies are observed for larger quantile values (upper right corner). However, these differences progressively vanish as the ED size increases to $N_{\textrm{ED}} = 1{,}500$. More importantly for the RBDO problem, the quantiles that saturate the constraint are accurately approximated, even at relatively small ED sizes.

In contrast, Kriging struggles to reproduce the correct quantile surfaces and fails to accurately identify the set that saturates the constraint, regardless of the ED size. This limitation is consistent with the previously observed deficiencies of Kriging in approximating the conditional response distributions, particularly in the tails that govern reliability-driven constraints. {\color{black}The HGP model also reproduces the constraint boundary accurately, with only minor discrepancies at the smallest experimental design size, confirming its suitability alongside the stochastic emulators.}
{\color{black}
	\begin{figure}[H]
		\centering
		\subfloat[HGP - $N_{\textrm{ED}} = 250$]{
			\includegraphics[width=0.48\textwidth]{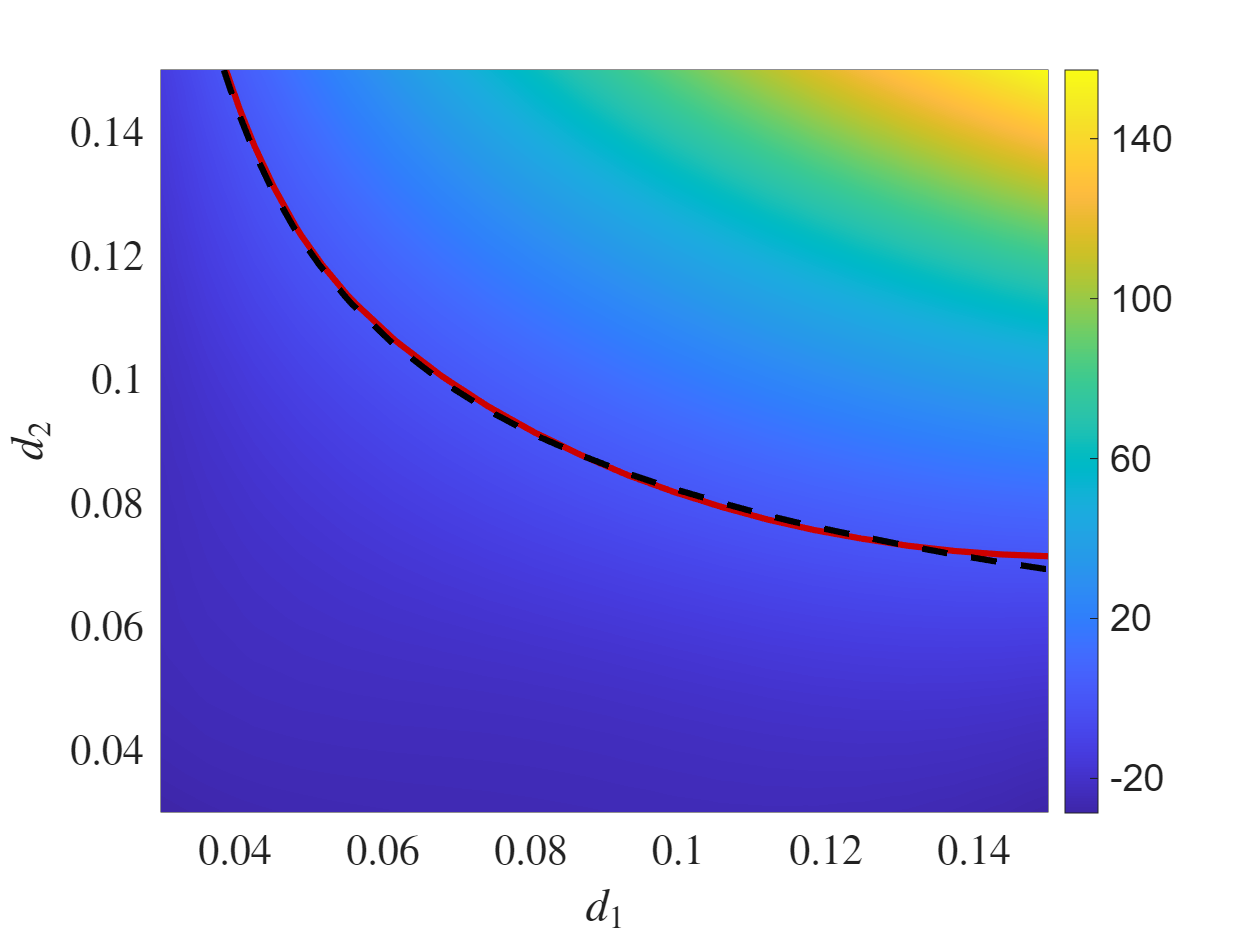}}
		\hfill
		\subfloat[HGP - $N_{\textrm{ED}} = 1{,}500$]{
			\includegraphics[width=0.48\textwidth]{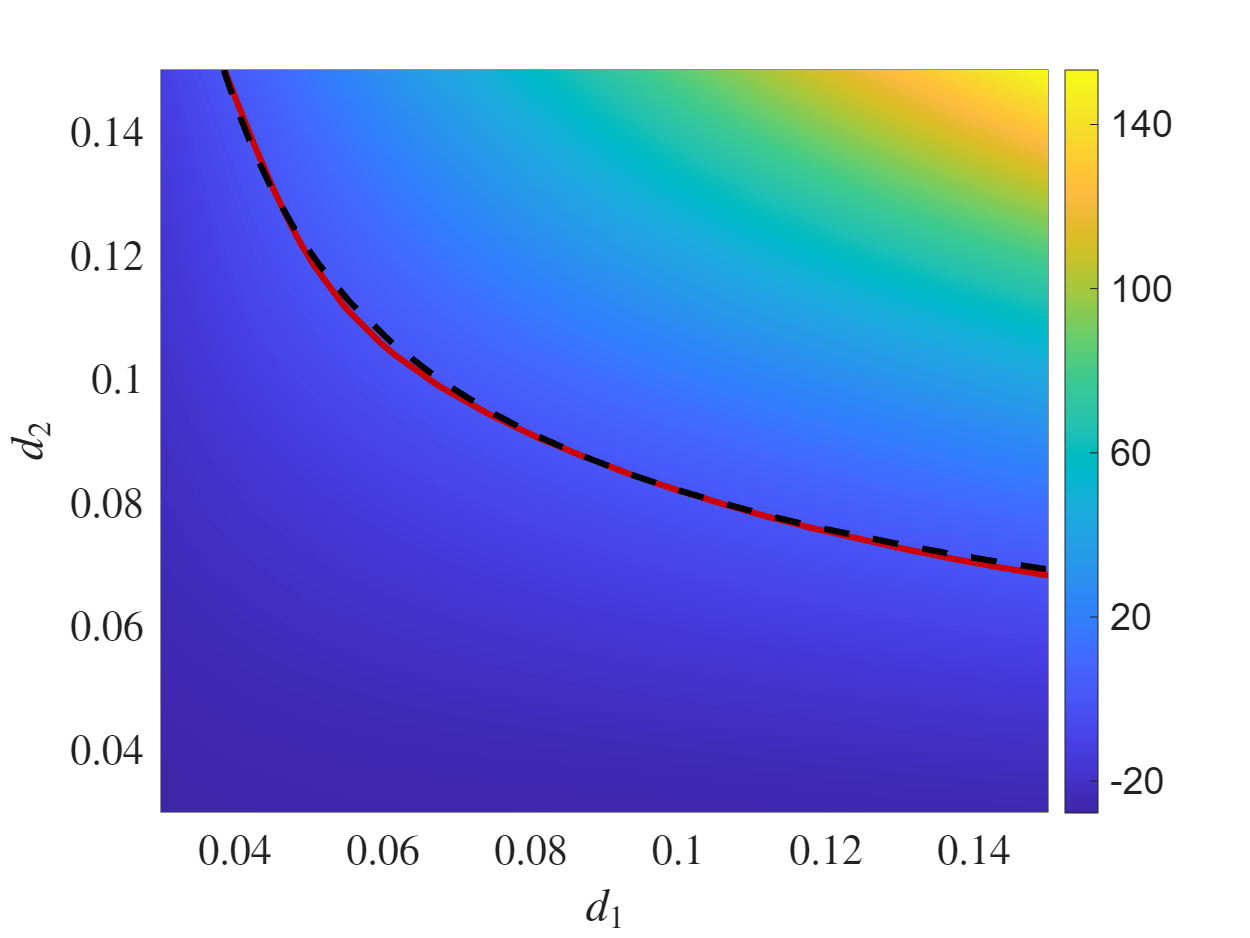}} \\
		\subfloat[Kriging - $N_{\textrm{ED}} = 250$]{
			\includegraphics[width=0.48\textwidth]{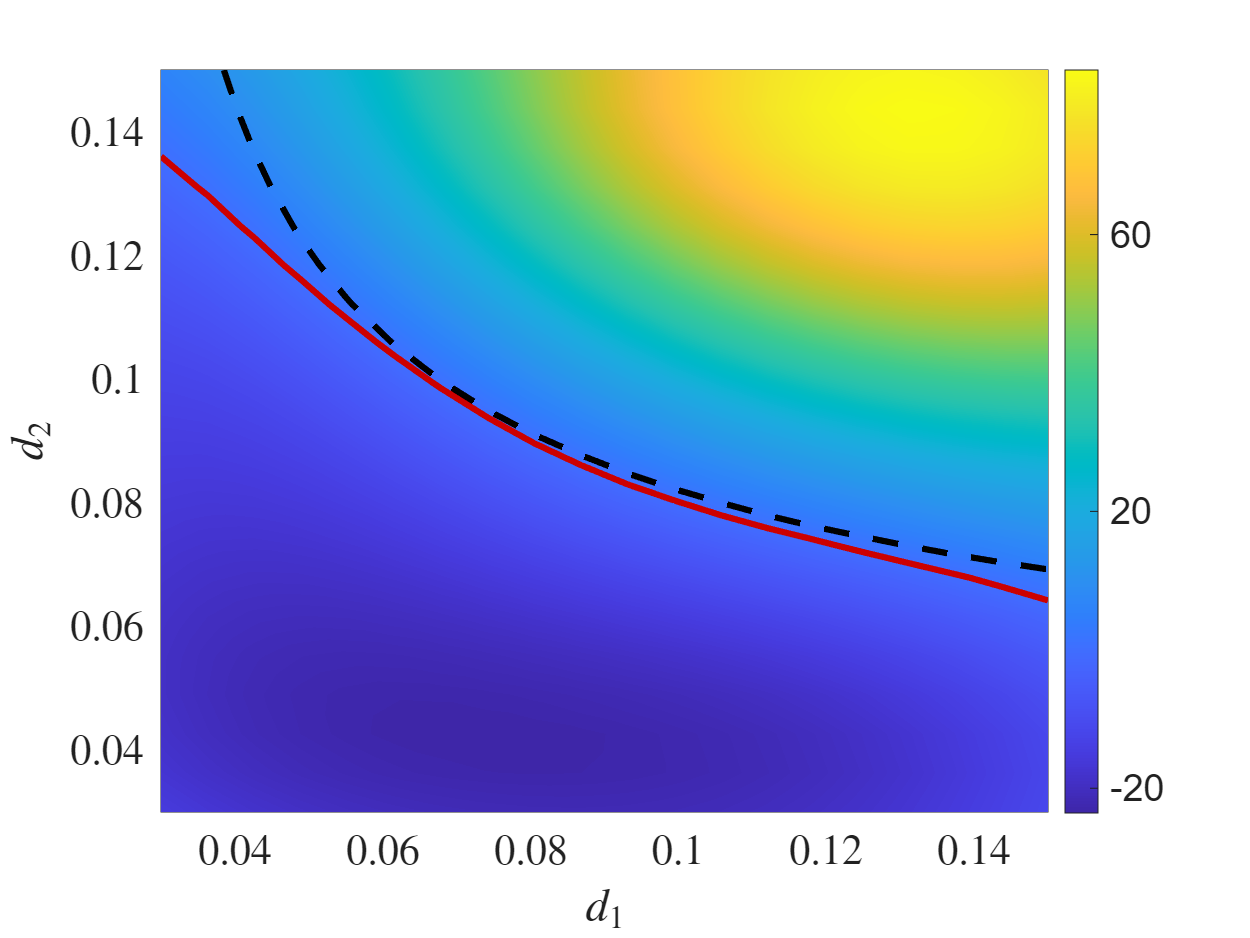}}
		\hfill
		\subfloat[Kriging - $N_{\textrm{ED}} = 1{,}500$]{
			\includegraphics[width=0.48\textwidth]{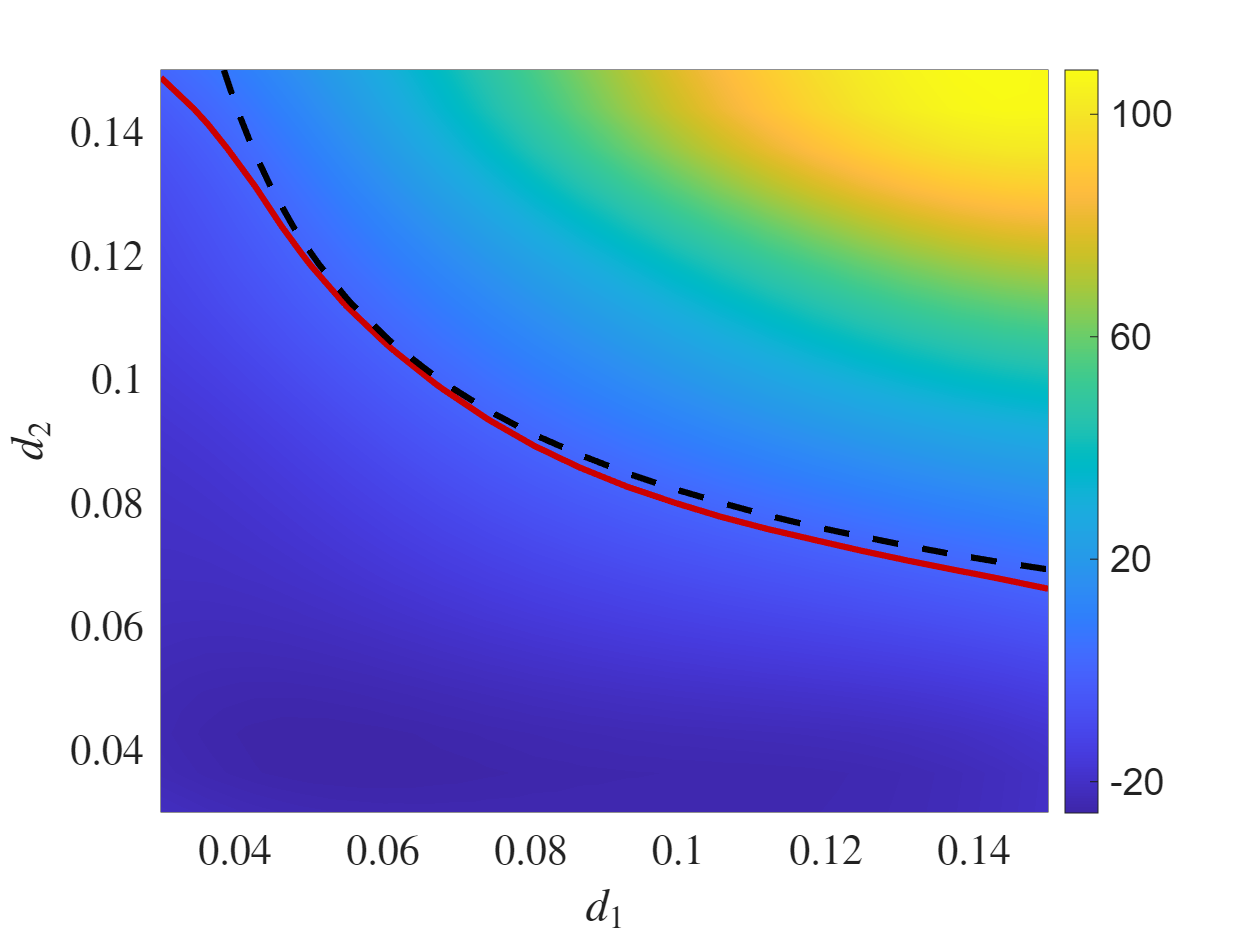}} \\
		\caption{\color{black} Corroded beam:  Quantiles $q_{0.05}\prt{\ve{d}}$ of the stochastic limit-state function evaluated over the design space using the HGP and Kriging surrogate models. The dashed black line corresponds to the zero value obtained from the original limit-state function. The red line corresponds to the surrogate.}
		\label{fig:Ex2:Quantiles:Sur2}
	\end{figure}
}

In addition to providing more accurate results than Kriging, the proposed stochastic-emulator–based strategies also offer substantial computational savings. Table~\ref{tab:Ex2:computtime} reports the average computational times, measured on a standard laptop, over five independent runs for the different methods and {\color{black}four} experimental design (ED) sizes, namely $N_{\textrm{ED}} = \acc{250,\, 500, \,1{,}000,\,1{,}500}$. As in the previous example, the optimization using GLaM converges in all cases within approximately $0.03$ seconds. The SPCE-based approach also converges in less than one second for all ED sizes. {\color{black}HGP, whose failure probability is likewise available in closed form, converges just as fast, in a few hundredths of a second.} In contrast, the Kriging-based approach takes two to three orders of magnitude more computational time. The computational cost increases as $N_{\textrm{ED}}$ becomes larger. It is even more dramatic when the MC sample size is larger. When using $N_{\textrm{MCS}} = 10^6$, the computational time ranges on average from approximately $104$ to $617$ seconds. This behavior is expected, as Kriging involves the manipulation of  dense matrices of size $N_{\textrm{ED}} \times N_{\textrm{MCS}}$ and inversion of matrices of size $N_{\textrm{ED}} \times N_{\textrm{ED}}$ during model evaluations, leading to poor scalability with increasing ED size.
\begin{table}[ht]
	\centering
	\begin{tabular}{lccccc}
		\hline
		$N_{\textrm{ED}}$ & $250$ & $500$ & $1000$ & $1500$  \\
		\hline
		GLaM 
		& $0.03$ 
		& $0.03$
		& $0.03$
		& $0.03$ \\
		SPCE 
		& $0.29$ 
		& $0.16$ 
		& $0.41$ 
		& $0.24$ \\
		\textcolor{black}{HGP} & \textcolor{black}{$0.05$} & \textcolor{black}{$0.02$} & \textcolor{black}{$0.07$} & \textcolor{black}{$0.11$} \\
		Kriging
		& $11.23$ 
		& $18.75$ 
		& $46.44$     
		& $69.02$ \\
		\hline
	\end{tabular}
	\caption{Corroded beam: Computational time (in seconds) of the optimization procedure for different experimental design sizes and surrogate modeling methods.}
	\label{tab:Ex2:computtime}
\end{table}

{\color{black}
	The corresponding training-time comparison is shown in Figure~\ref{fig:Ex2:traintime}. As in the previous example, GLaM is by far the cheapest surrogate to train, while SPCE is expensive but its training time grows only mildly with $N_{\textrm{ED}}$. In contrast, the training cost of Kriging and HGP increases rapidly with the size of the training set $N_{\textrm{ED}}$. Because SPCE scales well with $N_{\textrm{ED}}$, it becomes cheaper to train than both Kriging and HGP for the largest experimental designs. The relative training cost of the surrogates therefore depends on both the input dimensionality and the experimental design size.
	\begin{figure}[H]
		\centering
		\color{black}
		\includegraphics[width=0.6\textwidth]{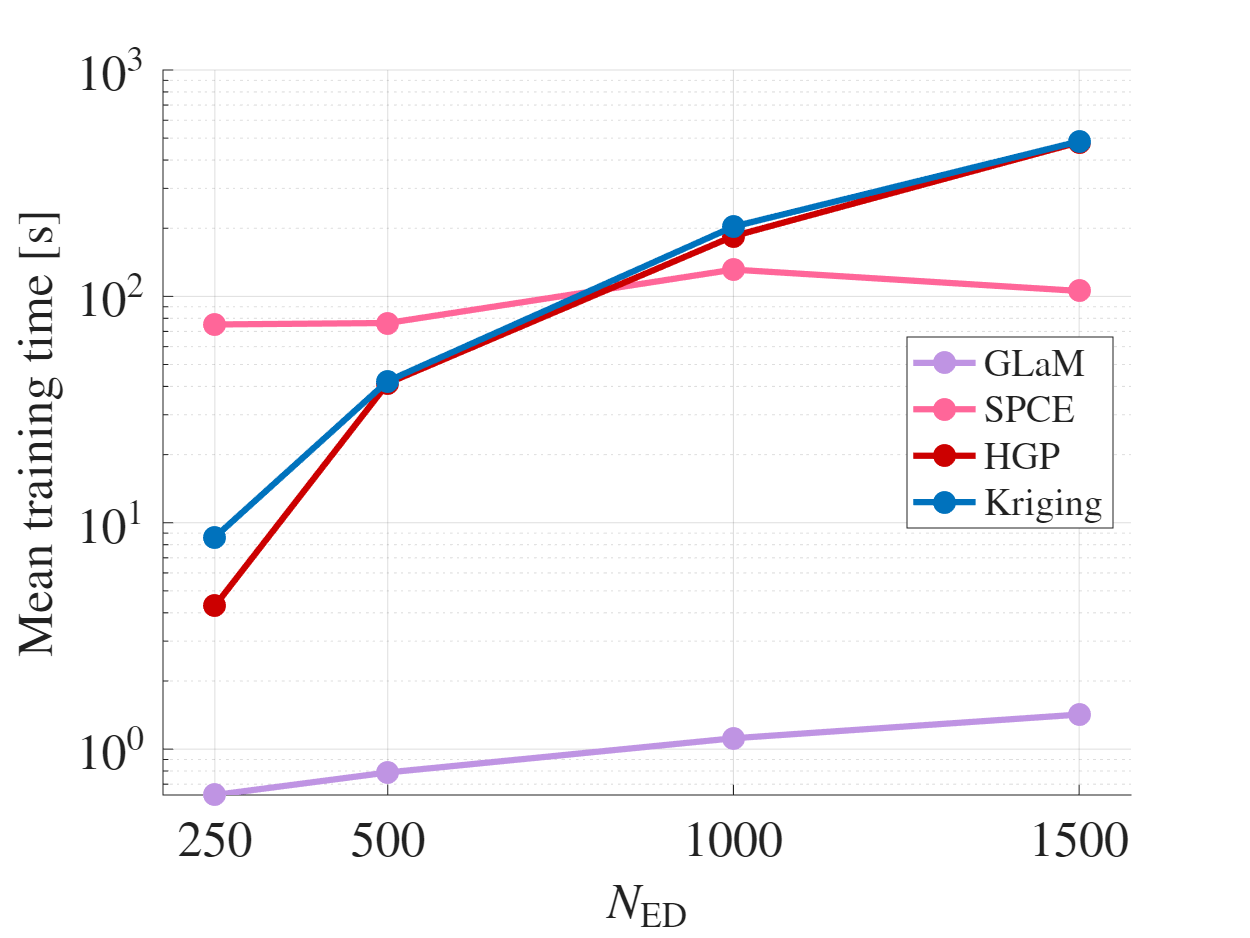}
		\caption{\color{black}Corroded beam: Mean surrogate training time (over $5$ repetitions) versus the experimental design size $N_{\textrm{ED}}$ for the four surrogate models.}
		\label{fig:Ex2:traintime}
	\end{figure}
}

{\color{black}
	\subsection{Truss frame structure}
	In this example, we consider the statically determinate four-panel Warren truss shown in Figure~\ref{fig:Ex4:sketch}, of span $8$~m, height $H = 2$~m and panel length $L_c = 2$~m. Its $15$ bars are lumped into four groups sharing a common circular cross-section, so that the design variables are the group diameters $\ve{d} = \prt{d_1, d_2, d_3, d_4}$: the lower chord ($d_1$), the upper chord ($d_2$), the two external diagonals ($d_3$) and the internal diagonals ($d_4$). The truss carries two statistically independent stochastic loads acting downward at the bottom nodes: $P_1(t)$ at the midspan node $B_3$ and $P_2(t)$ applied simultaneously at the quarter-span nodes $B_2$ and $B_4$.
	\begin{figure}[H]
		\centering
		\color{black}
		\includegraphics[width=0.85\textwidth]{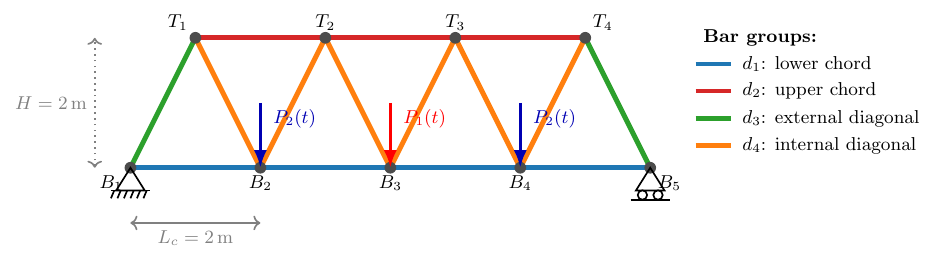}
		\caption{Warren truss: geometry, four bar groups and the two independent stochastic loads $P_1$ (midspan) and $P_2$ (quarter-spans).}
		\label{fig:Ex4:sketch}
	\end{figure}
	
	Because the truss is statically determinate and quasi-static loading is considered, the axial force in each bar is obtained by superposition of the two loads,
	\begin{equation}
		N_b(t) = \sum_{j=1}^{2} \beta_{bj}\, P_j(t), \qquad b = 1,\dots,15,
	\end{equation}
	where the influence coefficients $\beta_{bj}$ are computed once through the method of joints. 
	Failure occurs when the axial stress in any bar group exceeds its yield strength. The system limit-state function is therefore the minimum over the four groups of the corresponding stress margins:
	\begin{equation}\label{eq:Ex4:LS}
		g_s(\ve{d}, \ve{Z}) = \min_{g = 1,\dots,4} \prt{ S_g - \sigma_g^{\max} }, \qquad
		\sigma_g^{\max} = \max_{b \in g}\ \max_{t}\ \frac{\left| N_b(t) \right|}{A_g}, \qquad
		A_g = \frac{\pi d_g^2}{4},
	\end{equation}
	where $S_g$ is the yield strength of group $g$ and $A_g$ its cross-sectional area. 
	
	The objective is to minimize the structural volume
	\begin{equation}
		\fc(\ve{d}) = \sum_{g=1}^{4} n_g\, A_g\, L_g, \qquad \mathbb{D} = \bra{0.05,\, 0.09}^4~\textrm{m},
	\end{equation}
	where $n_g \in (4, 3, 2, 6)$ is the number of bars in each group and $L_g$ their length ($L_c = 2$~m for the chords, $L_d = 2.236$~m for the diagonals). The target failure probability is $\bar{p}_f = 0.01$.
	
	Each load $P_j(t)$ is modeled as a translation process \citep{Grigoriu1998}: an underlying zero-mean, unit-variance Gaussian process with exponential autocorrelation of length $l_{c,j}$ is mapped, through the probability integral transform, onto a Gumbel marginal with the mean and coefficient of variation reported in Table~\ref{tab:Ex4:Z}. The two loads share the same functional form but have markedly different correlation lengths ($l_{c,1} = 10$~yr, $l_{c,2} = 2$~yr). As a consequence, the effective load combination seen by each bar group depends on the design, and the conditional response cannot be reduced to a single peak-load variable, making the problem genuinely high-dimensional. Each process is discretized over a service life of $T = 50$~years on $N_t = 300$ time steps and truncated with a Karhunen--Lo\`eve (KL) expansion retaining $99\%$ of the variance, which requires $n_{\ve{\theta},1} = 94$ and $n_{\ve{\theta},2} = 232$ terms, i.e.\ $326$ KL coordinates in total.
	\begin{table}[H]
		\centering
		\color{black}
		\begin{tabular}{lccc}
			\hline
			Parameter & Distribution & Mean ($\mu$) & CoV ($\delta \%$)\\
			\hline
			Yield strength \( S_g \) [MPa] (per group) & Lognormal & $250$ & $10$ \\								KL coordinates \( \theta_1, \ldots, \theta_{326} \) [--] & Gaussian & $0$ & $1^{\ast}$ \\
			Load \( P_1 \) [kN] ($l_{c,1} = 10$~yr) & Gumbel & $200$ & $30$ \\
			Load \( P_2 \) [kN] ($l_{c,2} = 2$~yr) & Gumbel & $120$ & $40$ \\
			\hline
		\end{tabular}
		\caption{Warren truss: environmental variables $\ve{Z}$ (first two rows) and loads moments. $^\ast$corresponds to the standard deviation, rather than the CoV. The two loads are represented as translation processes with Gumbel marginals.}
		\label{tab:Ex4:Z}
	\end{table}
	
	Overall, the problem involves $n_{\ve{Z}} = 330$ environmental variables ($4$ yield strengths and $326$ KL coordinates) and $n_{\ve{d}} = 4$ design variables. GLaM, SPCE and HGP are constructed in the reduced design space of dimension $n_{\ve{d}} = 4$, whereas the Kriging surrogate is built in the full augmented space of dimension $n_{\textrm{tot}} = 334$ and embedded in a double-loop Monte Carlo reliability analysis. An isotropic Gaussian kernel is selected for Kriging to facilitate the training. 
	
	The reference solution is obtained by solving the RBDO problem directly on the original model, using a large Monte Carlo set of size $N_{\textrm{MC}} = 10^5$ samples to estimate the failure probability. It yields the optimal cost $\mathfrak{c}\prt{\ve{d}^{\ast,\textrm{ref}}} = 0.097$~m$^3$ at $\ve{d}^{\ast,\textrm{ref}} = (62.0, 66.7, 62.5, 60.4)$~mm, which sits on the constraint boundary. When using surrogates, experimental designs of sizes $N_{\textrm{ED}} = \acc{100, 200, 300, 400, 500}$ are considered, and the analysis is repeated $15$ times.
	
	Figure~\ref{fig:Ex4:Fstar} shows boxplots of the optimal costs obtained over the $15$ repetitions. The three design-space methods (GLaM, SPCE and HGP) cluster around the reference solution, whereas Kriging, built in the full $334$-dimensional augmented space, converges to a strongly biased and overly cheap design (median cost about $35\%$ below the reference), illustrating its breakdown in high dimensions. Among the design-space methods, HGP yields the most concentrated boxplots, but with an upward (conservative) bias. This is likely due to the non-{G}aussianity of the limit-state $g_s\prt{\ve{d};\omega}$ for each possible design $\ve{d}$, see below. SPCE progressively reduces both its bias and its variability as $N_{\textrm{ED}}$ increases, becoming the most accurate for the larger experimental designs. GLaM recovers the reference cost on the median but remains the most variable and does not improve monotonically with $N_{\textrm{ED}}$, a behavior consistent with the training difficulties of the stochastic emulators.
	\begin{figure}[H]
		\centering
		\color{black}
		\includegraphics[width=0.6\textwidth]{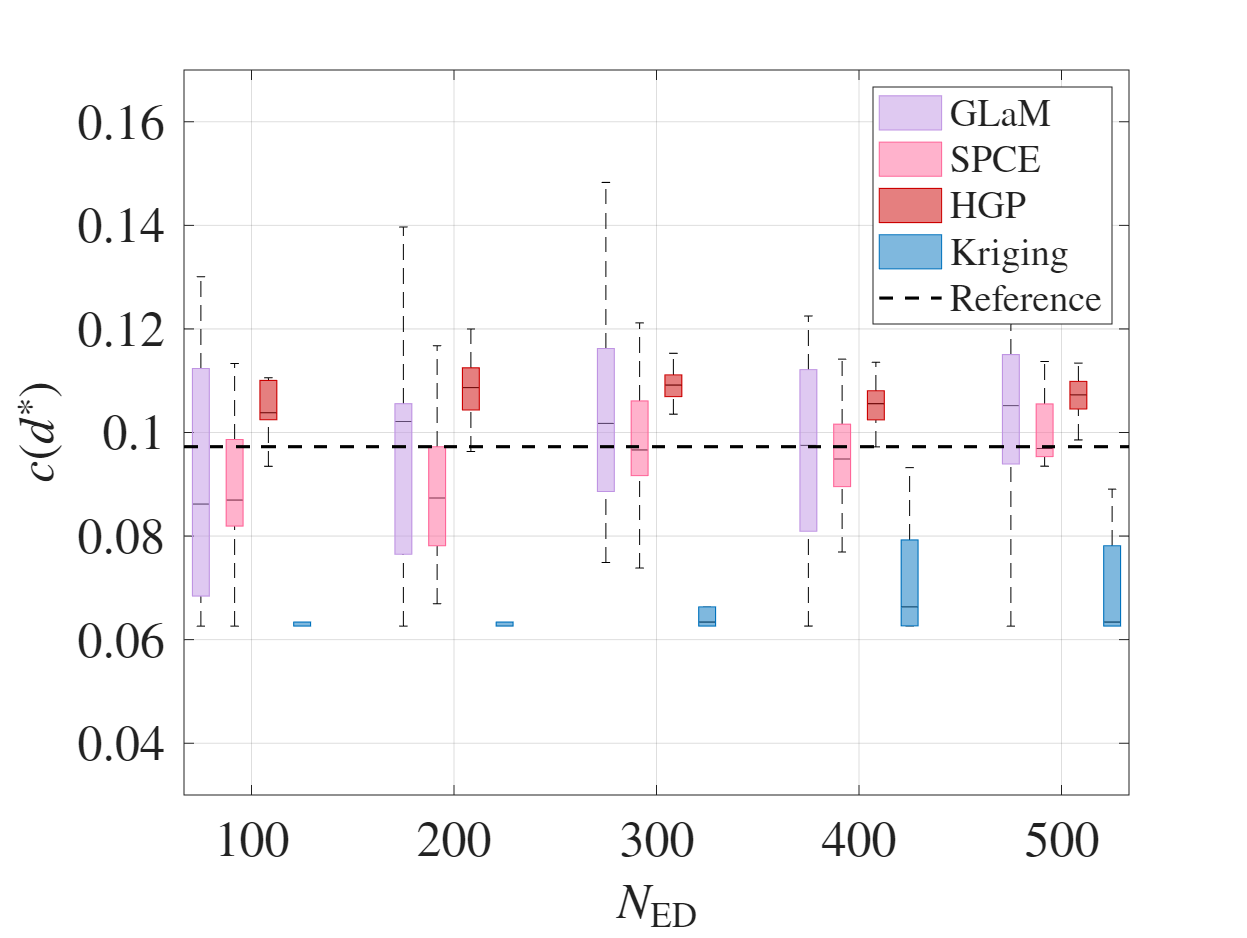}
		\caption{Warren truss: Boxplots of the optimal costs $\mathfrak{c}\prt{\ve{d}^{\ast}}$ for different methods and experimental design sizes. The dashed line is the reference optimal cost.}
		\label{fig:Ex4:Fstar}
	\end{figure}
	
	Table~\ref{tab:Ex4:error} reports the median relative errors on the optimal cost computed after Eq.~\eqref{eq:error}. SPCE improves steadily with the experimental design size, reaching a relative error of $3.9 \cdot 10^{-2}$ for $N_{\textrm{ED}} = 500$, while HGP remains stable around $10^{-1}$ and GLaM fluctuates without a clear trend. Kriging exhibits by far the largest errors ($\approx 3.5 \cdot 10^{-1}$ at all experimental design sizes), reflecting the biased solution already visible in the boxplots. These trends corroborate the boxplots: among the design-space methods, SPCE is the most accurate for sufficiently large experimental designs, whereas GLaM is the least robust while HGP produces consistently biased solutions.
	\begin{table}[H]
		\centering
		\color{black}
		\begin{tabular}{lccccc}
			\hline
			$N_{\textrm{ED}}$ & $100$ & $200$ & $300$ & $400$ & $500$ \\
			\hline
			GLaM    & $2.5 \cdot 10^{-1}$ & $1.6 \cdot 10^{-1}$ & $1.8 \cdot 10^{-1}$ & $1.8 \cdot 10^{-1}$ & $1.8 \cdot 10^{-1}$ \\
			SPCE    & $1.2 \cdot 10^{-1}$ & $1.2 \cdot 10^{-1}$ & $7.2 \cdot 10^{-2}$ & $6.3 \cdot 10^{-2}$ & $3.9 \cdot 10^{-2}$ \\
			HGP     & $6.7 \cdot 10^{-2}$ & $1.2 \cdot 10^{-1}$ & $1.2 \cdot 10^{-1}$ & $8.5 \cdot 10^{-2}$ & $1.0 \cdot 10^{-1}$ \\
			Kriging & $3.5 \cdot 10^{-1}$ & $3.6 \cdot 10^{-1}$ & $3.5 \cdot 10^{-1}$ & $3.2 \cdot 10^{-1}$ & $3.5 \cdot 10^{-1}$ \\
			\hline
		\end{tabular}
		\caption{Warren truss: Relative error on the optimal cost $\mathfrak{c}\prt{\ve{d}^{\ast}}$ after Eq.~\eqref{eq:error} for the median solution (over $15$ repetitions) for each method and experimental design size $N_{\textrm{ED}}$.}
		\label{tab:Ex4:error}
	\end{table}
	
	Figure~\ref{fig:Ex4:Density} shows the conditional distributions of the limit-state function at the reference optimal design $\ve{d}^{\ast,\textrm{ref}}$ for three experimental design sizes and the four surrogate models. For each method, the surrogate with the median error over the $15$ repetitions is reported. The reference density is obtained from $10^4$ evaluations of the original model. GLaM and SPCE are evaluated using $10^4$ realizations of their latent variables, HGP is sampled from the predicted Gaussian conditional distribution, and Kriging is evaluated using the same $10^4$ realizations of the input uncertainty vector $\ve{Z}$. Although the agreement with the reference varies across methods and experimental design sizes, the quality of the conditional approximation closely mirrors the optimization results reported in Fig.~\ref{fig:Ex4:Fstar}. Whenever the conditional density is accurately reproduced, the corresponding optimal cost is close to the reference, whereas larger discrepancies in the tail lead to larger deviations in the boxplots. This is particularly evident for GLaM, which provides an excellent approximation for $N_{\textrm{ED}}=300$ but a noticeably poorer one for $N_{\textrm{ED}}=500$, consistent with the degradation of the optimization results and highlighting the training difficulties discussed previously. SPCE captures the overall shape reasonably well but remains somewhat over-dispersed, leading to a slight overestimation of the failure probability. HGP exhibits a similar behaviour due to its Gaussian conditional assumption, which results in consistently conservative designs. In contrast, Kriging collapses the conditional response to an almost deterministic, strongly shifted density, assigning almost no probability mass to the failure region and therefore producing severely non-conservative solutions.
\begin{figure}[H]
	\centering
	
	\subfloat[GLaM -- $N_{\textrm{ED}} = 100$]{
		\includegraphics[width=0.31\textwidth]{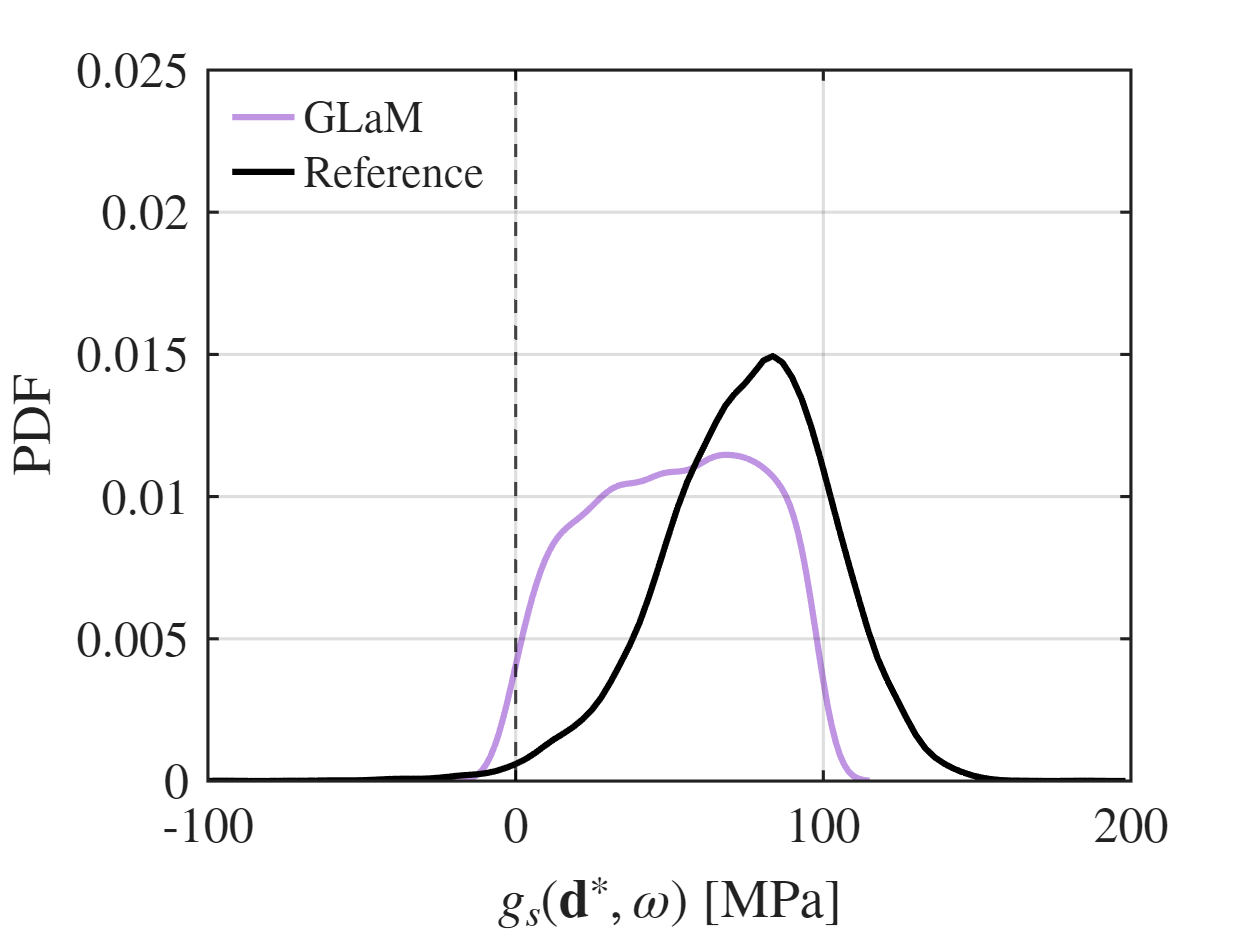}}
	\hfill
	\subfloat[GLaM -- $N_{\textrm{ED}} = 300$]{
	\includegraphics[width=0.31\textwidth]{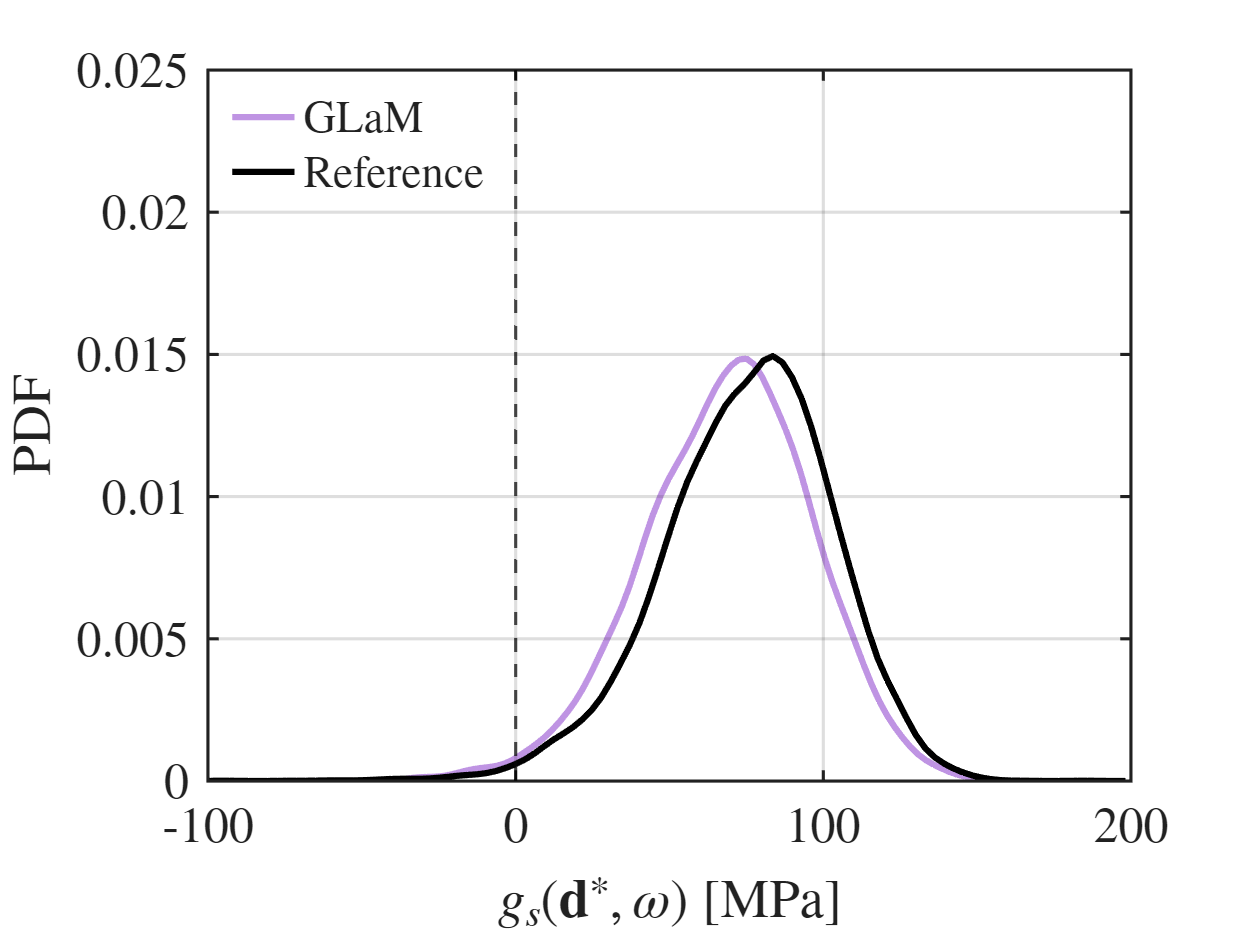}}
		\hfill
	\subfloat[GLaM -- $N_{\textrm{ED}} = 500$]{
		\includegraphics[width=0.31\textwidth]{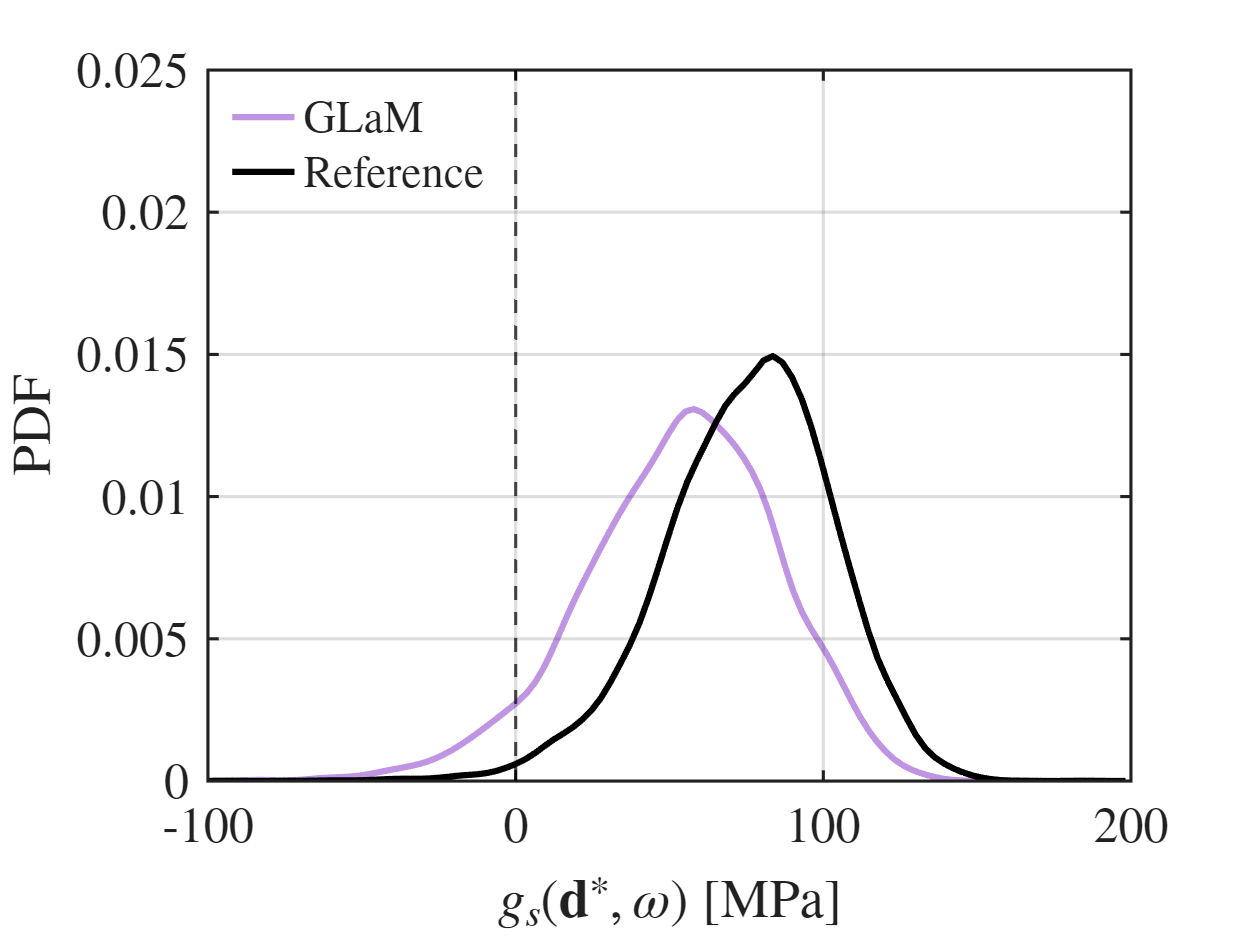}}

	\vspace{0.5em}
	
	\subfloat[SPCE -- $N_{\textrm{ED}} = 100$]{
		\includegraphics[width=0.31\textwidth]{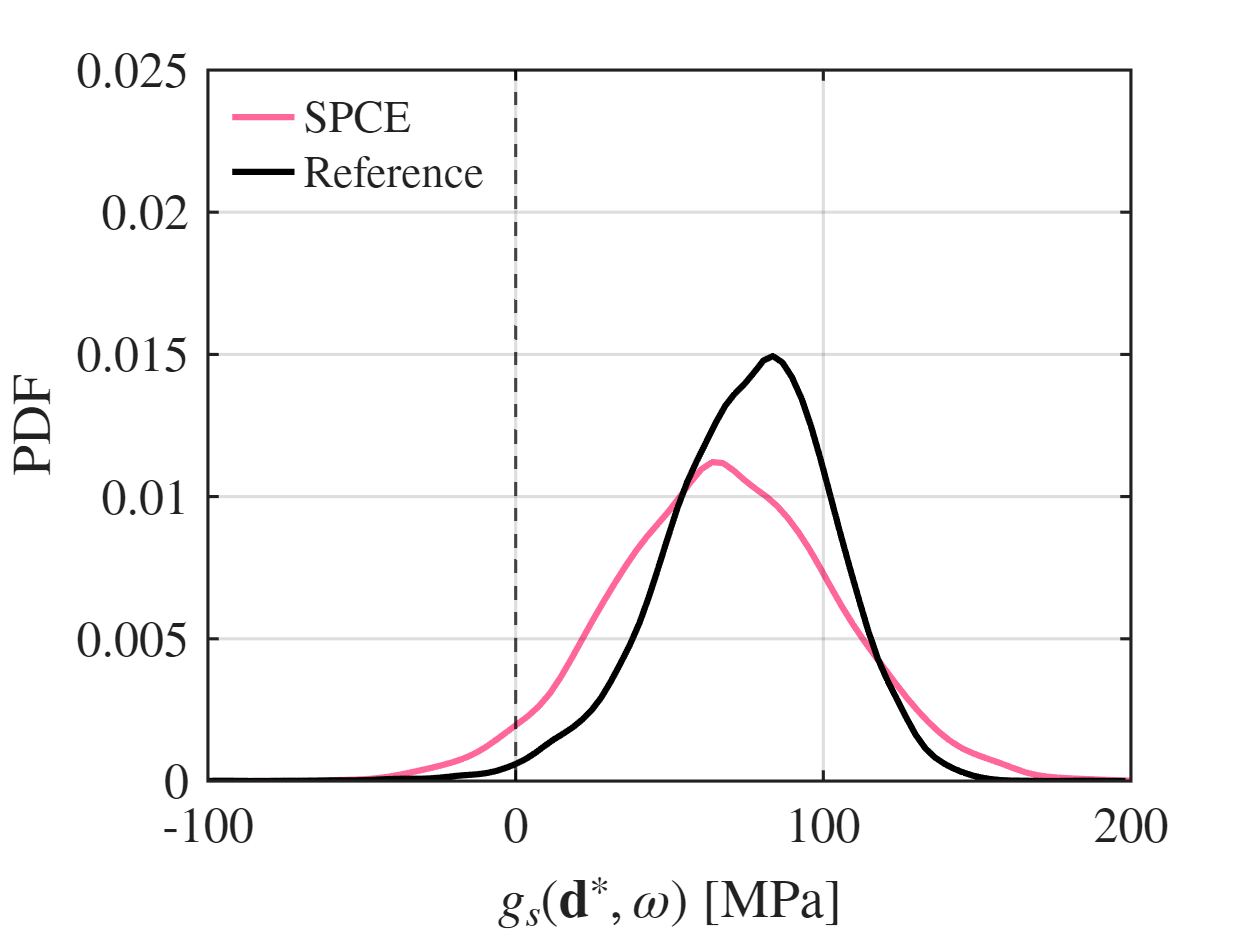}}
	\hfill
	\subfloat[SPCE -- $N_{\textrm{ED}} = 300$]{
	\includegraphics[width=0.31\textwidth]{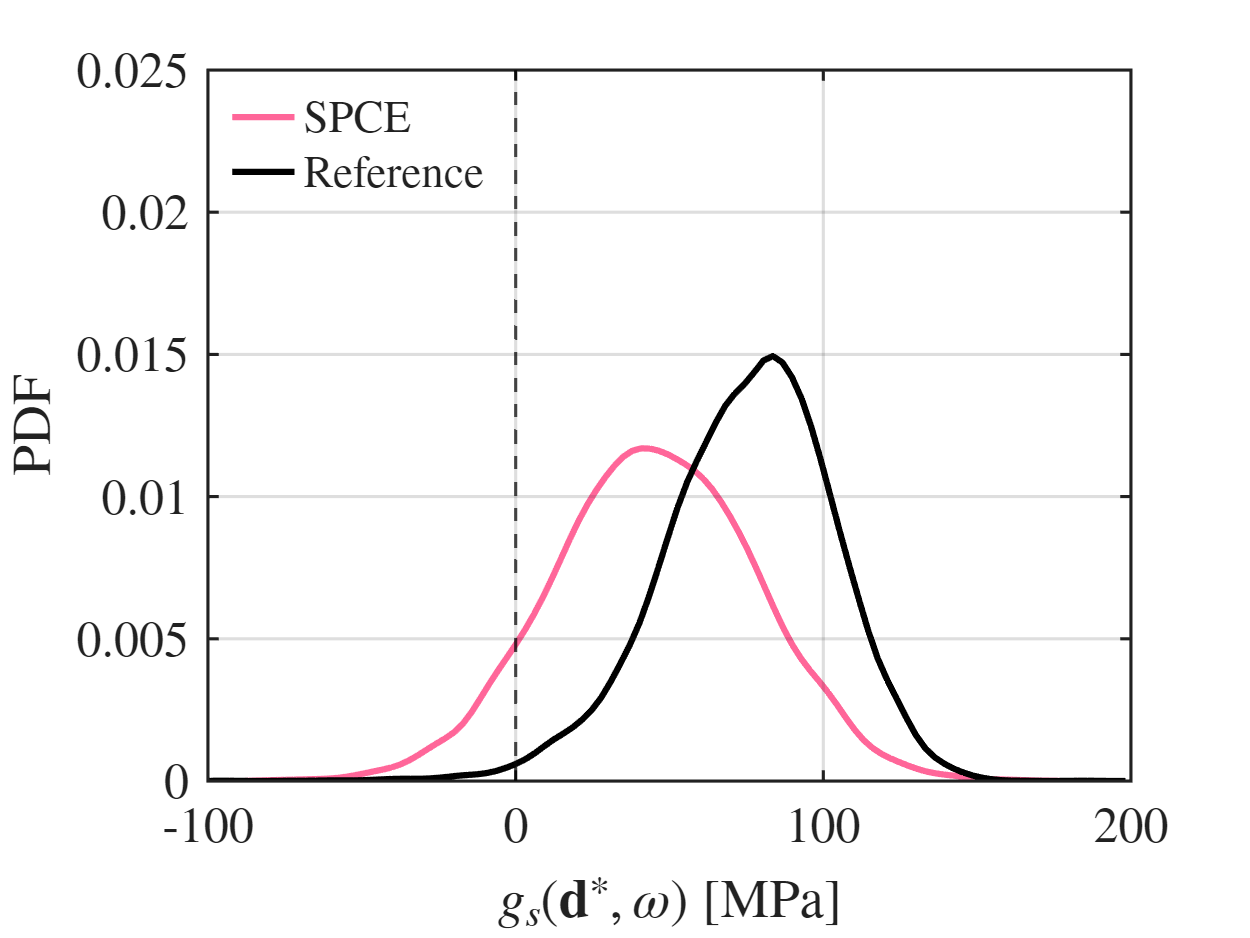}}
	\hfill
	\subfloat[SPCE -- $N_{\textrm{ED}} = 500$]{
		\includegraphics[width=0.31\textwidth]{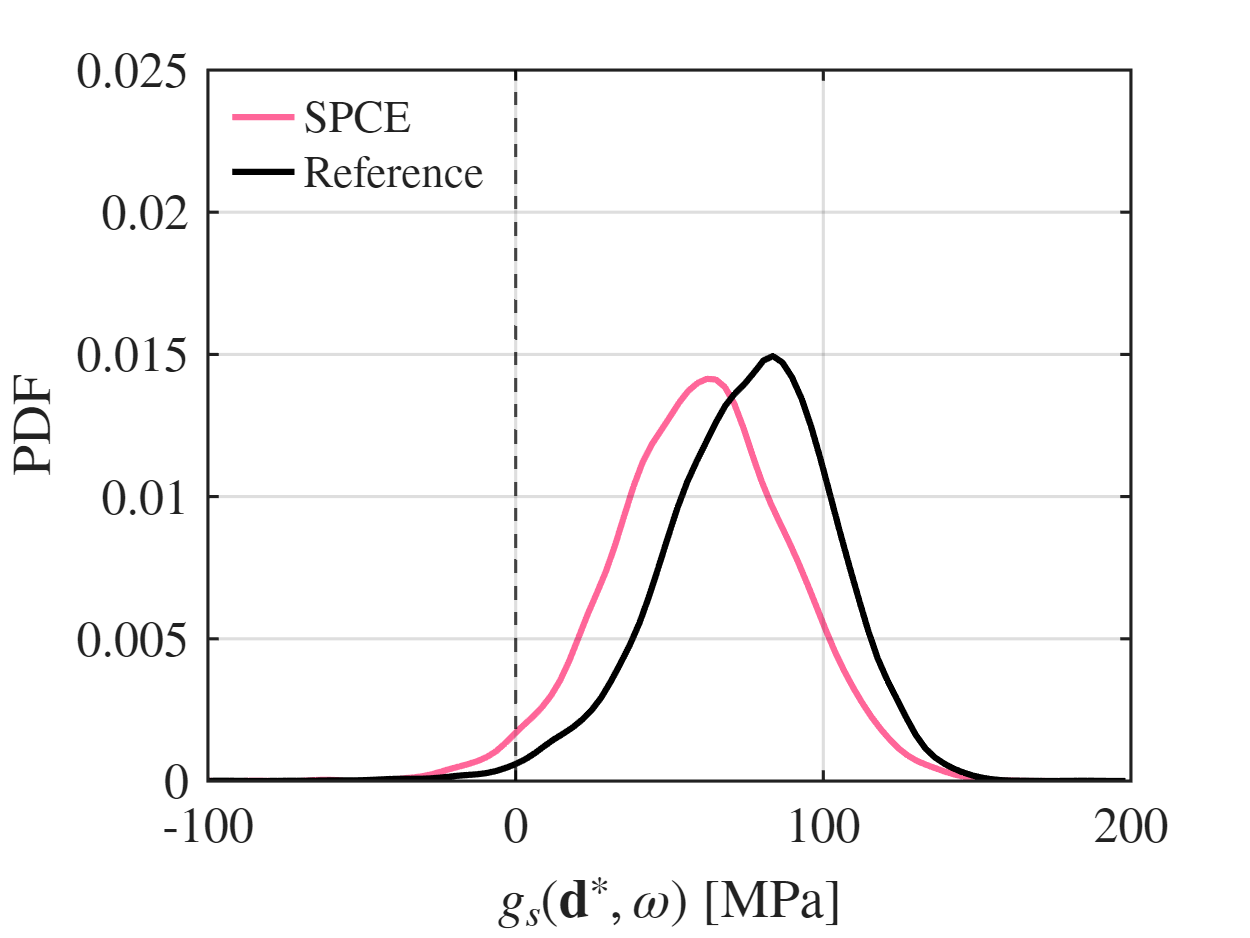}}
		
	\caption{Warren truss: Conditional probability density functions of the limit-state function at the reference optimal design $\ve{d}^{\ast,\textrm{ref}}$, for the median-error surrogate of each method and experimental design size. The dashed vertical line marks the failure threshold $g_s = 0$.}
	\label{fig:Ex4:Density}
\end{figure}

\begin{figure}[H]
	\ContinuedFloat
	\centering
	
	\subfloat[HGP -- $N_{\textrm{ED}} = 100$]{
		\includegraphics[width=0.31\textwidth]{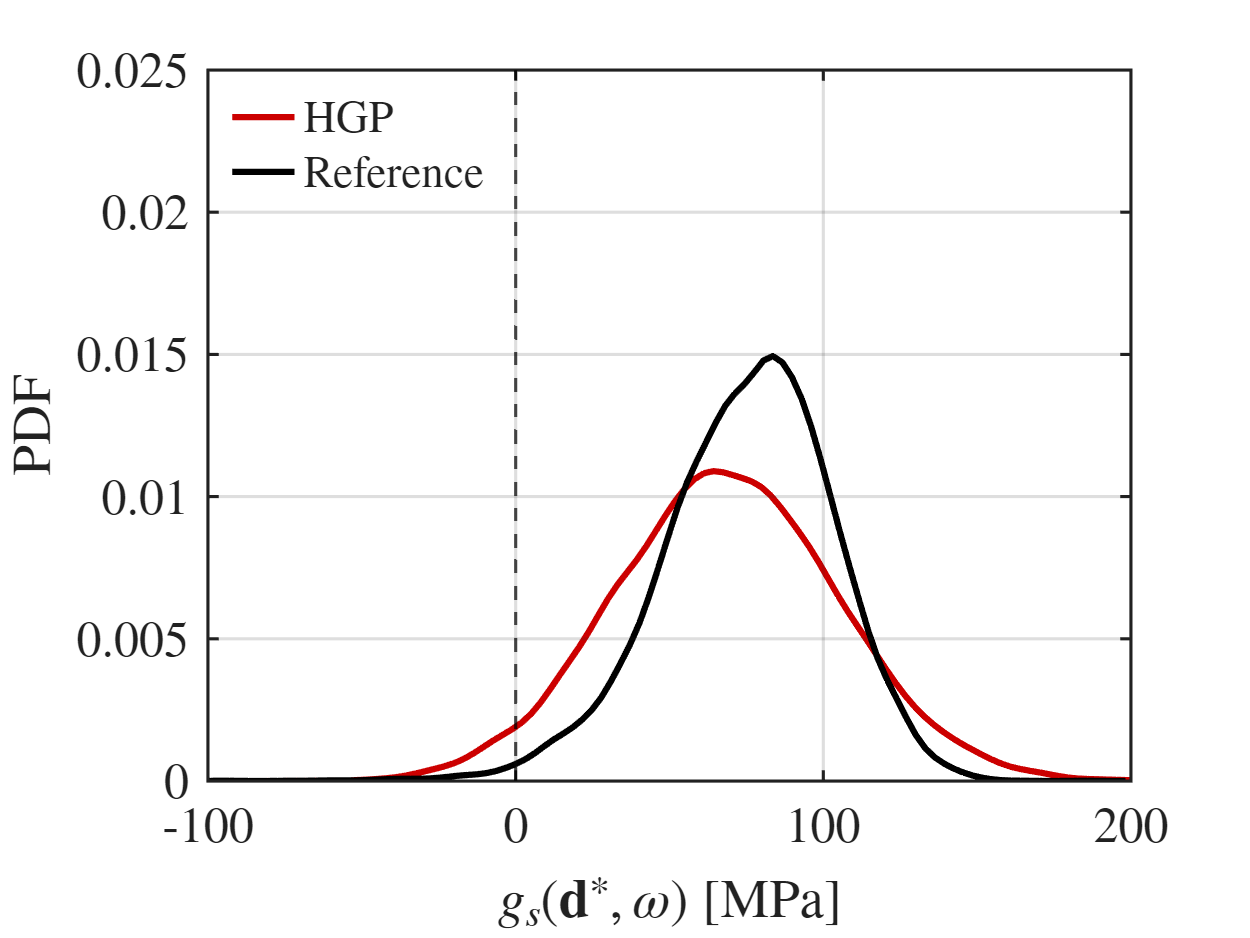}}
	\hfill
	\subfloat[HGP -- $N_{\textrm{ED}} = 300$]{
		\includegraphics[width=0.31\textwidth]{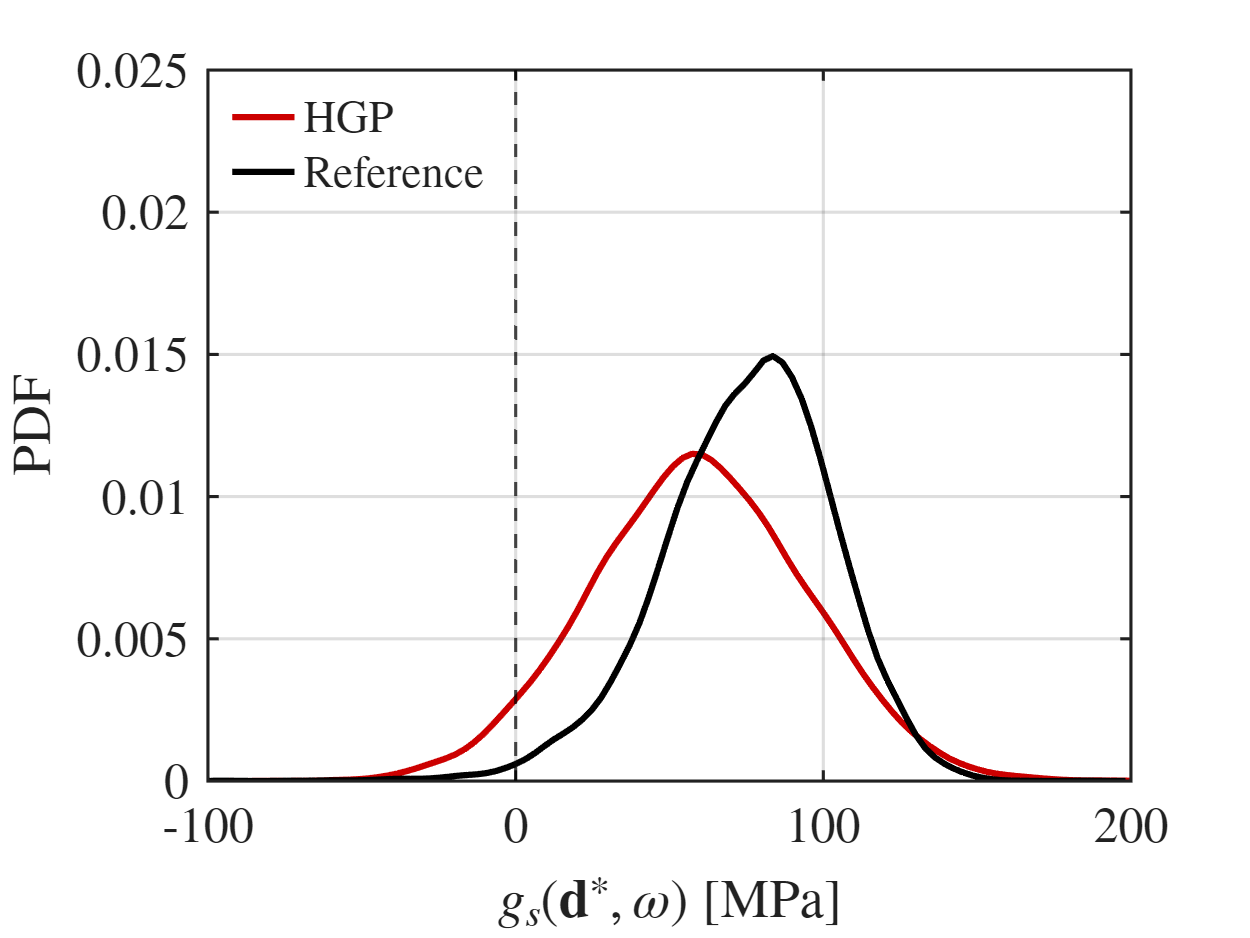}}
	\hfill
	\subfloat[HGP -- $N_{\textrm{ED}} = 500$]{
		\includegraphics[width=0.31\textwidth]{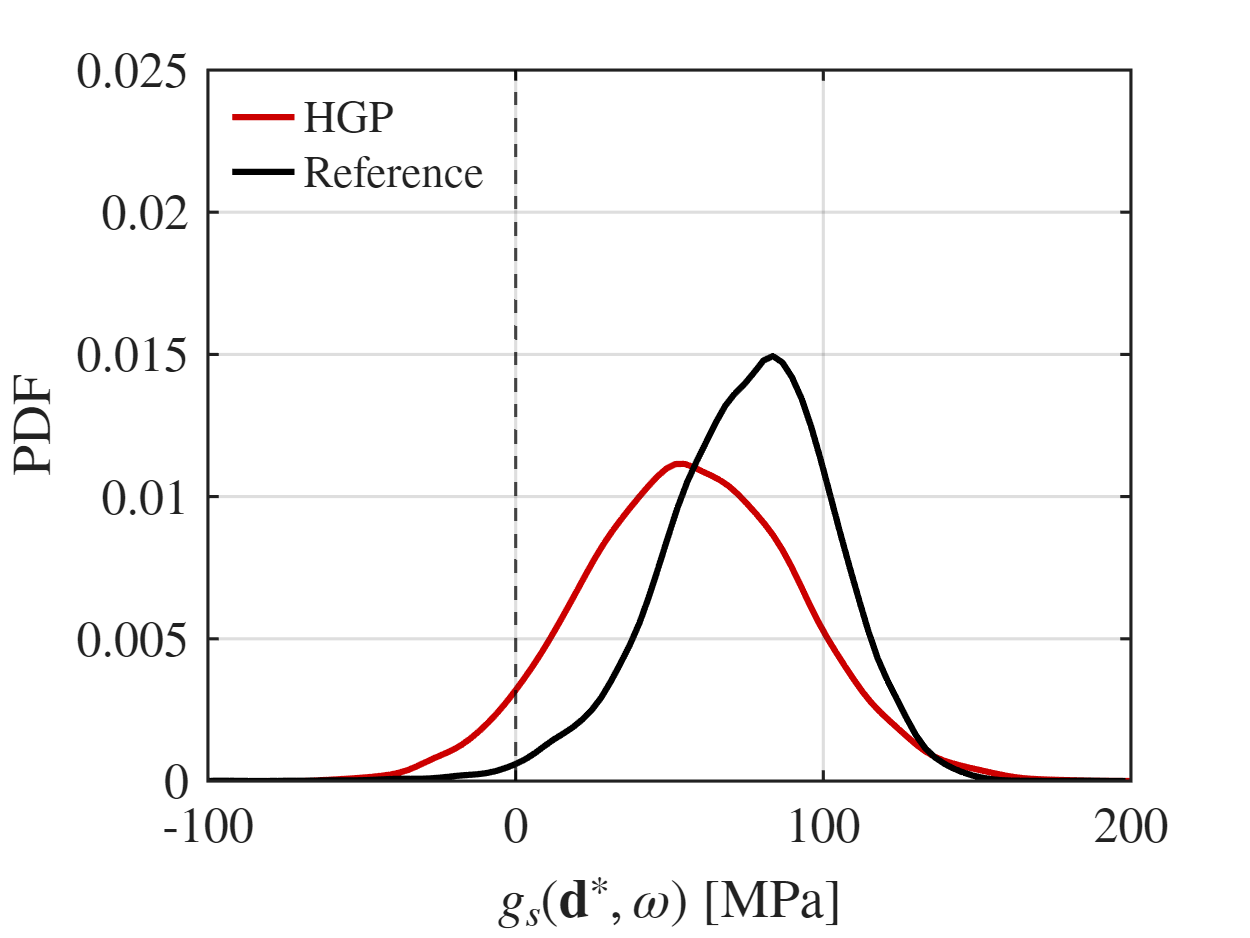}}
	
	\vspace{0.5em}

	\subfloat[Kriging -- $N_{\textrm{ED}} = 100$]{
		\includegraphics[width=0.31\textwidth]{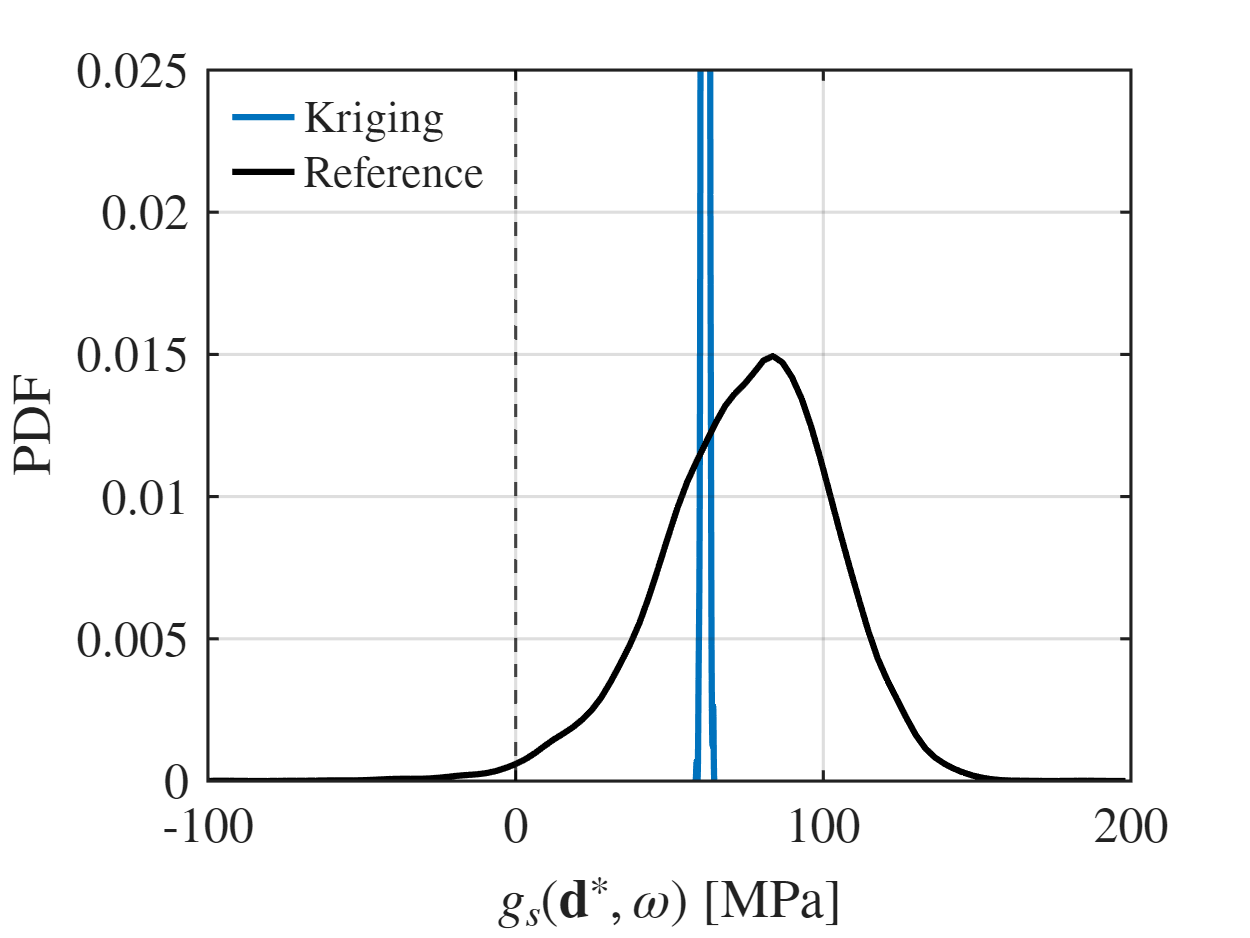}}
	\hfill
	\subfloat[Kriging -- $N_{\textrm{ED}} = 300$]{
		\includegraphics[width=0.31\textwidth]{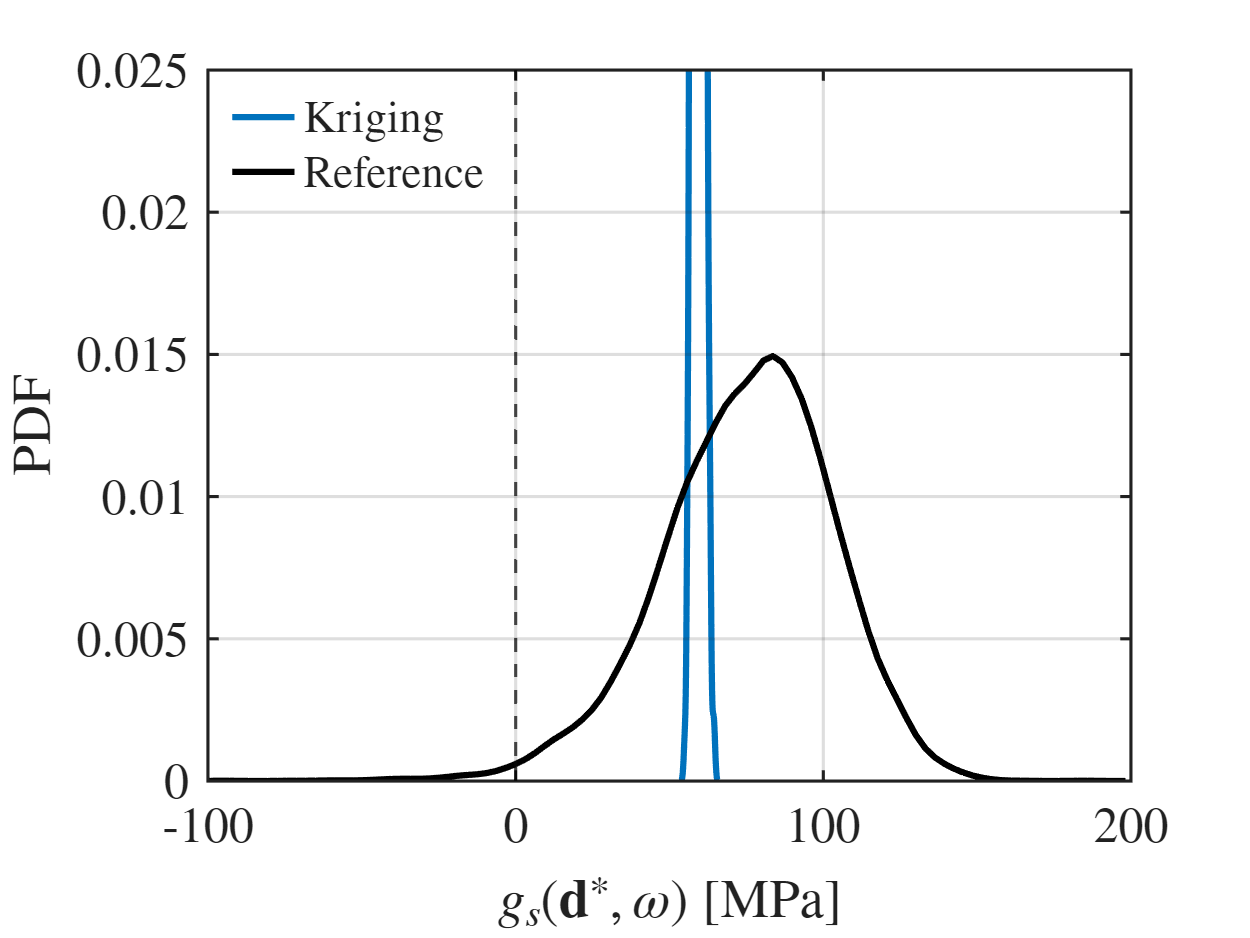}}
	\hfill
	\subfloat[Kriging -- $N_{\textrm{ED}} = 500$]{
		\includegraphics[width=0.31\textwidth,trim=0 0 33.12bp 0,clip]{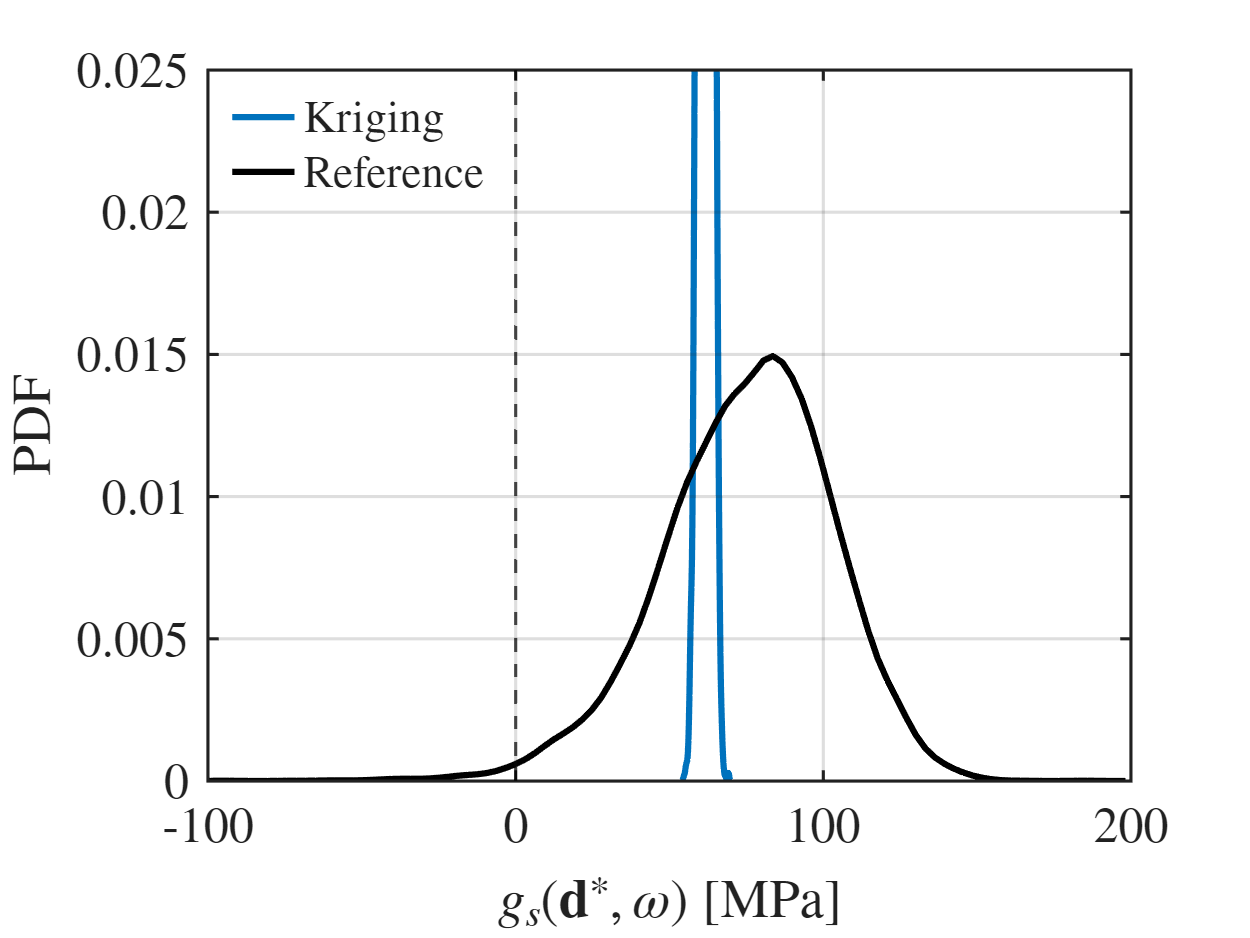}}
	
	\caption[]{Warren truss: Conditional probability density functions of the limit-state function at the reference optimal design $\ve{d}^{\ast,\textrm{ref}}$, for the median-error surrogate of each method and experimental design size (continued).}
\end{figure}
	
	Finally, the results highlight the computational advantage of building the surrogate directly in the design space. For GLaM and SPCE the failure probability is evaluated analytically at each optimization iteration through Eqs.~\eqref{eq:GLaM:q} and~\eqref{eq:SPCE:Pf}, and HGP relies on a closed-form Gaussian conditional distribution, so that in all three cases the optimization completes in a few seconds irrespective of $N_{\textrm{ED}}$. In contrast, the Kriging-based double loop performs a Monte Carlo reliability analysis in the $334$-dimensional augmented space at every candidate design, and requires between roughly one and five hours per run (from about $4.1 \cdot 10^{3}$~s at $N_{\textrm{ED}} = 100$ to $2.0 \cdot 10^{4}$~s at $N_{\textrm{ED}} = 500$), i.e.\ three to four orders of magnitude more than the design-space approaches, for an overall poor result. 
	
	
}

{\color{black}
	\subsection{Seismic shear frame}
	As a final, even higher-dimensional application, we consider the RBDO of the three-story shear frame sketched in Figure~\ref{fig:Ex5:sketch}, subjected to stochastic ground motion. The design variables are the three inter-story stiffnesses $\ve{d} = \prt{k_1, k_2, k_3}$ and the objective is to minimize the total stiffness $\fc(\ve{d}) = \sum_{i=1}^{3} k_i$ under a constraint on the peak inter-story drift.
	
	The frame is idealized as a lumped-mass shear building whose equation of motion under base excitation reads
	\begin{equation}
		\ve{M}\ddot{\ve{x}} + \ve{C}\dot{\ve{x}} + \ve{K}\ve{x} = -\ve{M}\,a_g(t),
	\end{equation}
	where $\ve{M}$ gathers the story masses $m_i$, $\ve{K}$ is the tridiagonal stiffness matrix assembled from the $k_i$ and $\ve{C}$ is a Rayleigh damping matrix (computed for a damping ratio $\zeta = 5\%$). More precisely, the mass matrix is diagonal, $\ve{M} = \mathrm{diag}\prt{m_1,\,m_2,\,m_3}$, while for a shear building the stiffness matrix is symmetric tridiagonal and assembled from the inter-story stiffnesses as
	\begin{equation}
		\ve{K} =
		\begin{bmatrix}
			k_1 + k_2 & -k_2      & 0    \\
			-k_2      & k_2 + k_3 & -k_3 \\
			0         & -k_3      & k_3
		\end{bmatrix}.
	\end{equation}
	The damping matrix follows the Rayleigh (proportional) model $\ve{C} = a_0\,\ve{M} + a_1\,\ve{K}$, whose coefficients are calibrated so that the target damping ratio $\zeta = 5\%$ is attained in the first two modes,
	\begin{equation}
		a_0 = \zeta\,\frac{2\,\omega_1\,\omega_2}{\omega_1 + \omega_2}, \qquad
		a_1 = \zeta\,\frac{2}{\omega_1 + \omega_2},
	\end{equation}
	where $\omega_1$ and $\omega_2$ are the first two undamped natural circular frequencies, obtained from the generalized eigenvalue problem $\prt{\ve{K} - \omega_j^2\,\ve{M}}\,\ve{\phi}_j = \ve{0}$~\citep{chopra2017,cloughpenzien1993}.
	
	The ground acceleration $a_g(t)$ is modeled as a Kanai--Tajimi filtered white-noise process~{\color{blue}\citep{kanai1957,tajimi1960}},
	\begin{equation}
		\ddot{x}_f + 2\zeta_g\omega_g\,\dot{x}_f + \omega_g^2\,x_f = w(t), \qquad
		a_g(t) = 2\zeta_g\omega_g\,\dot{x}_f + \omega_g^2\,x_f,
	\end{equation}
	with $\omega_g = 15$~rad/s, $\zeta_g = 0.6$ and white-noise spectral density $S_0 = 2.5\cdot10^{-3}$~m$^2$/s$^3$. The excitation is discretized over $T = 20$~s with $\Delta t = 0.01$~s, which yields $2{,}000$ independent standard Gaussian increments. The performance is measured by the peak inter-story drift ratio $\theta_{\max} = \max_{i,t} |u_i(t) - u_{i-1}(t)|/h$ (story height $h = 3$~m), and the limit-state function is
	\begin{equation}
		g_s(\ve{d}, \ve{Z}) = \theta_{\lim} - \theta_{\max}, \qquad \theta_{\lim} = 1/200.
	\end{equation}
	The design space is defined by $k_i \in \bra{k_{0,i},\, 1.7\,k_{0,i}}$ with $\ve{k}_0 = (6.0, 5.5, 5.0)\cdot10^7$~N/m, together with the soft (tapered-design) constraint $k_1 \ge k_2 \ge k_3$. The target failure probability is $\bar{p}_f = 5\%$.
	\begin{figure}[H]
		\centering
		\color{black}
		\subfloat[Shear frame]{\raisebox{2mm}{\includegraphics[width=0.5\textwidth]{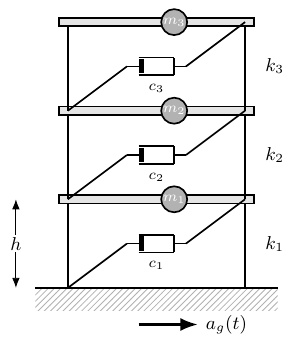}}}
		\\
		\subfloat[Ground acceleration $a_g(t)$]{\includegraphics[width=0.49\textwidth]{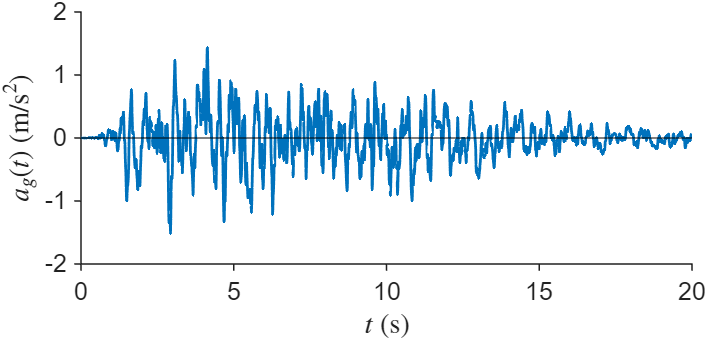}}
		\hfill
		\subfloat[Inter-story drift response]{\includegraphics[width=0.49\textwidth]{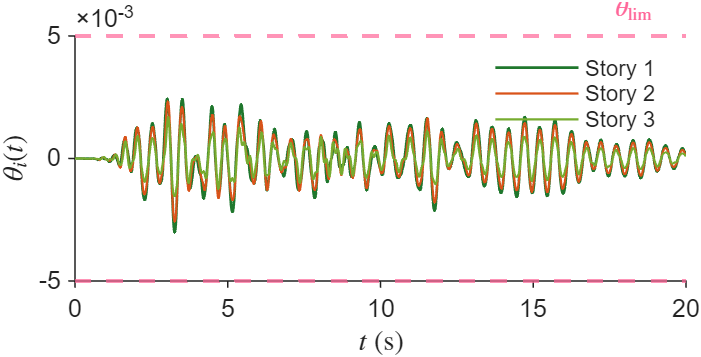}}
		\caption{Seismic frame: three-story shear building under stochastic Kanai--Tajimi ground motion, with a sample ground-acceleration realization and the corresponding inter-story drift response.}
		\label{fig:Ex5:sketch}
	\end{figure}
	\begin{table}[H]
		\centering
		\color{black}
		\begin{tabular}{lccc}
			\hline
			Parameter & Distribution & Mean ($\mu$) & CoV ($\delta \%$)\\
			\hline
			Story masses \( m_i \) [kg] & Lognormal & $10^5$ & $10$ \\
			Excitation \( W_1, \ldots, W_{2000} \) [--] & Gaussian & $0$ & $1^{\ast}$ \\
			\hline
		\end{tabular}
		\caption{Seismic frame: environmental variables $\ve{Z}$. $^\ast$corresponds to the standard deviation, rather than the CoV.}
		\label{tab:Ex5:Z}
	\end{table}
	
	Overall, the problem involves $n_{\ve{Z}} = 2{,}003$ environmental variables ($3$ story masses and $2{,}000$ standard normal variables modelling the excitation) and $n_{\ve{d}} = 3$ design variables, i.e.\ a total dimension $n_{\textrm{tot}} = 2{,}006$. At this scale a Kriging surrogate in the augmented space is entirely out of reach. We therefore compare GLaM and SPCE against HGP and its lognormal variant HGP$_{\log}$, which models $\log\theta_{\max}$ instead of $\theta_{\max}$ (motivated by the common lognormal assumption for seismic drift), all built in the three-dimensional design space. The reference solution, obtained with the original model and a Monte Carlo sample of size $10^5$, yields the optimal cost $\mathfrak{c}\prt{\ve{d}^{\ast,\textrm{ref}}} \approx 1.89\cdot10^8$~N/m. Experimental designs of sizes $N_{\textrm{ED}} = \acc{250,\, 500,\, 750,\, 1{,}000}$ are considered and the analysis is repeated $15$ times.
	
	Figure~\ref{fig:Ex5:Fstar} shows the boxplots of the optimal cost. GLaM, SPCE and HGP$_{\log}$ all converge towards the reference solution, whereas the plain (Gaussian) HGP systematically over-estimates the optimal cost. Table~\ref{tab:Ex5:error} confirms this: HGP$_{\log}$ and SPCE reach relative errors of a few percent at $N_{\textrm{ED}} = 1{,}000$, GLaM follows closely, while HGP stagnates around $1.3\cdot10^{-1}$.
	\begin{figure}[H]
		\centering
		\color{black}
		\includegraphics[width=0.6\textwidth]{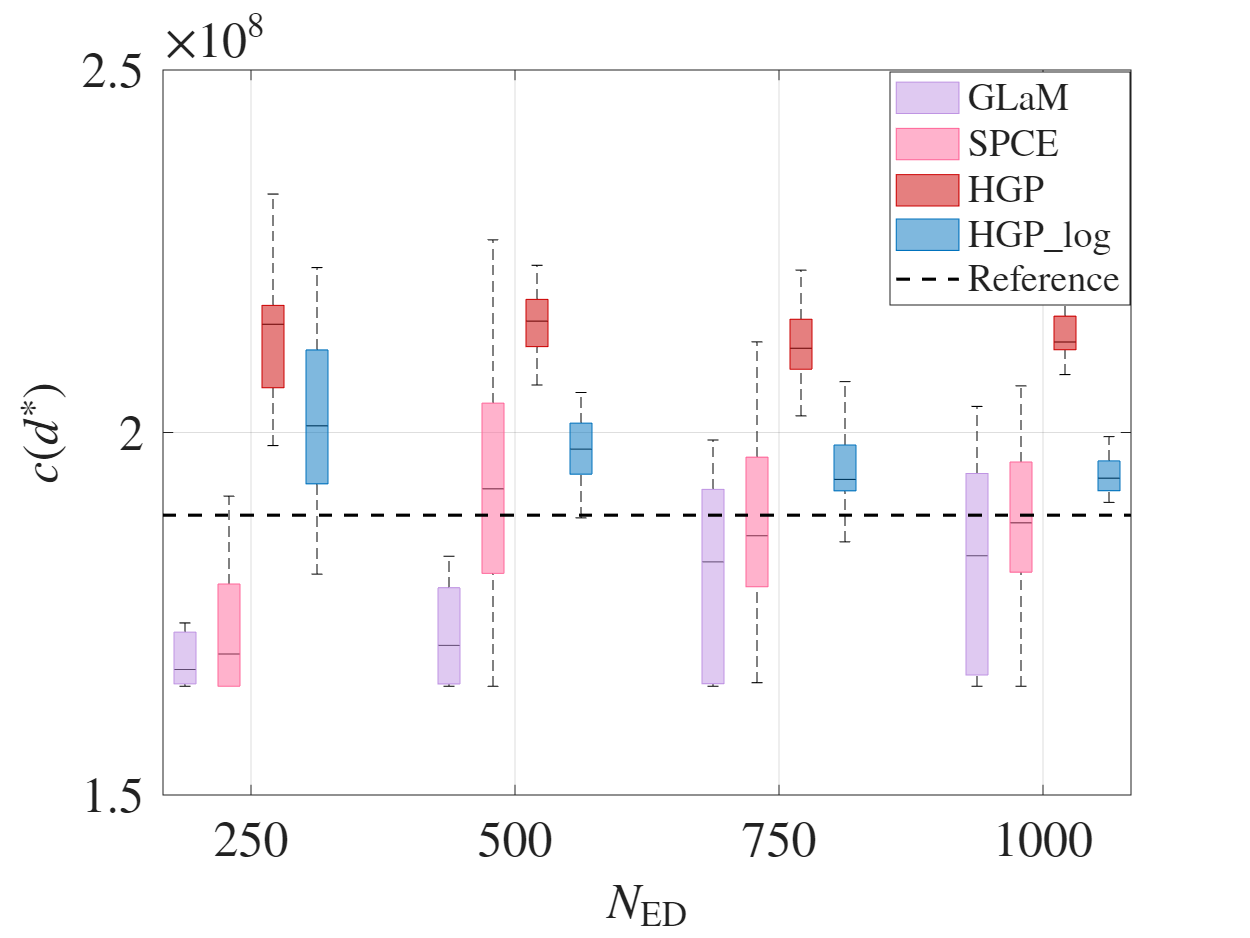}
		\caption{Seismic frame: Boxplots of the optimal costs $\mathfrak{c}\prt{\ve{d}^{\ast}}$ for different methods and experimental design sizes. The dashed line is the reference optimal cost.}
		\label{fig:Ex5:Fstar}
	\end{figure}
	\begin{table}[H]
		\centering
		\color{black}
		\begin{tabular}{lcccc}
			\hline
			$N_{\textrm{ED}}$ & $250$ & $500$ & $750$ & $1{,}000$ \\
			\hline
			GLaM         & $1.1 \cdot 10^{-1}$ & $9.5 \cdot 10^{-2}$ & $5.5 \cdot 10^{-2}$ & $5.1 \cdot 10^{-2}$ \\
			SPCE         & $1.0 \cdot 10^{-1}$ & $5.1 \cdot 10^{-2}$ & $4.6 \cdot 10^{-2}$ & $3.9 \cdot 10^{-2}$ \\
			HGP          & $1.4 \cdot 10^{-1}$ & $1.4 \cdot 10^{-1}$ & $1.2 \cdot 10^{-1}$ & $1.3 \cdot 10^{-1}$ \\
			HGP$_{\log}$ & $6.5 \cdot 10^{-2}$ & $4.8 \cdot 10^{-2}$ & $2.6 \cdot 10^{-2}$ & $2.7 \cdot 10^{-2}$ \\
			\hline
		\end{tabular}
		\caption{Seismic frame: Relative error on the optimal cost $\mathfrak{c}\prt{\ve{d}^{\ast}}$ after Eq.~\eqref{eq:error} for the median solution (over $15$ repetitions) for each method and experimental design size $N_{\textrm{ED}}$.}
		\label{tab:Ex5:error}
	\end{table}
	
	Figure~\ref{fig:Ex5:Density} shows the conditional distributions of the limit-state function at the reference design for $N_{\textrm{ED}} = 1{,}000$. The response is markedly positively skewed. GLaM, SPCE and HGP$_{\log}$ reproduce this skewness, and in particular the failure region $g_s < 0$ that governs the reliability constraint, whereas the plain HGP, restricted to a Gaussian conditional distribution, misrepresents the tail and consequently over-estimates the optimal cost. This example highlights that the assumed conditional-response family has a direct impact on the RBDO solution, and that the flexible emulators GLaM and SPCE capture the correct distribution without requiring such an assumption. As in the previous examples, the failure probability is available in closed form (or semi-analytically) for all four surrogates, so that each optimization completes in a few seconds despite the $2{,}006$-dimensional stochastic input.
	\begin{figure}[H]
		\centering
		\color{black}
		\subfloat[GLaM]{\includegraphics[width=0.48\textwidth]{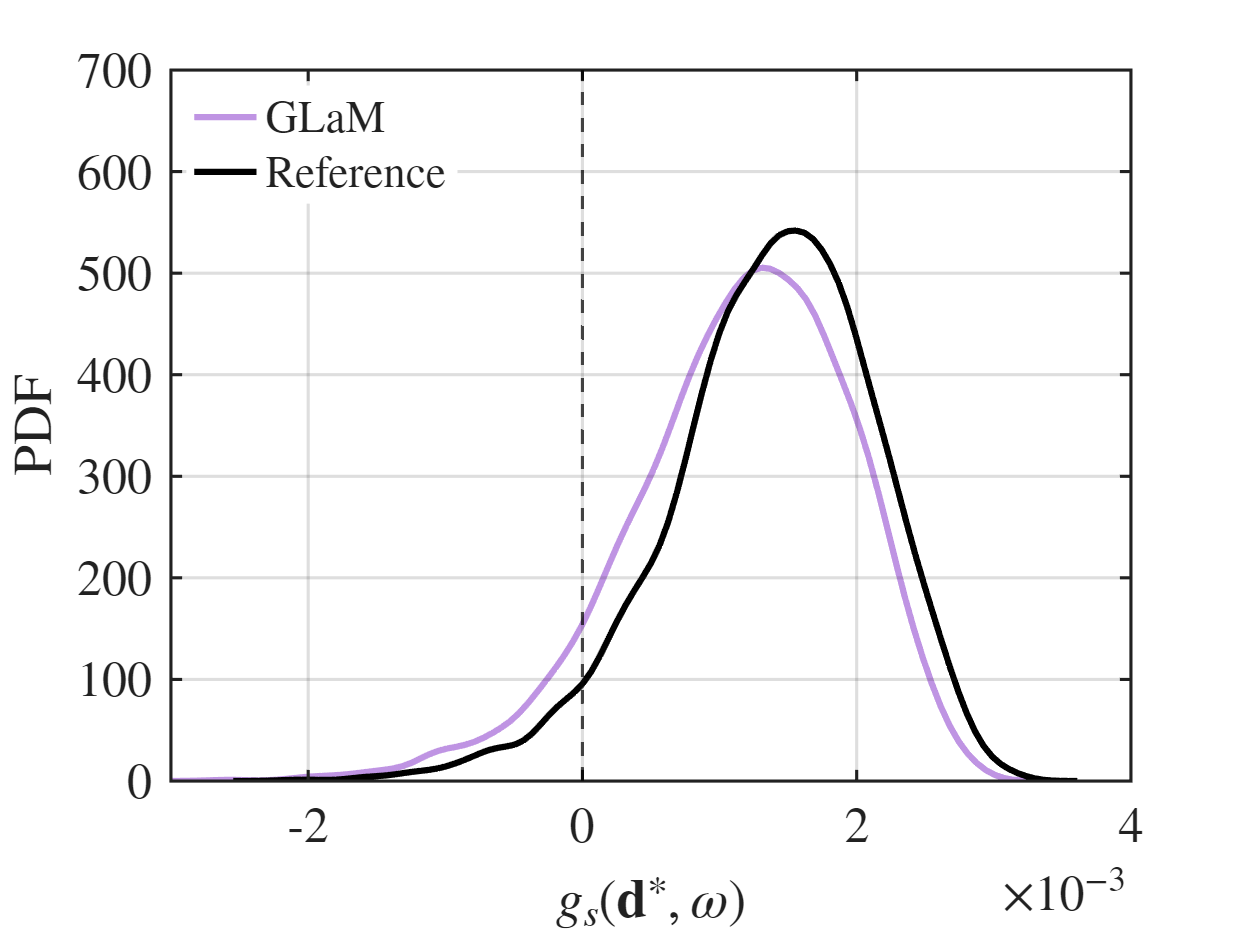}}
		\hfill
		\subfloat[SPCE]{\includegraphics[width=0.48\textwidth]{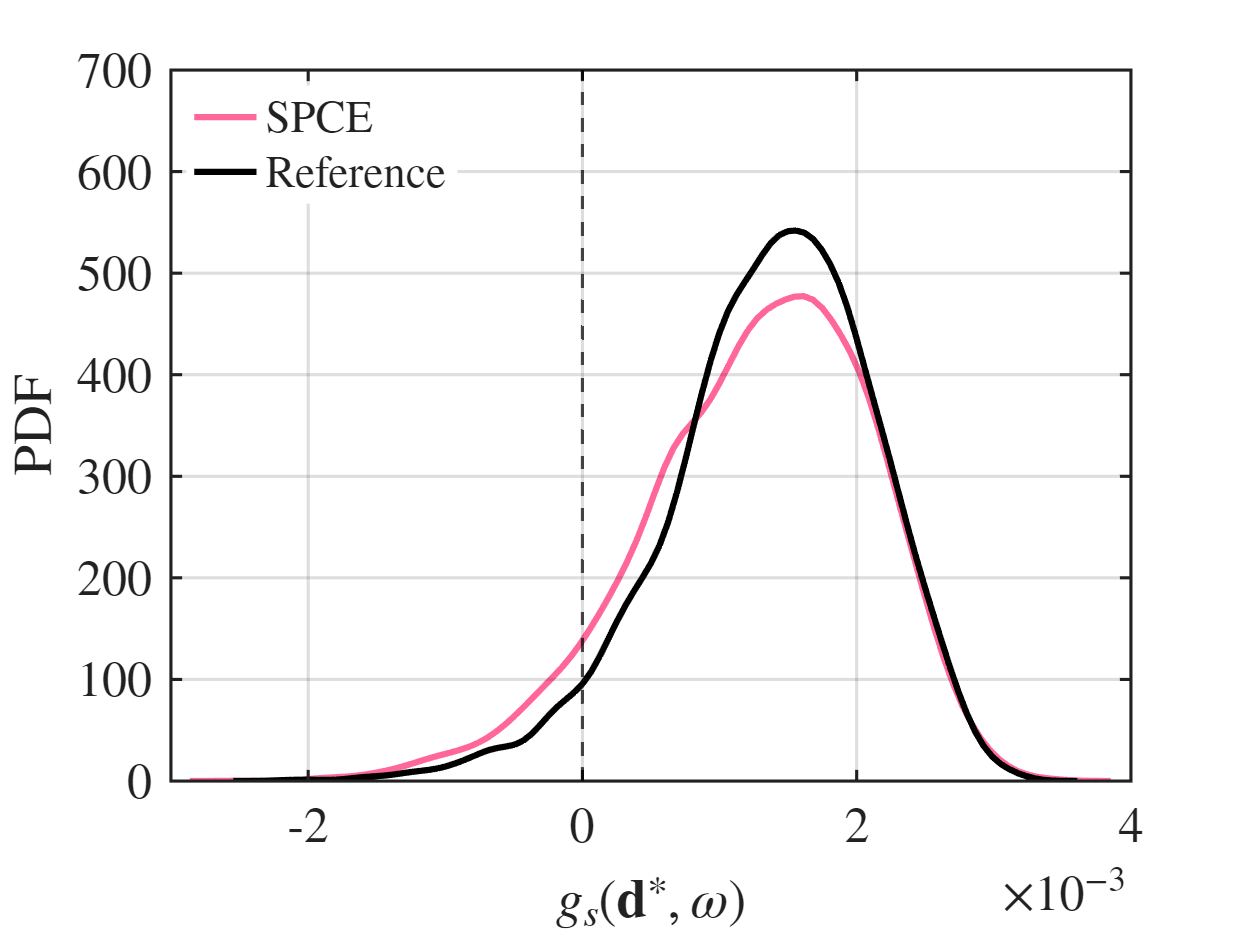}} \\
		\subfloat[HGP]{\includegraphics[width=0.48\textwidth]{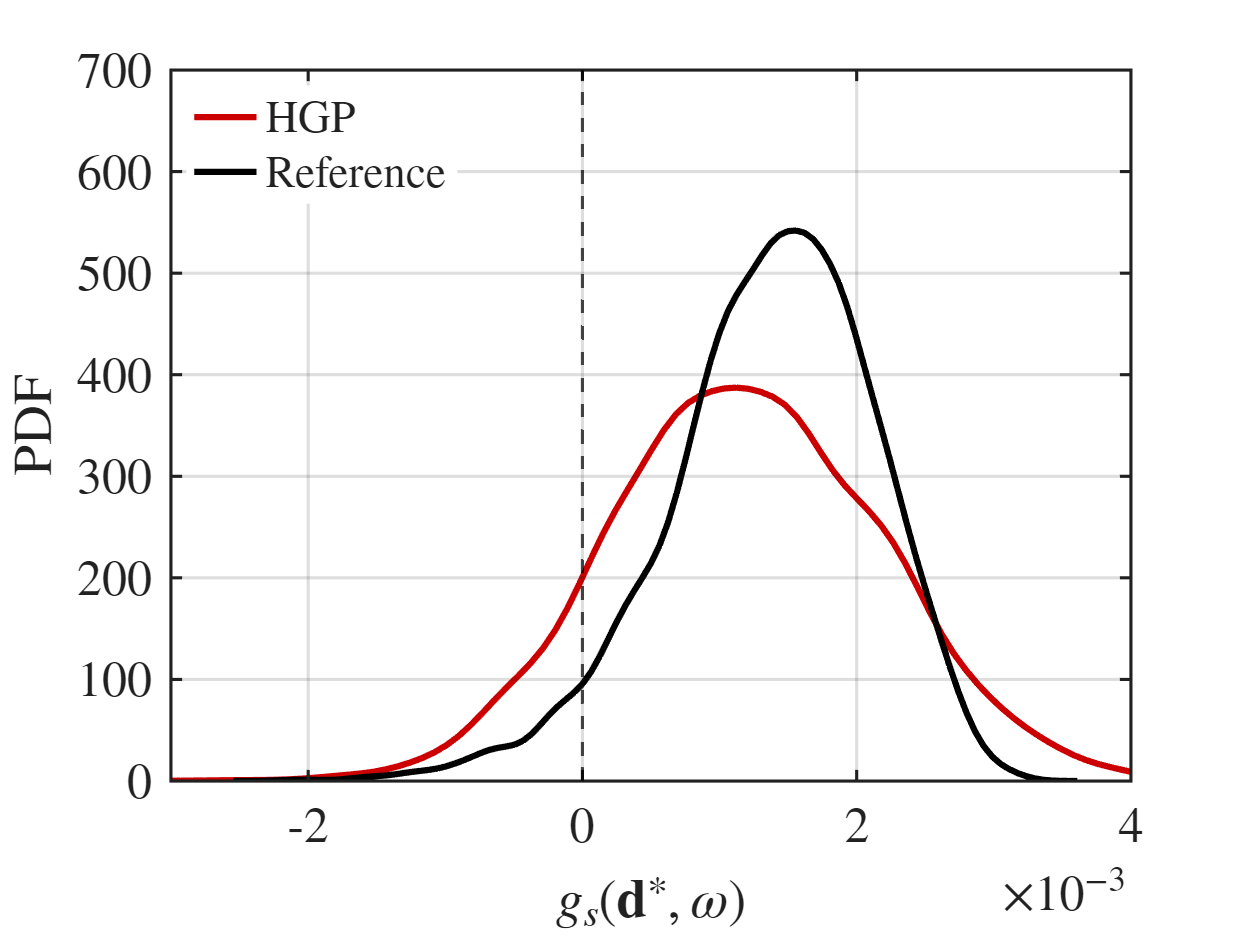}}
		\hfill
		\subfloat[HGP$_{\log}$]{\includegraphics[width=0.48\textwidth]{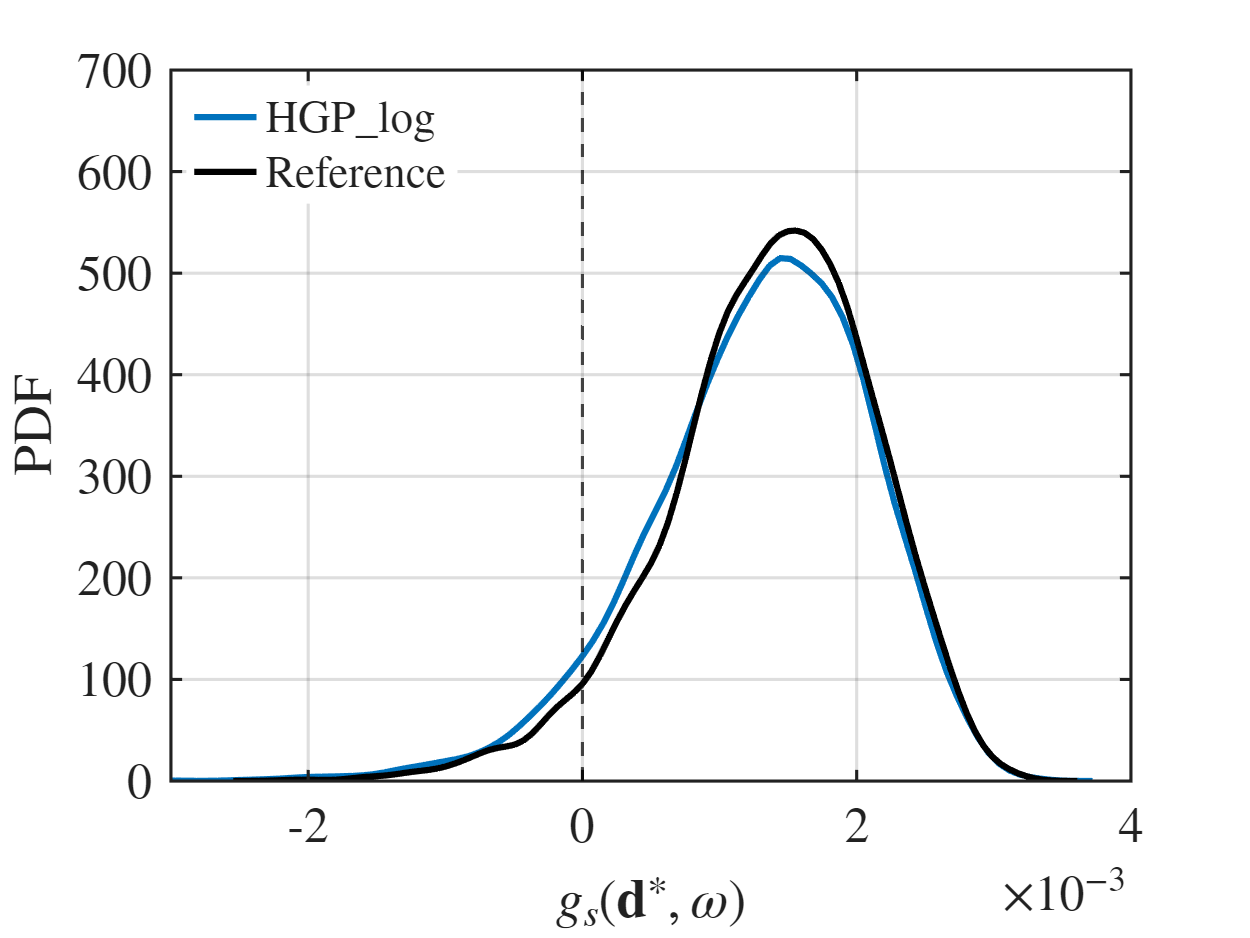}}
		\caption{Seismic frame: Conditional probability density functions of the limit-state function at the reference optimal design $\ve{d}^{\ast,\textrm{ref}}$ for $N_{\textrm{ED}} = 1{,}000$ (median-error surrogate over the $15$ repetitions). The dashed vertical line marks the failure threshold $g_s = 0$.}
		\label{fig:Ex5:Density}
	\end{figure}
}

{\color{black}		
	\section{Discussion}
	
	Despite the promising results, several limitations of the proposed approach must be acknowledged. Most notably, the quality of the RBDO solution is highly dependent on the accuracy of the stochastic emulator. Compared to Kriging, which exhibits relatively stable convergence behavior as the experimental design size increases, the stochastic emulators considered here are more challenging to train and appear less robust. In particular, increasing the size of the experimental design does not always lead to monotonic improvements in accuracy. Moreover, the influence of the limit-state function complexity and the nature of the input uncertainties on the training process is not yet fully understood. A more systematic investigation of these aspects is required and constitutes ongoing work on GLaM and SPCE methods that lies beyond the scope of the present paper, see e.g., \citet{HatstattSIAMUQ2026} for a Bayesian take on the training.
	
	Preliminary investigations suggest that the extent of the space in which the stochastic emulator is constructed has a significant impact on the quality of the resulting approximation. One promising strategy to improve robustness is therefore to restrict the design space initially and to expand it adaptively in a sequential manner to focus the modeling effort on the most relevant regions of the design space.
	
	Furthermore, alternative stochastic emulators developed in the deep learning community, such as conditional variational autoencoders \citep{Sohn2015,ZhangCVAE2021, Ramchadran2024} or normalizing flows \citep{Chen2018,Kobyzev2020,Liu2022}, as well as more recent mixture density networks developed for seismic response modeling \citep{Peng2025}, could in principle be employed. While these approaches may offer improved robustness and accuracy, they do not provide closed-form expressions for the conditional failure probabilities. As a result, the reliability constraints would need to be evaluated numerically, which may again increase the overall computational cost of the RBDO procedure and partially offset the benefits provided by SPCE and GLaM. Moreover, these mainstream machine learning techniques are known to work in the big data regime and typically require $\mathcal{O}(10^4-10^5)$ training points.
	
	Finally, the surrogate models considered in this work were constructed using a static experimental design. However, numerous studies have shown that active learning strategies can significantly enhance the efficiency of RBDO methods, particularly in high-dimensional settings where accurate global approximations are infeasible. By adaptively enriching the experimental design in regions that are most relevant for estimating conditional failure probabilities or quantiles, and where the design cost is favorable, active learning offers a natural and promising extension of the proposed framework. This direction will be explored in future work.
	
	\section{Conclusion}
	
	This paper introduced a novel framework for solving reliability-based design optimization (RBDO) problems by adopting a paradigm shift in which the deterministic limit-state function and the description of input uncertainties are merged into a unified stochastic simulator representation. Following this viewpoint, stochastic emulators are constructed from a limited number of model evaluations to approximate the conditional response of the limit-state function for any design encountered throughout the optimization process.
	
	The proposed framework is particularly suited for high-dimensional problems, where multiple and potentially complex sources of uncertainty must be accounted for. By constructing the emulator solely in the design space, the dimensionality of the surrogate modeling problem is substantially reduced. Moreover, the approach naturally accommodates stochastic simulators and situations where simulations are noisy or affected by uncertainties that cannot be explicitly modeled.
	
	In this work, two types of stochastic emulators were investigated, namely generalized lambda models and stochastic polynomial chaos expansions. For both approaches, the conditional failure probability or associated quantile can be evaluated analytically for a given design, resulting in a deterministic reliability constraint and eliminating the classical double-loop structure of RBDO. This enables the use of general-purpose optimization algorithms and leads to substantial computational savings compared with traditional Kriging-based double-loop approaches.
	
	The proposed methodology was demonstrated on four application examples with problem dimensionalities ranging from $5$ to $2{,}006$. No consistently
	superior scheme can be identified between GLaM and SPCE, as their relative
	performance varies with the size of the experimental design. In the higher-dimensional cases ($n_{\textrm{tot}} = 105$, $334$ and $2{,}006$), the proposed stochastic emulator-based strategies provide accurate solutions in settings where Kriging becomes either impractical or altogether infeasible: at $n_{\textrm{tot}} = 334$, a Kriging model can still be constructed but converges to a biased solution at a cost of several hours per run, whereas at $n_{\textrm{tot}} = 2{,}006$ it cannot be constructed at all. Both GLaM and SPCE successfully address these challenging cases, although their relative performance varies with the problem characteristics and the size of the experimental design, with neither method consistently outperforming the other.
	
	In the lower-dimensional example, the proposed approach remains viable, although its performance is not systematically superior to that of Kriging in terms of solution accuracy. Nevertheless, in all cases, the computational efficiency of the proposed method is significantly higher, with convergence achieved within fractions of a second to a few seconds, compared with tens of seconds to several hours for the Kriging-based double-loop approach.
	
	Overall, the results demonstrate the potential of stochastic emulators as a scalable alternative for reliability-based design optimization problems involving complex and high-dimensional uncertainties.
}

\nocite{SudretHDR}
\bibliographystyle{chicago}
\bibliography{RBDO_StoEmu}

\begin{thebibliography}{}

\bibitem[\protect\citeauthoryear{Agarwal and Renaud}{Agarwal and
  Renaud}{2004}]{Agarwal2004}
Agarwal, H. and J.~Renaud (2004).
\newblock Reliability-based design optimization using response surfaces in
  application to multidisciplinary systems.
\newblock {\em Engineering Optimization\/}~{\em 36\/}(3), 291--311.

\bibitem[\protect\citeauthoryear{Aoues and Chateauneuf}{Aoues and
  Chateauneuf}{2010}]{Aoues2010}
Aoues, Y. and A.~Chateauneuf (2010).
\newblock Benchmark study of numerical methods for reliability-based design
  optimization.
\newblock {\em Structural and Multidisciplinary Optimization\/}~{\em 41\/}(2),
  277--294.

\bibitem[\protect\citeauthoryear{Au and Beck}{Au and Beck}{2001}]{Au2001}
Au, S.~K. and J.~L. Beck (2001).
\newblock Estimation of small failure probabilities in high dimensions by
  subset simulation.
\newblock {\em Probabilistic Engineering Mechanics\/}~{\em 16\/}(4), 263--277.

\bibitem[\protect\citeauthoryear{Au and Patelli}{Au and
  Patelli}{2016}]{Au2016b}
Au, S.-K. and E.~Patelli (2016).
\newblock Rare event simulation in finite-infinite dimensional space.
\newblock {\em Reliability Engineering and System Safety\/}~{\em 148}, 67--77.

\bibitem[\protect\citeauthoryear{Billings}{Billings}{2013}]{Billings2013}
Billings, S.~A. (2013).
\newblock {\em {Nonlinear system identification: NARMAX methods in the time,
  frequency, and spatio-temporal domains}}.
\newblock John Wiley {\&} Sons.

\bibitem[\protect\citeauthoryear{Boroson and Missoum}{Boroson and
  Missoum}{2017}]{Boroson2017}
Boroson, E. and S.~Missoum (2017).
\newblock Stochastic optimization of nonlinear energy sinks.
\newblock {\em Structural and Multidisciplinary Optimization\/}~{\em 55},
  633--646.

\bibitem[\protect\citeauthoryear{Chateauneuf and Aoues}{Chateauneuf and
  Aoues}{2008}]{Chateauneuf2008}
Chateauneuf, A. and Y.~Aoues (2008).
\newblock {\em Structural design optimization considering uncertainties},
  Chapter~9, pp.\  217--246.
\newblock Taylor \& Francis.

\bibitem[\protect\citeauthoryear{Chen, Li, Chen, Wang, Pu, and Duke}{Chen
  et~al.}{2018}]{Chen2018}
Chen, C., C.~Li, L.~Chen, W.~Wang, Y.~Pu, and L.~C. Duke (2018).
\newblock Continuous-time flows for efficient inference and density estimation.
\newblock In {\em Proceedings of the 35th International Conference on Machine
  Learning}, Volume~80 of {\em Proceedings of Machine Learning Research}, pp.\
  824--833.

\bibitem[\protect\citeauthoryear{Chen, Hasselman, and Neil}{Chen
  et~al.}{1997}]{Chen1997}
Chen, X., K.~Hasselman, T., and D.~J. Neil (1997).
\newblock Reliability based structural design optimization for practical
  applications.
\newblock In {\em 38th Structures, Structural Dynamics, and Materials
  Conference}, pp.\  2724--2732.

\bibitem[\protect\citeauthoryear{Chopra}{Chopra}{2017}]{chopra2017}
Chopra, A.~K. (2017).
\newblock {\em Dynamics of Structures: Theory and Applications to Earthquake
  Engineering\/} (5th ed.).
\newblock Hoboken, NJ: Pearson.

\bibitem[\protect\citeauthoryear{Clark, Bae, and Forster}{Clark
  et~al.}{2020}]{Clark2020}
Clark, D.~L., H.~Bae, and E.~E. Forster (2020).
\newblock {G}aussian surrogate dimension reduction for efficient
  reliability-based design optimization.
\newblock In {\em AIAA SciTech Forum}, January 6-20, 2020, Orlanda, Florida,
  USA.

\bibitem[\protect\citeauthoryear{Clough and Penzien}{Clough and
  Penzien}{1993}]{cloughpenzien1993}
Clough, R.~W. and J.~Penzien (1993).
\newblock {\em Dynamics of Structures\/} (2nd ed.).
\newblock New York: McGraw-Hill.

\bibitem[\protect\citeauthoryear{Du and Chen}{Du and Chen}{2004}]{Du2004}
Du, X. and W.~Chen (2004).
\newblock Sequential optimization and reliability assessment method for
  efficient probabilistic design.
\newblock {\em Journal of Mechanical Design\/}~{\em 126\/}(2), 225--233.

\bibitem[\protect\citeauthoryear{Dubourg, Sudret, and Bourinet}{Dubourg
  et~al.}{2011}]{Dubourg2011}
Dubourg, V., B.~Sudret, and J.-M. Bourinet (2011).
\newblock Reliability-based design optimization using {Kriging} and subset
  simulation.
\newblock {\em Structural and Multidisciplinary Optimization\/}~{\em 44\/}(5),
  673--690.

\bibitem[\protect\citeauthoryear{Faes and Valdebenito}{Faes and
  Valdebenito}{2020}]{Faes2020}
Faes, M.~G. and M.~A. Valdebenito (2020).
\newblock Fully decoupled reliabiliy-based design optimization of structural
  systems subject to uncertain loads.
\newblock {\em Computer Methods in Applied Mechanics and Engineering\/}~{\em
  371}, 113313.

\bibitem[\protect\citeauthoryear{Grigoriu}{Grigoriu}{1998}]{Grigoriu1998}
Grigoriu, M. (1998).
\newblock Simulation of stationary non-{G}aussian translation processes.
\newblock {\em Journal of Engineering Mechanics\/}~{\em 124}, 121--126.

\bibitem[\protect\citeauthoryear{Hasofer and Lind}{Hasofer and
  Lind}{1974}]{Hasofer1974}
Hasofer, A.-M. and N.-C. Lind (1974, February).
\newblock Exact and invariant second moment code format.
\newblock {\em Journal of Engineering Mechanics (ASCE)\/}~{\em 100\/}(1),
  111--121.

\bibitem[\protect\citeauthoryear{Hatstatt, Moustapha, Marelli, and
  Sudret}{Hatstatt et~al.}{2026}]{HatstattSIAMUQ2026}
Hatstatt, A., M.~Moustapha, S.~Marelli, and B.~Sudret (2026).
\newblock Introducing {B}ayesian inference for stochastic polynomial chaos
  expansions.
\newblock In {\em SIAM Conf. on Uncertainty Quantification, Minneapolis (USA),
  March 22-25}.

\bibitem[\protect\citeauthoryear{Jerez, Jensen, and Beer}{Jerez
  et~al.}{2022}]{Jerez2022}
Jerez, D.~J., H.~A. Jensen, and M.~Beer (2022).
\newblock Reliability-based design optimization of structural systems under
  stochastic excitation: {A}n overview.
\newblock {\em Mechanical Systems and Signal Processing\/}~{\em 106}, 108397.

\bibitem[\protect\citeauthoryear{Jia and Taflanidis}{Jia and
  Taflanidis}{2013}]{Jia2013}
Jia, G. and A.~A. Taflanidis (2013).
\newblock Non-parametric stochastic subset optimization for optimal-reliability
  design problems.
\newblock {\em Comput. Struct.\/}~{\em 126}, 86--99.

\bibitem[\protect\citeauthoryear{Jiang, Zhang, Beer, Zhou, and Leng}{Jiang
  et~al.}{2024}]{Jiang2024}
Jiang, Y., X.~Zhang, M.~Beer, H.~Zhou, and Y.~Leng (2024).
\newblock An efficient method for reliability-based design optimization of
  structures under random excitation by mapping between reliability and
  operator norm.
\newblock {\em Reliability Engineering and Systems Safety\/}~{\em 245}, 109972.

\bibitem[\protect\citeauthoryear{Kanai}{Kanai}{1957}]{kanai1957}
Kanai, K. (1957).
\newblock Semi-empirical formula for the seismic characteristics of the ground.
\newblock {\em Bulletin of the Earthquake Research Institute, University of
  Tokyo\/}~{\em 35\/}(2), 309--325.

\bibitem[\protect\citeauthoryear{Kim, S., and Song}{Kim et~al.}{2024}]{Kim2024}
Kim, J., Y.~S., and J.~Song (2024).
\newblock Active learning-based optimization of structures under stochastic
  excitations with first-passage probability constraints.
\newblock {\em Engineering Structures\/}~{\em 307}, 117873.

\bibitem[\protect\citeauthoryear{Kim and Song}{Kim and Song}{2021}]{Kim2021}
Kim, J. and J.~Song (2021).
\newblock Reliability-based design optimization using quantile surrogates by
  adaptive gaussian process.
\newblock {\em Journal of Engineering Mechanics\/}~{\em 147}, 04021020--1 --
  16.

\bibitem[\protect\citeauthoryear{Kobyzev, Prince, and Brubaker}{Kobyzev
  et~al.}{2020}]{Kobyzev2020}
Kobyzev, I., S.~Prince, and M.~Brubaker (2020).
\newblock Normalizing flows: An introduction and review of current methods.
\newblock {\em IEEE transactions on pattern analysis and machine
  intelligence\/}~{\em 43\/}(11), 3964--3979.

\bibitem[\protect\citeauthoryear{Kroetz, Beck, Costa, Macedo, Marelli, and
  Sudret}{Kroetz et~al.}{2026}]{Kroetz2026}
Kroetz, H., A.~T. Beck, L.~G.~L. Costa, F.~C. Macedo, S.~Marelli, and B.~Sudret
  (2026).
\newblock Fragility analysis of structures under non-synoptic winds using
  stochastic emulators.
\newblock {\em Engineering Structures\/}~{\em 357}, 122515.

\bibitem[\protect\citeauthoryear{Kroetz, Moustapha, Beck, and Sudret}{Kroetz
  et~al.}{2020}]{KroetzRESS2020}
Kroetz, H., M.~Moustapha, A.~Beck, and B.~Sudret (2020).
\newblock A two-level {K}riging-based approach with active learning for solving
  time-variant risk optimization problems.
\newblock {\em Reliability Engineering \& System Safety\/}~{\em
  203\/}({107033}).

\bibitem[\protect\citeauthoryear{Lataniotis, Wicaksono, Marelli, and
  Sudret}{Lataniotis et~al.}{2026}]{UQdoc_22_105}
Lataniotis, C., D.~Wicaksono, S.~Marelli, and B.~Sudret (2026).
\newblock {UQLab user manual -- Kriging (Gaussian process modeling)}.
\newblock Technical report, Chair of Risk, Safety and Uncertainty
  Quantification, ETH Zurich, Switzerland.
\newblock Report UQLab-V2.2-105.

\bibitem[\protect\citeauthoryear{L\'azaro-Gredilla and
  Titsias}{L\'azaro-Gredilla and Titsias}{2011}]{Lazaro-Gredilla2011}
L\'azaro-Gredilla, M. and M.~K. Titsias (2011).
\newblock Variational heteroscedastic {G}aussian process regression.
\newblock In {\em 28th International Conference on Machine Learning, Bellevue,
  WA, USA, June 28 - July 2, 2011}.

\bibitem[\protect\citeauthoryear{Lee and Rahman}{Lee and
  Rahman}{2022}]{Lee2022}
Lee, D. and S.~Rahman (2022).
\newblock Reliability-based design optimization under dependent random
  variables by a generalized polynomial chaos expansion.
\newblock {\em Structural and Multidisciplinary Optimization\/}~{\em 65\/}(21).

\bibitem[\protect\citeauthoryear{Lee}{Lee}{2025}]{Lee2025}
Lee, U. (2025).
\newblock Reliability-based design optimization ({RBDO}) framework for
  high-energy laser weapons ({HELWs}) using artificial neural network ensemble.
\newblock {\em Expert Systems with Applications\/}~{\em 287}, 128202.

\bibitem[\protect\citeauthoryear{Li and Wang}{Li and Wang}{2022}]{Li2022}
Li, M. and Z.~Wang (2022).
\newblock Deep reliability learning with latent adaptation for design
  optimization under uncertainty.
\newblock {\em Computer Methods in Applied Mechanics and Engineering\/}~{\em
  397}, 115130.

\bibitem[\protect\citeauthoryear{Li, Schofield, and G\"onen}{Li
  et~al.}{2019}]{Li2019}
Li, Y., O.~Schofield, and M.~G\"onen (2019).
\newblock A tutorial on {D}irichlet process mixture modeling.
\newblock {\em Journal of Mathematical Psychology\/}~{\em 91}.

\bibitem[\protect\citeauthoryear{Liang, Mourelatos, and Tu}{Liang
  et~al.}{2004}]{Liang2004}
Liang, J., Z.~Mourelatos, and J.~Tu (2004).
\newblock A single-loop method for reliability-based design optimization.
\newblock In {\em {Proc. DETC'04 ASME 2004 Design engineering technical
  conferences and computers and information in engineering conference, Sept.28
  - Oct. 2, 2004, Salt Lake City, Utah, USA}}.

\bibitem[\protect\citeauthoryear{Ling, Lu, and Zhang}{Ling
  et~al.}{2021}]{Ling2021}
Ling, C., Z.~Lu, and W.~Zhang (2021).
\newblock Bayesian support vector regression for reliability-based design
  optimization.
\newblock {\em AIAA Journals\/}~{\em 55}, 5141--5157.

\bibitem[\protect\citeauthoryear{Liu, Gong, and Liu}{Liu
  et~al.}{2022}]{Liu2022}
Liu, X., C.~Gong, and Q.~Liu (2022).
\newblock Flow straight and fast: Learning to generate and transfer data with
  rectified flow.
\newblock In {\em Tenth International Conference on Learning Representations,
  ICLR2022, Virtual Event, April 25-29, 2022}.

\bibitem[\protect\citeauthoryear{L\"uthen}{L\"uthen}{2022}]{LuethenThesis}
L\"uthen, N. (2022).
\newblock {\em Sparse spectral surrogate models for deterministic and
  stochastic computer simulations}.
\newblock Ph.\ D. thesis, ETH Z\"urich, Z\"urich, Switzerland.

\bibitem[\protect\citeauthoryear{Lüthen, Zhu, Marelli, and Sudret}{Lüthen
  et~al.}{2026a}]{UQdoc_22_120}
Lüthen, N., X.~Zhu, S.~Marelli, and B.~Sudret (2026a).
\newblock {UQLab user manual - Generalized Lambda Model}.
\newblock Technical report, Chair of Risk, Safety and Uncertainty
  Quantification, ETH Zurich, Switzerland.
\newblock Report UQLab-V2.2-120.

\bibitem[\protect\citeauthoryear{Lüthen, Zhu, Marelli, and Sudret}{Lüthen
  et~al.}{2026b}]{UQdoc_22_121}
Lüthen, N., X.~Zhu, S.~Marelli, and B.~Sudret (2026b).
\newblock {UQLab user manual - Stochastic PCE}.
\newblock Technical report, Chair of Risk, Safety and Uncertainty
  Quantification, ETH Zurich, Switzerland.
\newblock Report UQLab-V2.2-121.

\bibitem[\protect\citeauthoryear{Marelli and Sudret}{Marelli and
  Sudret}{2014}]{Marelli2014a}
Marelli, S. and B.~Sudret (2014).
\newblock {UQLab}: a framework for uncertainty quantification in {MATLAB}.
\newblock In {\em MascotNum Annual Workshop, Zurich}.

\bibitem[\protect\citeauthoryear{Melchers and Beck}{Melchers and
  Beck}{2018}]{Melchers2018}
Melchers, R.~E. and A.~T. Beck (2018).
\newblock {\em Structural reliability analysis and prediction}.
\newblock John Wiley \& Sons.

\bibitem[\protect\citeauthoryear{Moustapha, Marelli, , and Sudret}{Moustapha
  et~al.}{2026}]{UQdoc_22_115}
Moustapha, M., S.~Marelli, , and B.~Sudret (2026).
\newblock {UQLab user manual -- Reliability-Based Design Optimization}.
\newblock Technical report, Chair of Risk, Safety and Uncertainty
  Quantification, ETH Zurich, Switzerland.
\newblock Report UQLab-V2.2-115.

\bibitem[\protect\citeauthoryear{Moustapha and Sudret}{Moustapha and
  Sudret}{2019}]{MoustaphaSMO2019}
Moustapha, M. and B.~Sudret (2019).
\newblock Surrogate-assisted reliability-based design optimization: a survey
  and a unified modular framework.
\newblock {\em Structural and Multidisciplinary Optimization\/}~{\em 60},
  2157--2176.

\bibitem[\protect\citeauthoryear{Moustapha, Sudret, Bourinet, and
  Guillaume}{Moustapha et~al.}{2016}]{MoustaphaSMO2016}
Moustapha, M., B.~Sudret, J.-M. Bourinet, and B.~Guillaume (2016).
\newblock Quantile-based optimization under uncertainties using adaptive
  {K}riging surrogate models.
\newblock {\em Structural and Multidisciplinary Optimization\/}~{\em 54\/}(6),
  1403--1421.

\bibitem[\protect\citeauthoryear{Nikolaidis and Burdisso}{Nikolaidis and
  Burdisso}{1988}]{Nikolaidis1988}
Nikolaidis, E. and R.~Burdisso (1988).
\newblock Reliability-based optimization: {A} safety index approach.
\newblock {\em Comput. Struct.\/}~{\em 28\/}(6), 781--788.

\bibitem[\protect\citeauthoryear{Papaioannou, Betz, Zwirglmaier, and
  Straub}{Papaioannou et~al.}{2015}]{Papaioannou2015}
Papaioannou, I., W.~Betz, K.~Zwirglmaier, and D.~Straub (2015).
\newblock {MCMC} algorithms for subset simulation.
\newblock {\em Probabilistic Engineering Mechanics\/}~{\em 41}, 89 -- 103.

\bibitem[\protect\citeauthoryear{Papaioannou, Papadimitriou, and
  Straub}{Papaioannou et~al.}{2016}]{Papaioannou2016}
Papaioannou, I., C.~Papadimitriou, and D.~Straub (2016).
\newblock Sequential importance sampling for structural reliability analysis.
\newblock {\em Struct. Saf.\/}~{\em 62}, 66--75.

\bibitem[\protect\citeauthoryear{Park and Lee}{Park and Lee}{2023}]{Park2023}
Park, J.~W. and I.~Lee (2023).
\newblock A new framework for efficient sequential sampling-based {RBDO} using
  space mapping.
\newblock {\em Journal of Mechanical Design\/}~{\em 145}, 031702.

\bibitem[\protect\citeauthoryear{Peng, Taflanidis, and Zhang}{Peng
  et~al.}{2026}]{Peng2025}
Peng, H., A.~A. Taflanidis, and J.~Zhang (2026).
\newblock Accelerating seismic response distribution estimation with scalable
  mixture density network stochastic surrogate models.
\newblock {\em Earthquake Engineering \& Structural Dynamics\/}~{\em 55},
  721--742.

\bibitem[\protect\citeauthoryear{Pires, Moustapha, Marelli, and Sudret}{Pires
  et~al.}{2025a}]{PiresSS2025b}
Pires, A.~V., M.~Moustapha, S.~Marelli, and B.~Sudret (2025a).
\newblock {AL-SPCE} - {R}eliability analysis for nondeterministic models using
  stochastic polynomial chaos expansions and active learning.
\newblock {\em Structural Safety\/}.
\newblock (submitted).

\bibitem[\protect\citeauthoryear{Pires, Moustapha, Marelli, and Sudret}{Pires
  et~al.}{2025b}]{PiresSS2025a}
Pires, A.~V., M.~Moustapha, S.~Marelli, and B.~Sudret (2025b).
\newblock Reliability analysis for nondeterministic limit-states using
  stochastic emulators.
\newblock {\em Structural Safety\/}~{\em 117}, 102621.

\bibitem[\protect\citeauthoryear{Ramchandran, Tikhonov, Lönnroth, Tiikkainen,
  and Lähdesmäki}{Ramchandran et~al.}{2024}]{Ramchadran2024}
Ramchandran, S., G.~Tikhonov, O.~Lönnroth, P.~Tiikkainen, and H.~Lähdesmäki
  (2024).
\newblock Learning conditional variational autoencoders with missing
  covariates.
\newblock {\em Pattern Recognition\/}~{\em 147}, 110113.

\bibitem[\protect\citeauthoryear{Rodrigues, Spence, and Beck}{Rodrigues
  et~al.}{2026}]{Rodrigues2026}
Rodrigues, I.~D., S.~M.~J. Spence, and A.~T. Beck (2026).
\newblock Incorporating statistical uncertainty into metamodel-driven
  performance-based risk optimization for seismic design.
\newblock {\em Engineering Structures\/}~{\em 348}, 121744.

\bibitem[\protect\citeauthoryear{Rubinstein and Kroese}{Rubinstein and
  Kroese}{2016}]{Rubinstein2016}
Rubinstein, R.~Y. and D.~P. Kroese (2016, November).
\newblock {\em Simulation and the {Monte} {Carlo} method}.
\newblock John Wiley {\&} Sons, Inc.

\bibitem[\protect\citeauthoryear{Sch\"ar, Marelli, and Sudret}{Sch\"ar
  et~al.}{2024}]{SchaerMSSP2024}
Sch\"ar, S., S.~Marelli, and B.~Sudret (2024).
\newblock Emulating the dynamics of complex systems using autoregressive models
  on manifolds {(mNARX)}.
\newblock {\em Mechanical Systems and Signal Processing\/}~{\em 208\/}(110956).

\bibitem[\protect\citeauthoryear{Schär, Marelli, and Sudret}{Schär
  et~al.}{2025}]{schaer_fnarx}
Schär, S., S.~Marelli, and B.~Sudret (2025).
\newblock Surrogate modeling with functional nonlinear autoregressive models
  ($\mathcal{F}$\nobreakdashes-narx).
\newblock {\em Reliability Engineering \& System Safety\/}~{\em 264\/}(111276).

\bibitem[\protect\citeauthoryear{Schär, Marelli, and Sudret}{Schär
  et~al.}{2026}]{schaer_mnarxp}
Schär, S., S.~Marelli, and B.~Sudret (2026).
\newblock mnarx\textsuperscript{+}: A surrogate model for complex dynamical
  systems using manifold-narx and automatic feature selection.
\newblock {\em Computer Methods in Applied Mechanics and Engineering\/}~{\em
  449\/}(118550).

\bibitem[\protect\citeauthoryear{Sohn, Lee, and Yan}{Sohn
  et~al.}{2015}]{Sohn2015}
Sohn, K., H.~Lee, and X.~Yan (2015).
\newblock Learning structured output representation using deep conditional
  generative models.
\newblock In C.~Cortes, N.~Lawrence, D.~Lee, M.~Sugiyama, and R.~Garnett
  (Eds.), {\em Advances in Neural Information Processing Systems}, Volume~28,
  pp.\  1 -- 9. Curran Associates, Inc.

\bibitem[\protect\citeauthoryear{Spall}{Spall}{2003}]{Spall2003}
Spall, J.~C. (2003).
\newblock {\em Introduction to stochastic search and optimization:
  {E}stimation, simulation and control}.
\newblock John Wiley \& Sons.

\bibitem[\protect\citeauthoryear{Sudret}{Sudret}{2007}]{SudretHDR}
Sudret, B. (2007).
\newblock {\em Uncertainty propagation and sensitivity analysis in mechanical
  models -- Contributions to structural reliability and stochastic spectral
  methods}.
\newblock Universit\'e Blaise Pascal, Clermont-Ferrand, France.
\newblock Habilitation \`a diriger des recherches, 173 pages.

\bibitem[\protect\citeauthoryear{Sudret}{Sudret}{2015}]{Sudret2015a}
Sudret, B. (2015).
\newblock {Polynomials chaos expansions and stochastic finite element methods}.
\newblock In K.-K. Phoon and J.~Ching (Eds.), {\em Risk Reliab. Geotech. Eng.},
  Chapter~6. Taylor and Francis.

\bibitem[\protect\citeauthoryear{Taflanidis and Beck}{Taflanidis and
  Beck}{2008}]{Taflanidis2008}
Taflanidis, A.~A. and J.~L. Beck (2008).
\newblock Stochastic subset optimization for optimal reliability problems.
\newblock {\em Prob. Eng. Mech\/}~{\em 23}, 324--338.

\bibitem[\protect\citeauthoryear{Tajimi}{Tajimi}{1960}]{tajimi1960}
Tajimi, H. (1960).
\newblock A statistical method of determining the maximum response of a
  building structure during an earthquake.
\newblock In {\em Proceedings of the 2nd World Conference on Earthquake
  Engineering}, Volume~2, Tokyo, Japan, pp.\  781--797.

\bibitem[\protect\citeauthoryear{Thedy and Liao}{Thedy and
  Liao}{2023}]{Thedy2023}
Thedy, J. and K.-W. Liao (2023).
\newblock Reliability‑based structural optimization using adaptive neural
  network multisphere importance sampling.
\newblock {\em Structural and Multidisciplinary Optimization\/}~{\em
  66\/}(119).

\bibitem[\protect\citeauthoryear{Thompson, McMullen, Nemani, Hu, and
  Hu}{Thompson et~al.}{2025}]{Thompson2025}
Thompson, T., M.~McMullen, V.~Nemani, Z.~Hu, and C.~Hu (2025).
\newblock A comparative study of acquisition functions for active learning
  {K}riging in reliability-based design optimization.
\newblock {\em Structural and Multidisciplinary Optimization\/}~{\em 68\/}(51).

\bibitem[\protect\citeauthoryear{Tu, Choi, and Park}{Tu et~al.}{1999}]{Tu1999}
Tu, J., K.~K. Choi, and Y.~H. Park (1999).
\newblock A new study on reliability-based design optimization.
\newblock {\em Journal of Mechanical Design\/}~{\em 121}, 557 -- 564.

\bibitem[\protect\citeauthoryear{Valdebenito and Schu\"eller}{Valdebenito and
  Schu\"eller}{2010}]{Valdebenito2010}
Valdebenito, A.~M. and G.~I. Schu\"eller (2010).
\newblock A survey on approaches for reliability-based optimization.
\newblock {\em Structural and Multidisciplinary Optimization\/}~{\em 42},
  645--663.

\bibitem[\protect\citeauthoryear{Wang and Song}{Wang and Song}{2016}]{Wang2016}
Wang, Z. and J.~Song (2016).
\newblock Cross-entropy-based adaptive importance sampling using von
  {M}ises-{F}isher mixture for high dimensional reliability.
\newblock {\em Structural Safety\/}~{\em 59}, 42--52.

\bibitem[\protect\citeauthoryear{Xiao, Yuan, and Zhan}{Xiao
  et~al.}{2022}]{Xiao2022}
Xiao, N.-C., K.~Yuan, and H.~Zhan (2022).
\newblock System reliability analysis based on dependent {K}riging predictions
  and parallel learning strategy.
\newblock {\em Reliability Engineering and System Safety\/}~{\em 218}, 108083.

\bibitem[\protect\citeauthoryear{Xiu and Karniadakis}{Xiu and
  Karniadakis}{2002}]{Xiu2002}
Xiu, D. and G.~E. Karniadakis (2002).
\newblock {The Wiener-Askey polynomial chaos for stochastic differential
  equations}.
\newblock {\em SIAM Journal on Scientific Computing\/}~{\em 24\/}(2), 619--644.

\bibitem[\protect\citeauthoryear{Zhang, Barabano, and Jin}{Zhang
  et~al.}{2020}]{ZhangCVAE2021}
Zhang, C., R.~Barabano, and B.~Jin (2020).
\newblock Conditional variational autoencoder for learned image reconstruction.
\newblock {\em Computation\/}~{\em 9\/}(114), 1--23.

\bibitem[\protect\citeauthoryear{Zhang, Aoues, Lemosse, and {Souza de
  Cursi}}{Zhang et~al.}{2021}]{Zhang2021}
Zhang, H., Y.~Aoues, D.~Lemosse, and E.~{Souza de Cursi} (2021).
\newblock A single loop approach with adaptive sampling and surrogate {K}riging
  for reliability-based design optimization.
\newblock {\em Engineering Optimization\/}~{\em 53}, 1450 -- 1466.

\bibitem[\protect\citeauthoryear{Zhang, Taflanidis, and Medina}{Zhang
  et~al.}{2017}]{Zhang2017}
Zhang, J., A.~A. Taflanidis, and J.~C. Medina (2017).
\newblock {Sequential approximate optimization for design under uncertainty
  problems utilizing Kriging metamodeling in augmented input space}.
\newblock {\em Computer Methods in Applied Mechanics and Engineering\/}~{\em
  315}, 369--395.

\bibitem[\protect\citeauthoryear{Zhang, Liu, Wu, Xu, and Jiang}{Zhang
  et~al.}{2024}]{Zhang2024}
Zhang, Z., H.~Liu, T.~Wu, J.~Xu, and C.~Jiang (2024).
\newblock A novel reliability-based design optimization method through
  instant-based transfer learning.
\newblock {\em Computer Methods in Applied Mechanics and Engineering\/}~{\em
  432}, 117388.

\bibitem[\protect\citeauthoryear{Zhou and Lu}{Zhou and Lu}{2019}]{Zhou2019AIAA}
Zhou, Y. and Z.~Lu (2019).
\newblock Active polynomial chaos expansion for reliability-based design
  optimization.
\newblock {\em AIAA Journals\/}~{\em 57}, 5431--5446.

\bibitem[\protect\citeauthoryear{Zhu}{Zhu}{2023}]{ZhuThesis}
Zhu, X. (2023).
\newblock {\em Surrogate modeling for stochastic simulators using statistical
  approaches}.
\newblock Ph.\ D. thesis, ETH Z\"urich, Z\"urich, Switzerland.

\bibitem[\protect\citeauthoryear{Zhu, Broccardo, and Sudret}{Zhu
  et~al.}{2023}]{ZhuPEM2023}
Zhu, X., M.~Broccardo, and B.~Sudret (2023).
\newblock Seismic fragility analysis using stochastic polynomial chaos
  expansions.
\newblock {\em Probabilistic Engineering Mechanics\/}~{\em 72\/}(103413),
  1--13.

\bibitem[\protect\citeauthoryear{Zhu and Sudret}{Zhu and
  Sudret}{2020}]{Zhu2020}
Zhu, X. and B.~Sudret (2020).
\newblock Replication-based emulation of the response distribution of
  stochastic simulators using generalized lambda distributions.
\newblock {\em International Journal for Uncertainty Quantification\/}~{\em
  10\/}(3), 249--275.

\bibitem[\protect\citeauthoryear{Zhu and Sudret}{Zhu and
  Sudret}{2021}]{ZhuSIAM2021}
Zhu, X. and B.~Sudret (2021).
\newblock Emulation of stochastic simulators using generalized lambda models.
\newblock {\em SIAM/ASA Journal on Uncertainty Quantification\/}~{\em 9\/}(4),
  1345--1380.

\bibitem[\protect\citeauthoryear{Zhu and Sudret}{Zhu and
  Sudret}{2023}]{ZhuStoPCE2023}
Zhu, X. and B.~Sudret (2023).
\newblock Stochastic polynomial chaos expansions to emulate stochastic
  simulators.
\newblock {\em International Journal for Uncertainty Quantification\/}~{\em
  13\/}(2), 31--52.

\end{thebibliography}
\end{document}